\documentclass[prx,aps,groupedaddress,twocolumn,superscriptaddress,longbibliography]{revtex4-1}
\usepackage{graphicx}
\usepackage{amsmath}
\usepackage{amssymb}

\usepackage[dvipsnames]{xcolor}
\usepackage{changepage}
\usepackage{tcolorbox}
\usepackage{pifont}
\usepackage{enumitem}
\usepackage{multirow}
\usepackage[export]{adjustbox}
\usepackage{braket}

\usepackage[T1]{fontenc}

\newlist{todolist}{itemize}{2}
\setlist[todolist]{label=$\square$}

\usepackage[colorlinks=true,linkcolor=blue]{hyperref}
\usepackage{cleveref}
\hypersetup{
	colorlinks,
	linkcolor={blue!75!black},
	citecolor={blue!75!black},
	urlcolor={blue!75!black},
}

\newcommand{\Opot}{\mathcal{O}_{\mathrm{pot}}}
\newcommand{\Okins}{\mathcal{O}_{\mathrm{kin},\square}}
\newcommand{\Opots}{\mathcal{O}_{\mathrm{pot},\square}}

\graphicspath{{figures/}}
\begin{document}
\title{Fate of many-body localization in an Abelian lattice gauge theory}

\author{Indrajit Sau}
\email{tpis2@iacs.res.in}
\affiliation{School of Physical Sciences, Indian Association for the Cultivation of Science, Jadavpur, Kolkata 700032, India}

\author{Debasish Banerjee}
\email{D.Banerjee@soton.ac.uk}
\affiliation{Theory Division, Saha Institute of Nuclear Physics, 1/AF Bidhannagar, Kolkata 700064, India}
\affiliation{School of Physics and Astronomy, University of Southampton, University Road, SO17 1BJ, UK}

\author{Arnab Sen}
\email{tpars@iacs.res.in}
\affiliation{School of Physical Sciences, Indian Association for the Cultivation of Science, Jadavpur, Kolkata 700032, India}

\date{\today}
\begin{abstract}
  We address the fate of many-body localization (MBL) of mid-spectrum eigenstates of a matter-free $U(1)$ quantum-link gauge theory Hamiltonian with random couplings on ladder geometries. Apart from level spacing distribution indicators like disorder-averaged mean level spacing, we also consider an intensive estimator $\mathcal{D} \in [0,1/4]$, which acts as a measure of elementary plaquettes on the lattice that are active or inert in mid-spectrum eigenstates as well as the concentration of these eigenstates in Fock space, with $\mathcal{D}$ equal to its maximum value of $1/4$ for Fock states in the electric flux basis. We calculate its distribution, $p(\mathcal{D})$, for $L_x \times L_y$ lattices, with $L_y=2$ and $4$, as a function of (a dimensionless) disorder strength $\alpha$ ($\alpha=0$ implies zero disorder) using exact diagonalization in many disorder realizations. Although finite-size estimators based on level spacings do not give a reliable critical disorder strength, $\alpha_c(L_y)$, beyond which MBL prevails as $L_x \rightarrow \infty$; a different estimator based on the skewness of $p(\mathcal{D})$ gives $\alpha_c(L_y=2)=31.04 \pm 0.54$ using data for $L_x \leq 14$ due to faster convergence. $p(\mathcal{D})$ for wider ladders with $L_y=4$ show a lower tendency to localize, suggesting a lack of MBL in two dimensions. A remarkable observation is the resolution of the (monotonic) infinite-temperature autocorrelation function of single plaquette diagonal operators in typical high-energy Fock states into a plethora of emergent timescales of increasing spatio-temporal heterogeneity as the disorder is increased. At intermediate $\alpha$ as well as for $\alpha$ slightly below $\alpha_c (L_y)$, a fraction of randomly selected initial Fock states display striking oscillatory temporal behavior of such plaquette operators in spatial regions formed out of connected plaquettes.
\end{abstract}

\maketitle

\section{Introduction}
Interacting many-body lattice models with finite-dimensional local Hilbert spaces are expected to follow the eigenstate thermalization hypothesis (ETH) which states that individual energy eigenstates of such systems have ``thermal'' expectation values for local observables~\cite{Deutsch1991, Srednicki1994, Rigol2008, Alessio2016} with the corresponding temperature determined by the energy density of the particular eigenstate. ETH also provides an explanation for how the rest of the system acts as a bath for a subsystem~\cite{Dymarsky2018} and causes local equilibration under its own unitary dynamics. It is of great conceptual importance to understand under what conditions ETH might be violated in generic non-integrable systems without the need for fine-tuning to an integrable limit where a macroscopic number of conservation laws emerge~\cite{SutherlandBook} that rule out conventional thermalization. 

Such a robust mechanism (i.e., stable with respect to small perturbations in the Hamiltonian) is possibly provided by many-body localization (MBL) where, in the presence of sufficiently strong disorder, interacting systems can resist thermalization~\cite{Basko1, Basko2, Huse1, Nandkishore2015, Abanin2019}. MBL can be viewed as the localization~\cite{AndersonLocalization} of the mid-spectrum eigenstates in a many-body Fock space~\cite{Aletmultifractal}, where the many-particle Fock states are eigenstates in an infinitely strong disorder. On the other hand, ETH posits that such eigenstates should be completely extended in this Fock space. The stability of MBL was argued not to be fine-tuned due to an \emph{emergent integrability} that arises from the presence of an extensive number of local conservation laws given by operators, dubbed as \emph{l-bits}, that mutually commute with each other, with these l-bits changing as a function of disorder in the many-body localized phase~\cite{SerbynMBL2013, HuseMBL2014}. The random field XXZ $S=1/2$ model on finite chains has been the workhorse for MBL~\cite{Pal2010, LuitzMBL2015} with several unique features characterizing MBL, such as area-law entanglement of midspectrum eigenstates, Poisson level statistics of the energy eigenvalues, as well as a logarithmic growth of entanglement between two parts of the system with time for quantum quenches from generic unentangled initial states, observed in numerical studies~\cite{Bauer_2013, SerbynMBL2013, MBLdynamics1_2008, MBLdynamics2_2012}.

Although strong arguments exist in favor of MBL in one dimension~\cite{MBLRG1, MBLRG2, MBLRG3, Imbrie1, Imbrie2} and its absence in higher dimensions~\cite{MBLavalanche1, MBLavalanche2} for short-range models, a rigorous proof of the same is still lacking. Numerical studies based on exact diagonalization (ED) have strong drifts in finite-size estimators, which makes locating the critical disorder strength of the MBL transition challenging. Techniques that work directly in the thermodynamic limit, such as numerical linked cluster expansion techniques~\cite{Devakul2015}, indicate that the MBL phase may be overestimated in ED studies suggesting that currently accessible system sizes may be too small to see a many-body localized phase, and instead one might be in a \emph{MBL regime} which crosses over to a thermal phase on much longer length scales. While some works have suggested the absence of MBL in one dimension~\cite{NoMBL2020}, more recent works~\cite{MBLhighdisorder1D2022a, MBLhighdisorder1D2022b} argued that much higher disorder may be needed to actually stabilize MBL in the thermodynamic limit (see Ref.~\onlinecite{sierant2024manybody} for a recent review on related aspects).

Recently, thermalization properties of short-ranged interacting models with constrained Hilbert spaces have received a great deal of attention due to the striking observation of persistent many-body revivals in a kinematically-constrained chain of $51$ Rydberg atoms~\cite{Bernien2017} when initialized in a N\'eel state while other high-energy initial states thermalized rapidly, as expected from ETH. A minimal model with a constrained Hilbert space to incorporate strong Rydberg blocking, the PXP model~\cite{Sachdev2002, Lesanovsky2012}, revealed that this ergodicity-breaking mechanism is due to the presence of some highly-athermal ETH-violating eigenstates~\cite{Turner2018a, Turner2018b}, dubbed quantum many-body scars, embedded in a spectrum that satisfies ETH. More recent work on infinite-temperature energy transport shows a novel superdiffusive regime~\cite{PXPtransport2023} in PXP chains.

It is interesting to ask whether such kinematically-constrained theories can exhibit MBL in the presence of quenched disorder. While studies on disordered PXP chains~\cite{MBLPXP2018} as well as on other constrained systems~\cite{MBLmodelconstrained2024} including disordered quantum dimer models on two-dimensional lattices~\cite{MBLQDM2020,Pietracaprina_2021} suggested the possibility of MBL in such models, an analysis of a family of generalized PXP models with quenched randomness in one-dimensional chains~\cite{SchardicchioPRL2021} gives evidence for the absence of MBL in the thermodynamic limit. In particular, \cite{MBLQDM2020} investigated the same Hamiltonian as here, but in a more constraining superselection sector, while \cite{Pietracaprina_2021} explored the same Hamiltonian  on a different lattice. The aim of both investigations was to use constrained Hilbert spaces to maximize the physical sizes for which MBL could be detected. Both investigations concluded the existence of ergodic and localized regimes on lattices involving 64 and 78 sites, and at below and above moderately large disorder strengths. However, no clear transition could be identified.

Constrained Hilbert spaces also arise naturally in Hamiltonian formulations of lattice gauge theories (LGTs)~\cite{Kogut1975} since physical (gauge-invariant) states satisfy an appropriate Gauss law. Refs.~\onlinecite{NandkishoreMBL2017, NandkishoreMBL2018} argued for the presence of MBL in lattice-regularized versions of quantum electrodynamics with dynamical matter in one and higher dimensions. In this article, we undertake a systematic study of the nature of the mid-spectrum eigenstates of a particular pure $U(1)$ lattice gauge theory without any dynamical matter as a function of the disorder strength when one of the non-commuting terms in the Hamiltonian is made random. We consider a $U(1)$ quantum link model (QLM) with the gauge degrees of freedom being quantum spins $S=1/2$~\cite{Chandrasekharan1997} that live on the links of ladders of a fixed width $L_y=2$ or $4$ and length $L_x$ and restrict to the Gauss law sector with zero charge at each vertex of the lattice. We consider the most local Hamiltonian in real space, consistent with the Gauss law, where the potential (kinetic) terms are defined on elementary plaquettes and are diagonal (off-diagonal) in the electric flux basis. Quenched disorder is introduced by making the coefficients of the diagonal terms to be random, where the degree of randomness is characterized by a dimensionless parameter, $\alpha$, that equals zero for no randomness and increases monotonically with increasing disorder. To probe MBL, apart from using standard diagnostics like level spacing distributions of the energy eigenvalues, we have considered the probability distribution $p(\mathcal{D})$ of an intensive estimator, $\mathcal{D} \in [0,1/4]$, that simultaneously acts as a measure of elementary plaquettes of the lattice being active or inert in a mid-spectrum eigenstate as well as its spread in Fock space. While estimators based on level spacing distributions show a strong finite-size drift, using the finite size scaling behavior of the skewness of $p(\mathcal{D})$, we could estimate the critical disorder needed to drive an MBL transition for thin ladders with $L_y=2$ based on data for $L_x \leq 14$. We also analyze the autocorrelation functions of single plaquette diagonal operators in a given disordered sample starting from typical high-energy Fock states
and see evidence of dynamic heterogeneity, for disorder strengths $\alpha/\alpha_c \approx 0.4$ and $\alpha/\alpha_c \approx 0.95$, i.e., much before MBL sets in as well as close to the transition (where $\alpha_c$ is the estimate of the ETH-MBL transition) for finite-sized ladders accessible via ED. In particular, the temporal behavior of diagonal plaquette operators in a single disorder realization show striking oscillatory dynamics which are dominated only by a few frequencies from certain randomly selected Fock states whose average energy lies close to the peak of the density of states as a function of energy. However, averaging over Fock states in a single disorder realization to obtain an infinite temperature ensemble result washes out these dynamic heterogeneities. In particular, the temporal oscillations of local diagonal operators are concentrated in spatial regions formed out of connected elementary plaquettes, with these plaquettes either oscillating in phase or out of phase with respect to each other or sharing some common frequencies despite having different temporal dynamics. This feature highlights the unusual quantum dynamics present in a disordered kinematically-constrained interacting system even before MBL sets in. 

The rest of the article is arranged as follows. We define the model and its symmetries in Sec.~\ref{sec:def}. We discuss the level statistics of the energy eigenstates in Sec.~\ref{sec:levelstat} by using data for many disorder realizations as a function of disorder strength and ladder dimensions. We show that the finite-size estimators based on level statistics have a strong drift with system size for numerically accessible ladder dimensions. We introduce the quantity $\mathcal{D}$ for mid-spectrum eigenstates in Sec.~\ref{sec:defineD}, and show how it is related to both the concentration of an eigenstate in Fock space (Sec.~\ref{subsec:Fockspace}) as well as whether elementary plaquettes in the lattice are active or inert (Sec.~\ref{subsec:activeornot}). In Sec.~\ref{subsec:FSS}, we analyze the distribution function, $p(\mathcal{D})$, obtained after using many disorder realizations, as a function of disorder and ladder dimensions which allows us to estimate the disorder strength beyond which MBL is stabilized in a reliable manner for thin ladders with $L_y=2$. We analyze the autocorrelation functions for single plaquette diagonal operators for a given disorder realization in Sec.~\ref{sec:dyn} as a function of disorder strength. Signatures of thermalization at small disorder are discussed in Sec.~\ref{subsec:dyntherm} while the emergence of spatio-temporal heterogeneity at intermediate and strong disorder are discussed starting from typical Fock states in Sec.~\ref{subsec:dynFock}. In particular, certain randomly selected Fock states, whose fraction becomes more significant with increasing disorder as it is tuned from $\alpha/\alpha_c \approx 0.4 $ to $\alpha/\alpha_c \approx 0.95$ (i.e., much before the ETH-MBL transition sets in and close to the ETH-MBL transition, respectively), display emergent \emph{oscillatory} behavior of these local diagonal operators in real time in regions formed out of connected elementary plaquettes on the lattice. We finally conclude and discuss some open issues in Sec.~\ref{sec:con}.

\section{Disordered U(1) QLM on ladders}
\label{sec:def}
\begin{figure}[h]
    \centering
    \includegraphics[scale=0.2]{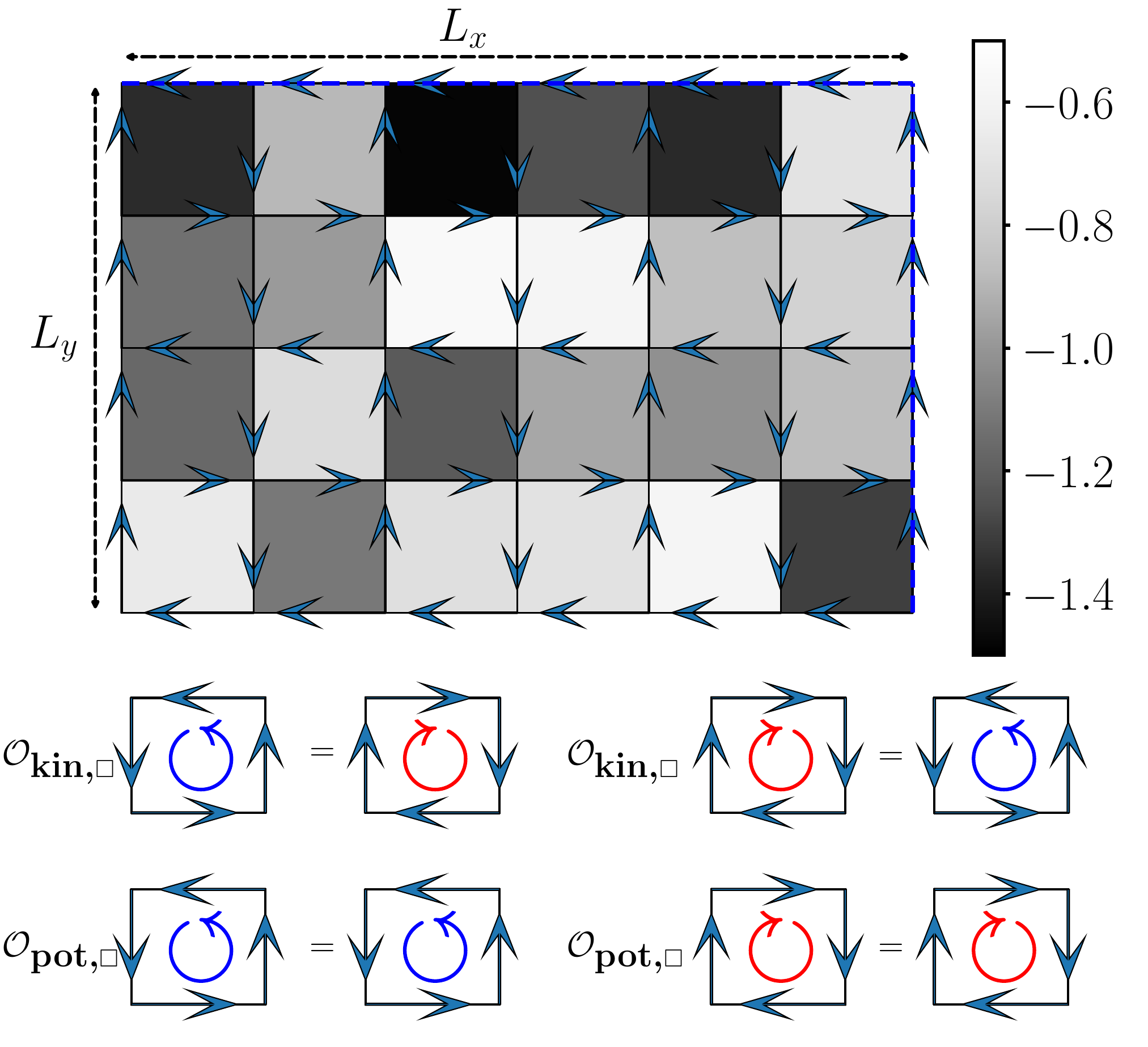}
    \caption{(Top panel) An electric flux configuration for a $L_x \times L_y = 6 \times 4$ lattice with periodic boundary conditions in both directions. The shading on the elementary plaquettes denote the different values of $- (1+\alpha R_\square)$ (see Eq.~\ref{eq:Hran}) for one particular disorder realization where $\alpha=1$ and $R_\square$ is an independently chosen random number at each plaquette from the uniform distribution $[-1/2,1/2]$. (Bottom panel) Action of $\Okins$ and $\Opots$ shown for elementary flippable plaquettes. Here, clockwise (anti-clockwise) circulation of electric fluxes around a plaquette is marked in red (blue) inside the plaquette.}
    \label{fig:latticesetup}
\end{figure}
We consider a disordered $U(1)$ QLM with gauge degrees of freedom being quantum spins $S=1/2$ living on the links $\mathbf{r},\hat{\mu}$ connecting two neighbouring sites $\mathbf{r}$ and $\mathbf{r}+\hat{\mu}$ (where $\hat{\mu} = \hat{i}, \hat{j}$) of a ladder whose width equals $L_y$ and length equals $L_x$ and take periodic boundary conditions in both directions (see Fig.~\ref{fig:latticesetup}, top panel). A $U(1)$ quantum link, $U_{\mathbf{r},\hat{\mu}} = S_{\mathbf{r},\hat{\mu}}^+$ is a raising operator of the electric flux $E_{\mathbf{r},\hat{\mu}} = S_{\mathbf{r},\hat{\mu}}^z$. We specifically consider even $L_x$ and $L_y$ with the following Hamiltonian:
\begin{equation}
    \mathcal{H}_{\rm dis}   = -\sum_\square \Okins  - \sum_\square (1+\alpha R_\square) \Opots
\label{eq:Hran}
\end{equation}
where each $R_\square$ is an independently chosen random number from the uniform distribution $[-1/2,1/2]$ whose specification on all the elementary plaquettes defines a single disorder realization of $\mathcal{H}_{\rm dis}$, $\alpha \geq 0$ and is a dimensionless characterization for the strength of disorder, with $\alpha=0$ ($\alpha \rightarrow \infty$) representing zero (infinite) disorder. The operator $\Okins$ changes the orientation of the electric flux loops around an elementary plaquette from clockwise to anticlockwise and vice versa (Fig.~\ref{fig:latticesetup}, bottom panel), and annihilates non-flippable plaquettes. $\Opots$ is a diagonal counting  operator in the electric flux basis, where each flippable (non-flippable) plaquette is counted as $1$ ($0$) (Fig.~\ref{fig:latticesetup}, bottom panel). Written explicitly, $ \Okins =(U_\square+U^\dagger_{\square})$ where $U_\square=U_{\mathbf{r},\hat{i}} U_{\mathbf{r}+\hat{i},\hat{j}} U^\dagger_{\mathbf{r}+\hat{j},\hat{i}}U^\dagger_{\mathbf{r},\hat{j}}$ which equals $U_\square=S^+_{\mathbf{r},\hat{i}} S^+_{\mathbf{r}+\hat{i},\hat{j}} S^-_{\mathbf{r}+\hat{j},\hat{i}}S^-_{\mathbf{r},\hat{j}}$ in terms of $S=1/2$ operators, while $\Opots = (U_\square+U_\square^\dagger)^2$ on each elementary plaquette.
Note that in the absence of disorder, this is the Hamiltonian of the celebrated Rokshar-Kivelson model \cite{Rokhsar1988} which has also been used to study the low-temperature physics of quantum spin-ice on the checkerboard lattice \cite{Shannon2004}. In the high-energy physics context, this model shows unconventional confinement \cite{Banerjee2013}.

This Hamiltonian has a local $U(1)$ symmetry generated by the Gauss law $G_{\mathbf{r}} = \sum_{\mu} (E_{\mathbf{r},\hat{\mu}} - E_{\mathbf{r-\hat{\mu}},\hat{\mu}})$. The physical states $|\psi\rangle$ satisfy $G_{\mathbf{r}}|\psi\rangle=0$ which implies that in-coming and out-going electric fluxes add up to zero on each site (see Fig.~\ref{fig:latticesetup}, top panel for an example of such an electric flux configuration), resulting in no background charge at any site, and providing a constrained Hilbert space.  The total electric flux winding around the lattice in a given periodic direction is a conserved quantity as well, related to a $U(1)$ center symmetry, and causes the Hilbert space to break up into distinct topological sectors, characterized by a pair of integer winding numbers $(W_x, W_y)$. We restrict ourselves to the largest such sector with $(W_x, W_y)=(0,0)$.

The model, without any disorder ($\alpha=0$), has a host of discrete symmetries, including translations by one lattice unit in both directions, discrete rotations and reflections, as well as an internal symmetry of charge conjugation which reverses all the electric fluxes. In the presence of disorder ($\alpha \neq 0$), only the internal symmetry survives and the Hilbert space can be block diagonalized into two sectors with an equal number of states, with the charge conjugation quantum number being $C = \pm 1$ using the basis states
\begin{eqnarray}
  |F_i\rangle_{\pm}=(|F_i\rangle \pm C_E|F_i\rangle)/\sqrt{2},
  \label{eq:FockCbasis}
\end{eqnarray}
where $|F_i\rangle$ denotes a Fock state in the electric flux basis and $C_E|F_i\rangle$ denotes another Fock state obtained by reversing all the electric fluxes of $|F_i \rangle$.

While an unconstrained Hamiltonian with $S=1/2$ degrees of freedom on the links of a $L_x \times L_y$ ladder contains $2^{2L_x L_y}$ configurations, the added local constraint of in- and out-going electric fluxes adding up to zero on each site dramatically decreases the number of allowed states in the Hilbert space (it still scales exponentially in $L_x L_y$, but with a lower coefficient in the exponent). Furthermore, restricting to the largest topological sector with $(W_x,W_y)=(0,0)$ and using the charge conjugation symmetry reduces the allowed number of configurations even further, as shown in Table.~\ref{tab:HSD}. For the rest of the article, we present results from ED for $L_x \times L_y$ ladders with $L_y=2$ and $L_x=8, 10, 12, 14$ as well as $6\times 4$ ladders. 
\begin{table}
    \centering
    \begin{tabular}{|c|c|}
        \hline
        Lattice & HSD in $C=\pm 1$ sector for $(W_x,W_y)=(0,0)$ \\
        \hline
        $8\times2$ & $1107$\\
        \hline
        $10\times2$ & $8953$\\
        \hline
        $12\times2$ & $73789$\\
        \hline
        $14\times2$ & $616227$\\
        \hline
         $16\times2$ & $5196627$\\
        \hline
        $4\times4$ & $495$\\
        \hline
        $6\times4$ & $16405$\\
        \hline
        $8\times4$ & $579583$\\
        \hline
        $6\times6$ & $2741358$\\
        \hline
    \end{tabular}
    \caption{Hilbert space dimension (HSD) for different ladders in charge conjugation resolved sector $C=+1/-1$ for the zero winding number topological sector with $(W_x,W_y)=(0,0)$.}
    \label{tab:HSD}
\end{table}

\begin{figure}[h]
    \centering
    \includegraphics[scale=0.27]{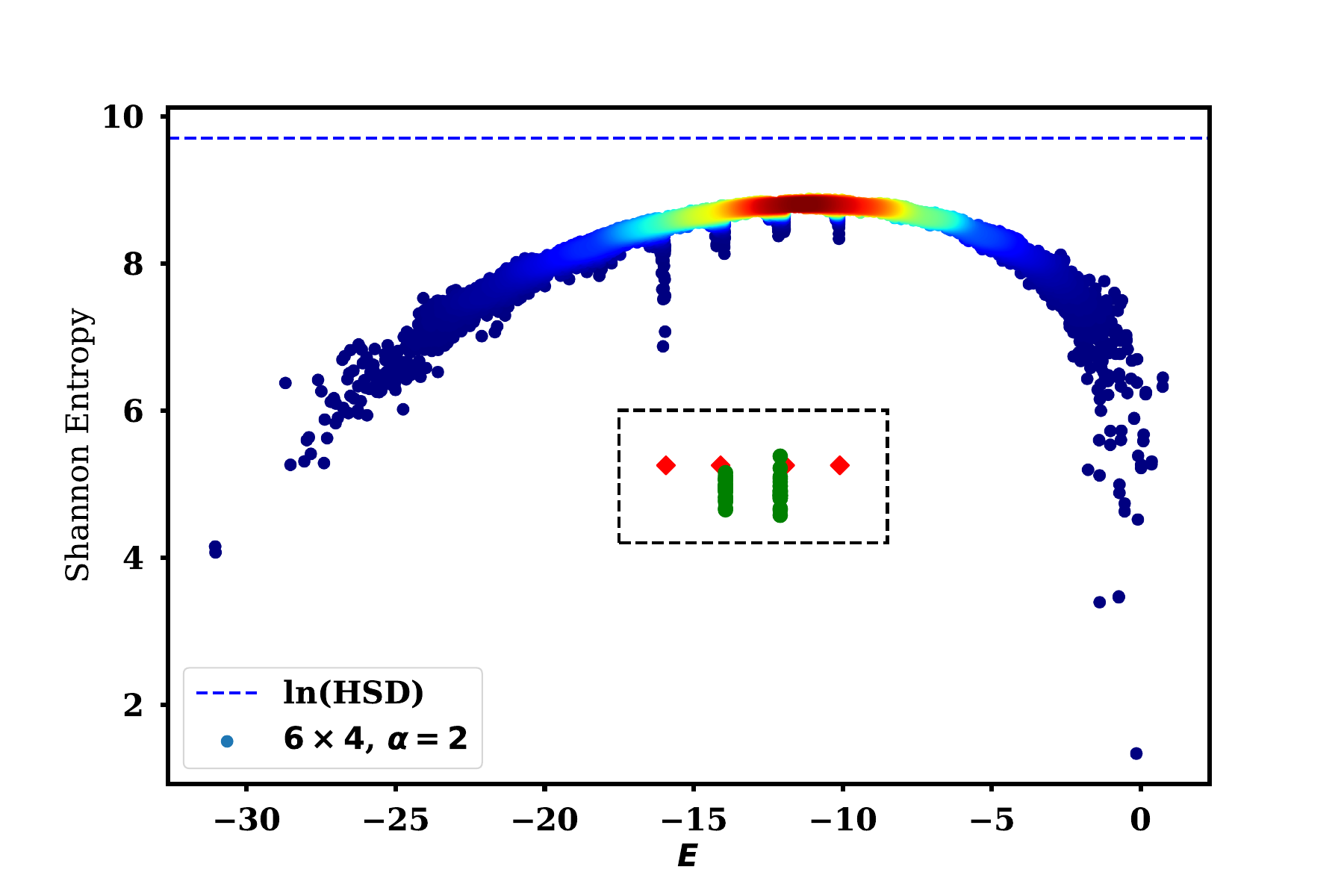}
    \caption{Shannon entropy $S_1$ (Eq.~\ref{eq:Shannon}) for the energy eigenstates of a single disorder realization of a $6 \times 4$ lattice with $\alpha=2$. The data for $C = \pm 1$ is shown together in the same plot and the density of states is indicated by a color map where warmer color corresponds to higher density of states. The sublattice scars are shown by a different point font and are enclosed by a box composed of dotted lines for clarity.}
    \label{fig:shannon6times4}
\end{figure}

The weakly disordered $U(1)$ QLM on ladders is expected to be non-integrable and thus satisfy ETH in the topological sector $(W_x,W_y)=(0,0)$ as already discussed in Refs.~\cite{Banerjee2021, Biswas:2022env}. Since mid-spectrum eigenstates in such a situation are expected to be completely delocalized in Fock space, where the Fock states are defined as in Eq.~\ref{eq:FockCbasis} in the $C=\pm 1$ sector, calculating the Shannon entropy defined as
\begin{eqnarray}
  S_1(|\Psi\rangle_{\pm})=-\sum_i |\psi_i|^2 \ln |\psi_i|^2
  \label{eq:Shannon}
  \end{eqnarray}
for any eigenstate $|\Psi\rangle_{\pm} =\sum_{i=1}^{\mathrm{HSD}}\psi_i|F_i\rangle_{\pm}$ (where the subscript $\pm$ denotes the charge conjugation sector) should yield values close to $\ln (\mathrm{HSD})$ for the mid-spectrum eigenstates. In Fig.~\ref{fig:shannon6times4}, we see that this expectation is true both for the eigenstates in $C=+1$ and $C=-1$ for a single disorder realization of a $6 \times 4$ lattice with $\alpha=2$, though the maximum value of Shannon entropy is somewhat lower than expected of a completely delocalized state~\cite{Masud2022finitesize, midspectrumETH2023}. At finite disorder, no two energy eigenstates are expected to be degenerate within these symmetry resolved sectors for any typical disorder realization. The only exception to this statement is provided by certain anomalous eigenstates $|\psi_{\mathrm{sub}}\rangle$, called \emph{sublattice scars}~\cite{Sausublattscars2024}, that are simultaneous eigenkets of $\sum_{\square} \Okins$ with eigenvalues $0$ or $\pm 2$ as well as of $\Opots$ with eigenvalue $1$ ($0$) on one (the other) sublattice (for even $L_x, L_y$, the lattice is bipartite with elementary plaquettes on one sublattice sharing edges with plaquettes of the other sublattice) with there being a equal number of sublattice scars which have eigenvalue $\Opots=1$ on one sublattice or the other. These sublattice scars are eigenstates of $H$ for any arbitrary $\alpha$ with energies $E_{0,\mathrm{even/odd}}=-\sum_{\square,\mathrm{even/odd}}(1+\alpha R_\square)$ for the states with eigenvalues $0$ for $\sum_{\square} \Okins$ and $\Opots=1$ for even (odd) sublattice of elementary plaquettes, and with energies $E_{\pm 2,\mathrm{even/odd}} =   E_{0,\mathrm{even/odd}} \pm 2$ for the states with eigenvalues $\pm 2$ for $\sum_{\square} \Okins$ and $\Opots=1$ for even (odd) sublattice of elementary plaquettes. For a $6 \times 4$ lattice, there are exactly $23$ such sublattice scars with energy $E_{0,\mathrm{even/odd}}$ and $1$ sublattice scar with energy $E_{\pm 2,\mathrm{even/odd}}$~\cite{Sausublattscars2024}. This degeneracy is clearly reflected in Fig.~\ref{fig:shannon6times4}. The anomalous nature of these eigenstates can be seen from the fact that these have significantly lower Shannon entropy than their neighboring eigenstates (Fig.~\ref{fig:shannon6times4}). The number of such sublattice scars is, however, a vanishing fraction of the total HSD and does not affect various statistical indicators of ETH versus MBL that we will discuss in Sec.~\ref{sec:levelstat} and Sec.~\ref{sec:defineD}. We will, nonetheless, show data for $C=-1$ for $8 \times 2$, $C=+1$ for $10 \times 2$ and $C=-1$ for $12 \times 2$ ladders since these sectors do not have any sublattice scars (compared to $4$ sublattice scars in the other sector) for these ladder dimensions and $C=-1$ for $6 \times 4$ ladders since this sector has only $4$ sublattice scars compared to $46$ sublattice scars in $C=+1$ for the same ladder dimension~\cite{Sausublattscars2024}. 

\section{Level spacing distribution}
\label{sec:levelstat}
The distribution of energy level spacings in a finite-sized system~\cite{Oganesyan2007} provides an important diagnostic for whether the model is non-integrable or not, as expected for MBL due to an emergent integrability. Here, we construct the distribution of consecutive level spacing ratios $\tilde{r} \in [0,1]$ of the Hamiltonian $\mathcal{H}_{\mathrm{dis}}$ at finite $\alpha$ after resolving in a sector with $C=+1$ or $C=-1$. The level spacing ratios, $r$, are defined as
\begin{eqnarray}
  r=\mathrm{min} \left \{ r_n, \frac{1}{r_n}\right \} \leq 1, r_n=\frac{s_n}{s_{n-1}}, s_n = E_{n+1}-E_n,
  \label{eq:levelr}
  \end{eqnarray}
where $E_n$ denotes an energy eigenvalue with $E_{n+1} > E_n$. When the model satisfies ETH, one expects the level spacing distribution, $P(r)$, to follow an appropriate Wigner-Dyson distribution (Gaussian orthogonal ensemble (GOE) distribution for the case in hand), while a Poisson distribution is expected for MBL~\cite{Atas2013}, where:
\begin{eqnarray}
  P_{\mathrm{GOE}}(r) = \frac{27}{4} \frac{r+r^2}{(1+r+r^2)^{5/2}}; \mbox{~~}P_\mathrm{P}(r)=\frac{2}{(1+r)^2}.
  \label{eq:universalD}
\end{eqnarray}
The mean level spacing ratio $\langle r \rangle$ also changes from $0.5307(1)$ for the GOE distribution~\cite{Atas2013} to $2\ln(2)-1 \approx 0.3863$ for the Poisson distribution, thus providing another related means to distinguish between ETH and MBL.
\begin{figure}[h]
  \centering
  \includegraphics[scale=0.25]{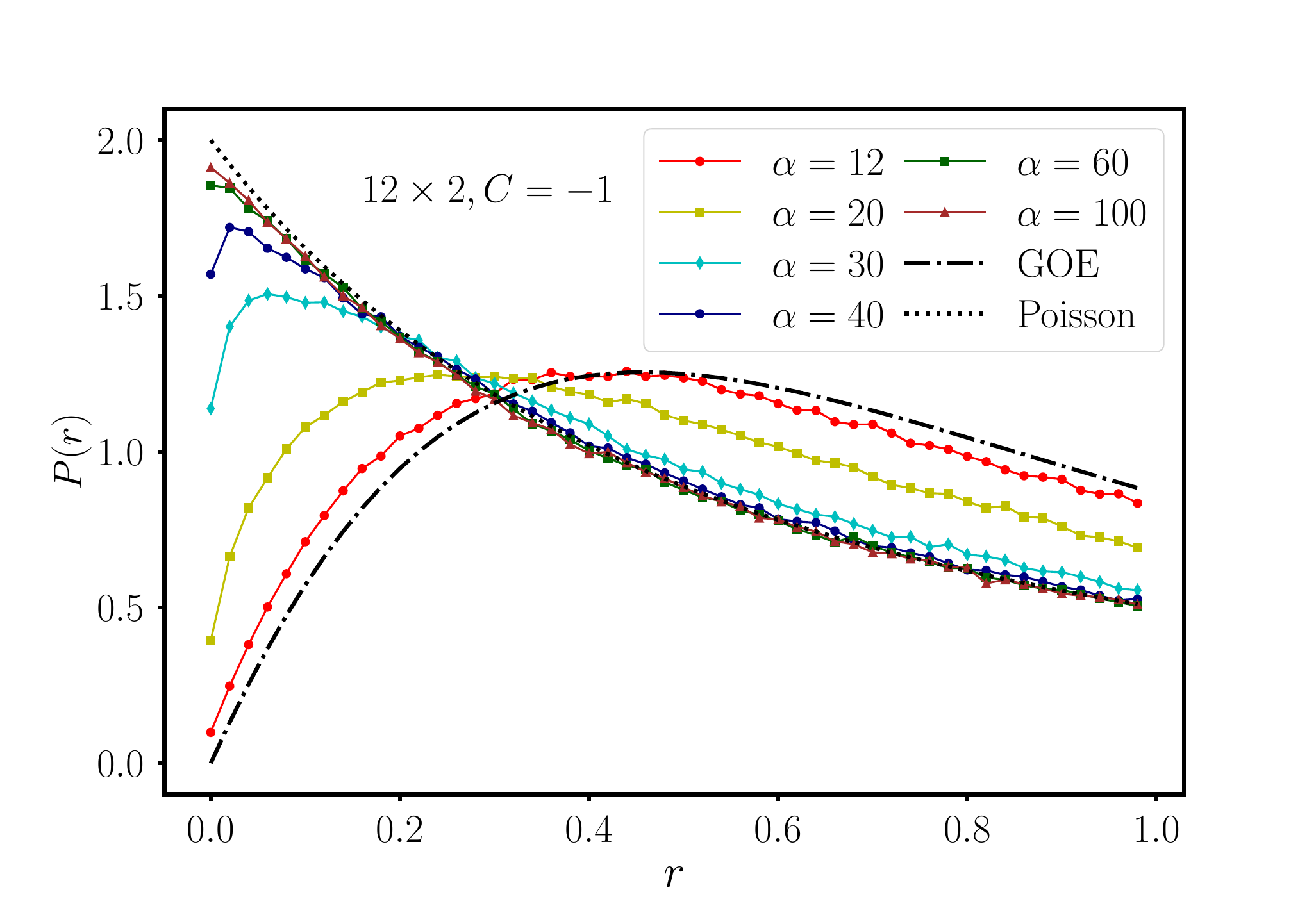}
    \includegraphics[scale=0.26]{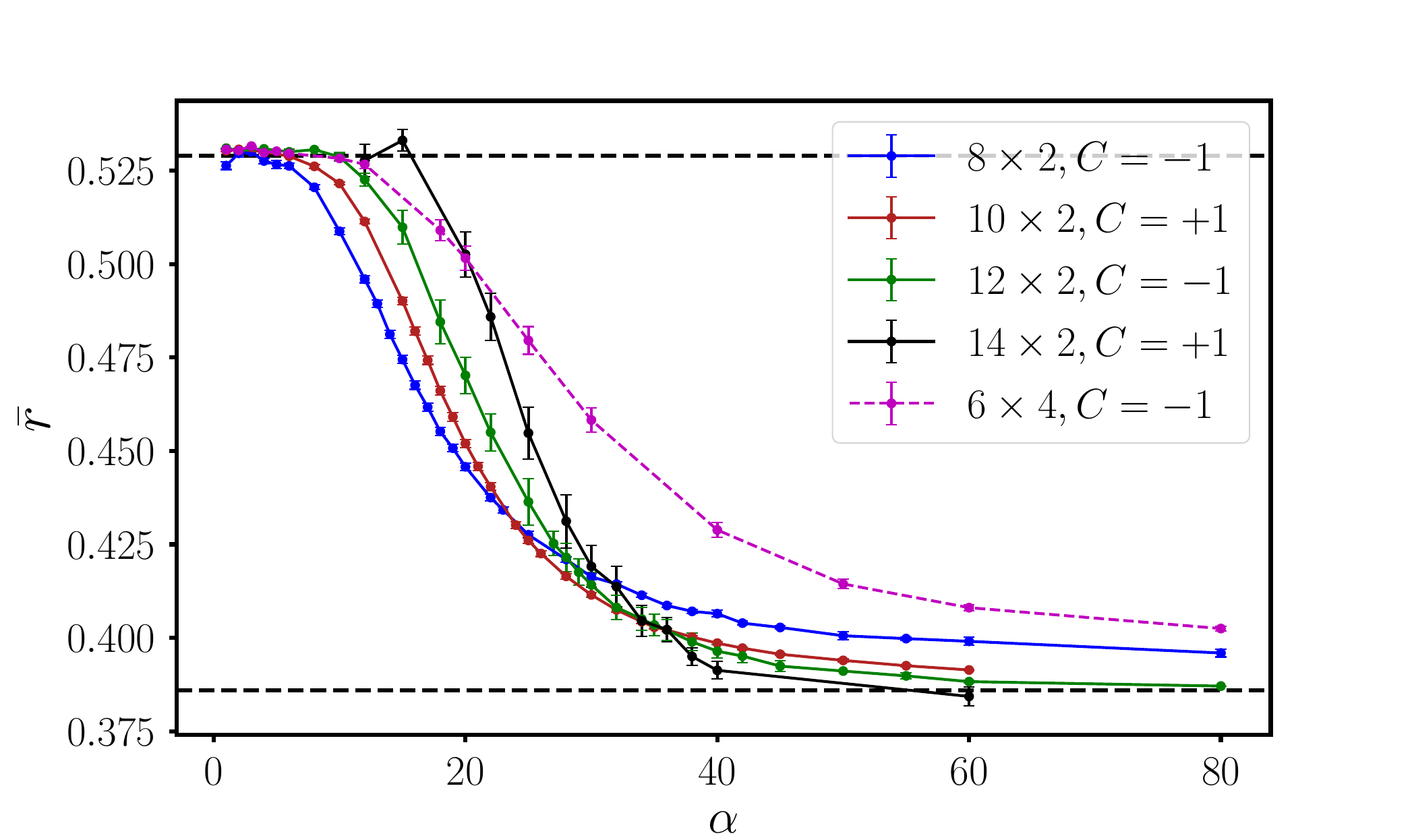}
    \caption{(Top panel) Disorder-averaged distribution $P(r)$ shown for a ladder of dimension $12 \times 2$ for various values of $\alpha$. The universal distribution functions $P_{\mathrm{GOE}}(r)$ and $P_\mathrm{P}(r)$ (Eq.~\ref{eq:universalD}) are also shown for comparison. (Bottom panel) Disorder-averaged mean level spacing $\bar{r} $ shown as a function of disorder strength $\alpha$ for various ladder dimensions. The dotted horizontal lines at $\bar{r} \approx 0.5307$ and $\bar{r} \approx 0.3863$ are the universal values for $P_{\mathrm{GOE}}(r)$ and $P_\mathrm{P}(r)$, respectively, and are shown here for comparison.}
    \label{fig:levelspacings}
\end{figure}


For the disordered $U(1)$ QLM (Eq.~\ref{eq:Hran}), we collect data for many independent disorder realizations ($500$ realizations for $8 \times 2$ and $10 \times 2$, $50$ realizations for $6 \times 4$, $10$ realizations at low disorder and $20$ to $30$ disorder realizations for high disorder for $12 \times 2$, {$10$ disorder realizations for $\alpha \le 20$ and $20$ realizations for $\alpha>20$ for $14\times 2$}) at each $\alpha$ to obtain the disorder-averaged distribution $P(r)$ and disorder-averaged mean level spacing {$\bar{r}=\overline{\braket{r}}$, where $\overline{\cdots}$ denotes the average over many disorder realizations} (Fig.~\ref{fig:levelspacings}). We also use shift-invert techniques (see Appendix~\ref{app:midspectrum} for details) to extract representative mid-spectrum eigenstates of $14 \times 2$ ladders from which level statistics can be extracted. In Fig.~\ref{fig:levelspacings} (top panel), we display the results for $P(r)$ as a function of $\alpha$ for $12 \times 2$ ladders that have the largest HSD (which equals $73789$) that we could access in our ED studies without needing shift-invert techniques. While even for $\alpha=12$, $P(r)$ remains close to $P_{\mathrm{GOE}}(r)$, it seems to smoothly crossover to $P_\mathrm{P}(r)$~\cite{levelspacing2001, levelspacingMBL2019, levelspacingMBL2020} as the disorder is increased to $\alpha=100$ suggesting a possible MBL at very large disorder. In Fig.~\ref{fig:levelspacings} (bottom panel), we show the results for the disorder-averaged mean level spacing $\bar{r}$ for different ladder dimensions as a function of disorder strength $\alpha$. $\bar{r}$ smoothly interpolates from the value expected from a GOE distribution at low $\alpha$ to the one expected from a Poisson distribution at large $\alpha$. Comparing the disorder-averaged mean level spacing $\bar{r}$ for a wider ladder with dimension $6 \times 4$ to that of the $12 \times 2$ ladder (Fig.~\ref{fig:levelspacings} (bottom panel)) clearly shows that a wider ladder, composed of the same number of elementary plaquettes, resists MBL more effectively with increasing disorder.

The crossing point of the curves for $L_x \times 2$ ladders (see Fig.~\ref{fig:levelspacings} (bottom panel)) can, in principle, be used to estimate the critical disorder strength, $\alpha_c$, needed to stabilize MBL for thin ladders. The crossing point shows a drift towards stronger disorder with increasing $L_x$ which makes such an estimation difficult without determining the crossing at $L_x$ and $aL_x$ where $a>1$ is a fixed number and then extrapolating to $L_x \rightarrow \infty$ for a fixed $L_y$. However, the limited range of sizes we have makes this procedure inapplicable and we resort to a different estimator which does not require data both at $L_x$ and $aL_x$ in the following.

\begin{figure}[h]
  \centering
  \includegraphics[scale=0.25]{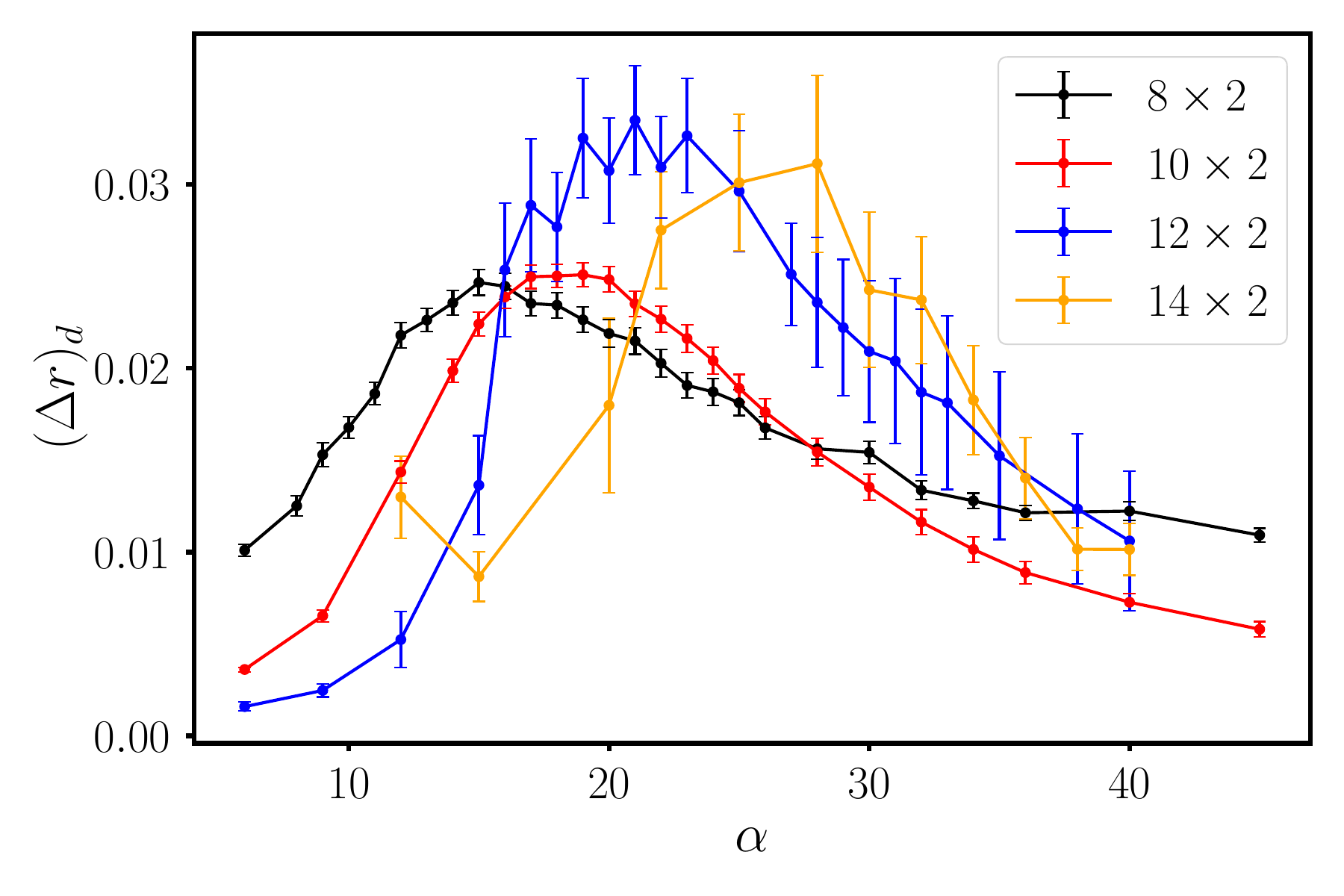}
    \includegraphics[scale=0.25]{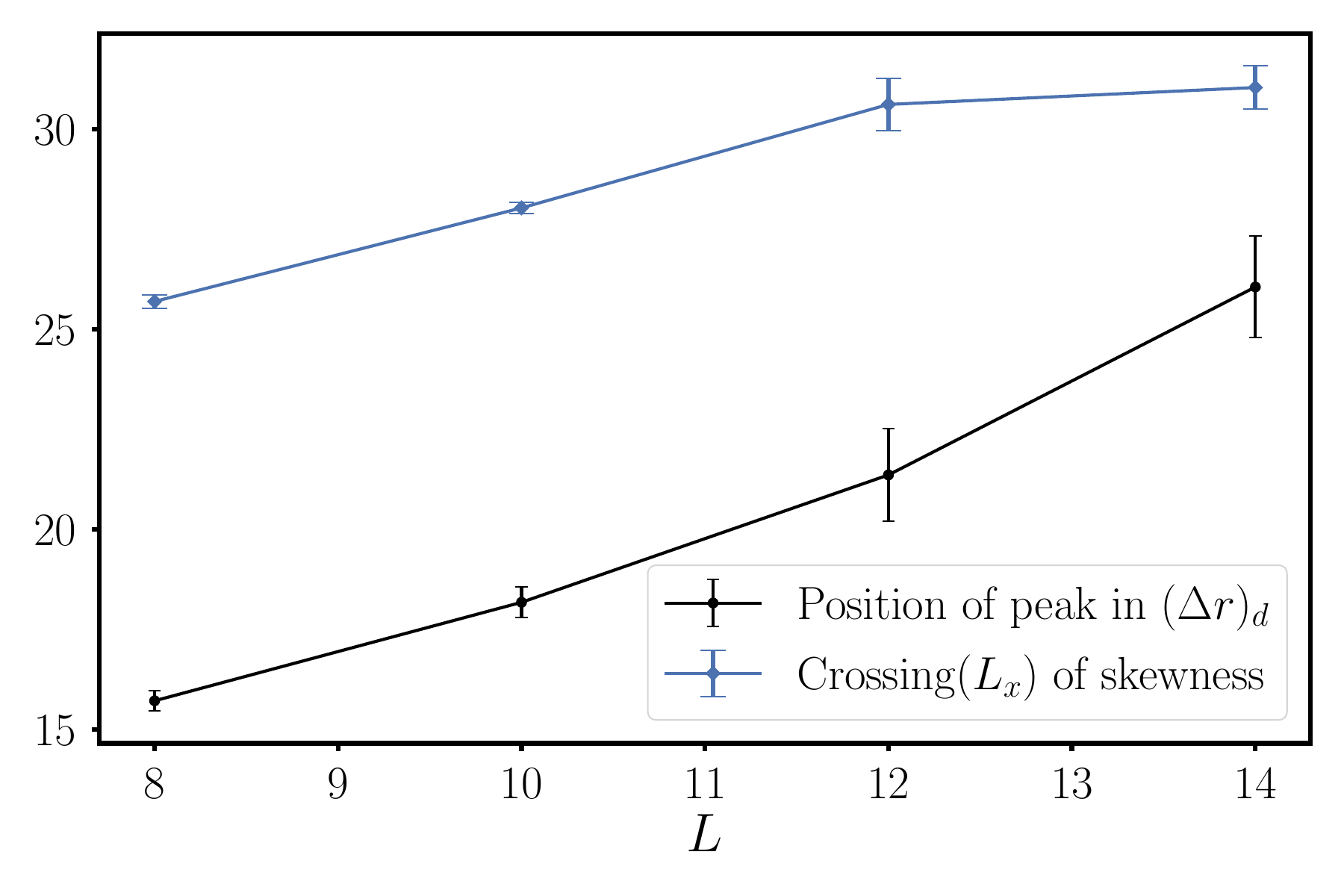}
    \caption{(Top panel) Disorder-averaged $(\Delta r)_d$ shown for thin ladders with $L_y=2$ and $8 \leq L_x \leq 14$ for various values of $\alpha$. (Bottom panel) The peak location for each such curve has been extracted and this finite-size estimator for $\alpha_c$ is shown as a function of $L_x$. Another estimator, extracted from the skewness of $p(\mathcal{D})$ and discussed in Sec.~\ref{subsec:FSS}, is shown in the same plot for comparison.}
    \label{fig:levelspacingscriticalcoupling}
\end{figure}

Consider the variance of the variable {$\braket{r}$ (Eq.~\ref{eq:levelr}), which is defined as $(\Delta r)_d^2 = \overline{\braket{r}^2-\bar{r}^2}$.} For the case of random disorder, the disorder-averaged $(\Delta r)_d$ shows a peak as a function of the disorder strength at a fixed system size. The location of this peak has been argued to act as a finite-size estimator for the MBL transition in Ref.~\onlinecite{PRBdelr}. We display the behavior of $(\Delta r)_d$ for thin ladders with $L_y=2$ for $L_x \leq 14$ in Fig.~\ref{fig:levelspacingscriticalcoupling} (top panel). The peak location of this quantity is then extracted and plotted as a function of $L_x$ in Fig.~\ref{fig:levelspacingscriticalcoupling} (bottom panel). {To extract the peak location, we first generate $500$ synthetic data set of $(\Delta r)_d$ for each $L_x$ and fit the data to a third order polynomial near the peak. From the fitting of those synthetic data we find the mean peak locations and the errors. For this analysis we have used $500$ realizations for $8 \times 2$ and $10 \times 2$ ladders, $30$ realizations for $12 \times 2$ ladder and  $20$ realizations for $14 \times 2$ ladder.} This estimator, however, shows a significant drift with increasing system size which implies that systems larger than $L_x=14$ are required to have an accurate determination of the critical disorder strength $\alpha_c$ in the thermodynamic limit. In the same plot, we display {\it another estimator} which has a much better finite-size convergence for the numerical accessible system sizes and we discuss it in detail in Sec.~\ref{subsec:FSS}.

\section{Probing nature of mid-spectrum eigenstates via $\mathcal{D}$}
\label{sec:defineD}
For the disordered $U(1)$ QLM (Eq.~\ref{eq:Hran}), the \emph{disorder field} $(1+\alpha R_\square)$ couples linearly to the local operator $\Opots$. This suggests that the nature of the mid-spectrum eigenstates may be probed more fruitfully using measures based on the behavior of $\langle \Opots \rangle$ in these eigenstates, where the expectation $\langle \rangle$ is taken with respect to that particular state. For operational reasons, we define the mid-spectrum eigenstates in any particular disorder realization by dividing the bandwidth $(E_{\mathrm{max}}-E_{\mathrm{min}})$, where $E_{\mathrm{max}}$ ($E_{\mathrm{min}}$) refers to the maximum (minimum) energy eigenvalue for that disorder realization, in $25$ equally sized bins and then labelling the states from the bin that contains the maximum number of eigenstates (thus maximizing the density of states as a function of energy) to be mid-spectrum.

For small disorder strength $\alpha \ll 1$ where ETH definitely holds~\cite{Banerjee2021}, the form of the Hamiltonian becomes irrelevant for mid-spectrum eigenstates since these locally mimic infinite-temperature thermal states and $\langle \Opots \rangle \rightarrow \langle \Opots \rangle_{\mathrm{th}}$, where $\langle \Opots \rangle_{\mathrm{th}}$ denotes the corresponding infinite-temperature expectation value. For $\alpha \rightarrow \infty$ (infinite disorder limit), the electric flux configurations $|F_i\rangle$, as well as $|F_i\rangle_{\pm}$ constructed from them (Eq.~\ref{eq:FockCbasis}), become eigenstates of $\mathcal{H}_{\mathrm{dis}}$ and thus $\langle \Opots \rangle$ equals $1$ or $0$ in each plaquette for every mid-spectrum eigenstate. Assuming that MBL exists when $\alpha \gg 1$ but finite, we expect $\langle \Opots \rangle$ to typically be pinned close to its extremal values of $0$ or $1$ on each plaquette (since $\Opots$ does not commute with $\mathcal{H}_{\mathrm{dis}}$ for finite $\alpha$, quantum fluctuations make $\langle \Opots \rangle$ deviate from its extreme values) for mid-spectrum eigenstates since this phase should be adiabatically connected~\cite{Imbrie2} to the infinite disorder point ($\alpha \rightarrow \infty$).

We define the following intensive estimator, $\mathcal{D} \in [0,1/4]$, where $N_p$ denotes the total number of elementary plaquettes $L_x L_y$ for a $L_x \times L_y$ ladder:
\begin{eqnarray}
  \mathcal{D} = \frac{1}{N_p} \sum_\square \mathcal{D}_\square \mbox{~~} \mathrm{where} \mbox{~~} \mathcal{D}_\square = \left(\langle \Opots \rangle -\frac{1}{2} \right)^2.
  \label{eq:formulaD}
  \end{eqnarray}
Assuming ETH, the value of $\mathcal{D}$ for mid-spectrum eigenstates should equal $\mathcal{D}_{\mathrm{th}}$ where $\langle \Opots \rangle_{\mathrm{th}}$ is obtained by using $\mathrm{Trace}[\Opots]/ \mathrm{HSD}$ where the trace can be directly carried over the electric flux Fock states in the zero winding number sector that are obtained from direct enumeration. The results for certain ladder dimensions are displayed in Table.~\ref{tab:Opot}. In all cases, $\mathcal{D}_{\mathrm{th}} \ll 1$. In particular, using a low-order polynomial extrapolation in $1/L_x$ for $\langle \Opots \rangle_{\rm th}$ (Table.~\ref{tab:Opot}) results in $\mathcal{D}_{\mathrm{th}} \approx 0.0031 (0.0070)$ for $L_x \rightarrow \infty$ for infinitely long ladders of width $L_y=2 (4)$. On the other hand, for $\alpha \rightarrow \infty$, $\mathcal{D}_\square \rightarrow 1/4$ from below. In particular, for electric flux Fock states $|F_i\rangle$, as well as for basis states of the form $|F_i\rangle_{\pm}$ (Eq.~\ref{eq:FockCbasis}), $\mathcal{D}=1/4$, as $\Opots$ remains unchanged under charge conjugation.

\begin{table}
    \centering
    \begin{tabular}{|c|c|c|}
        \hline
        Lattice & $\langle \Opots \rangle_{\mathrm{th}}$ & $\mathcal{D}_{\mathrm{th}}$\\
         \hline
        $6\times2$ & $0.496454$ & $0.0000125$\\
        \hline
        $8\times2$ & $0.482384$ & $0.0003103$\\
        \hline
        $10\times2$ & $0.474254$ & $0.00066286$\\
        \hline
        $12\times2$ & $0.468986$ & $0.000961868$\\
        \hline
        $14\times2$ & $0.465299$ & $0.0012041$\\
        \hline
         $16\times2$ & $0.462576$ & $0.00140056$\\
        \hline
        $4\times4$ & $0.460606$ & $0.00155189$ \\
        \hline
        $6\times4$ & $0.4457$ & $0.0029485$ \\
        \hline
        $8\times4$ & $0.438232$ & $0.00381529$\\
        \hline
        $10\times4$ & $0.433827$ & $0.0043789$\\
        \hline
        $6\times6$ & $0.435854$ & $0.004115$\\
        \hline
    \end{tabular}
    \caption{The values of $\langle \Opots \rangle_{\mathrm{th}}$ and $\mathcal{D}_{\mathrm{th}}$ obtained from direct enumeration of electric flux Fock states in the zero winding numbers sector shown for certain $L_x \times L_y$ ladders.}
    \label{tab:Opot}
\end{table}

\subsection{$\mathcal{D}$ as estimator of concentration of eigenstate in Fock space}
\label{subsec:Fockspace}

\begin{widetext}

    \begin{figure}
        \centering
        \includegraphics[scale=0.205]{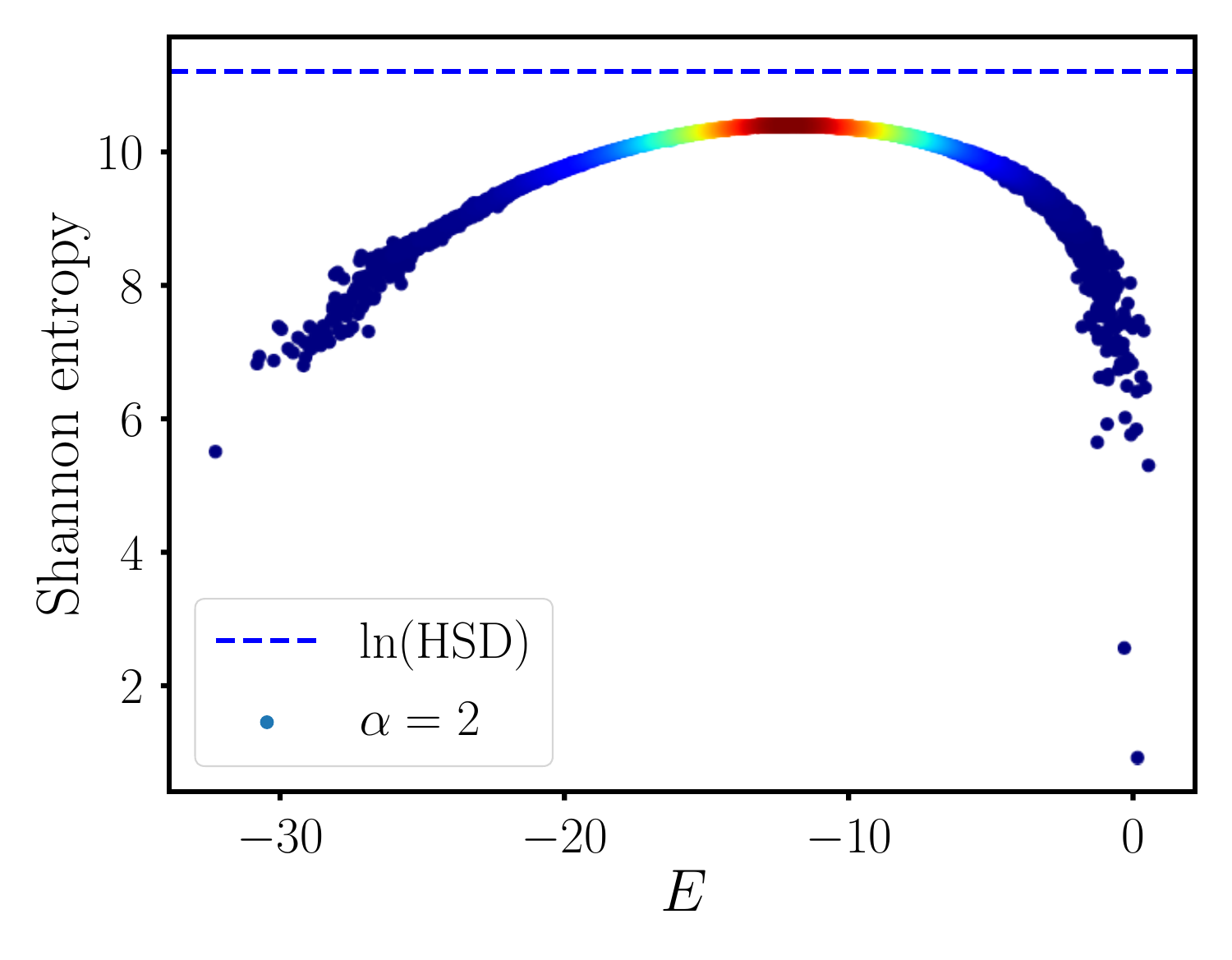}
        \includegraphics[scale=0.2]{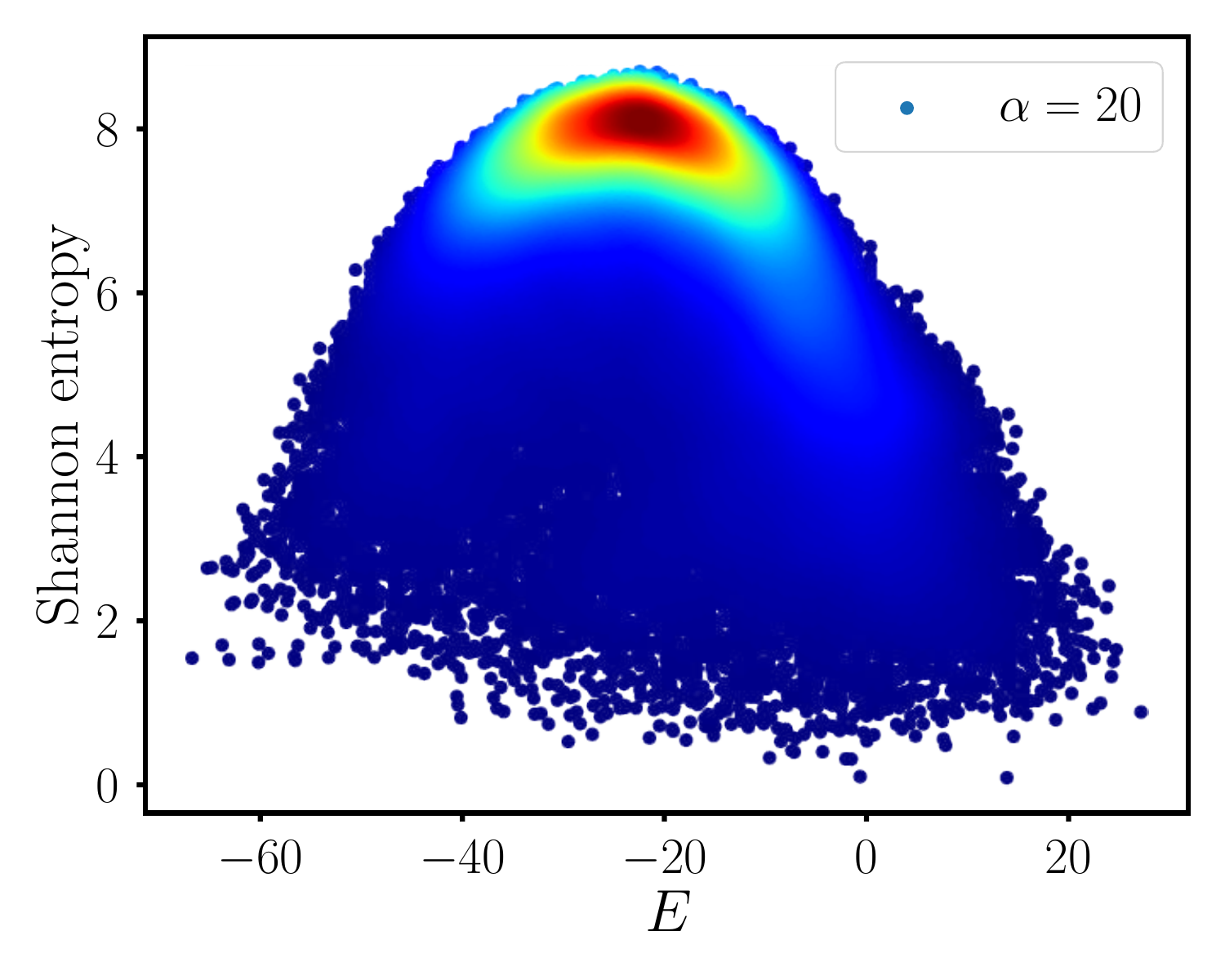}
        \includegraphics[scale=0.2]{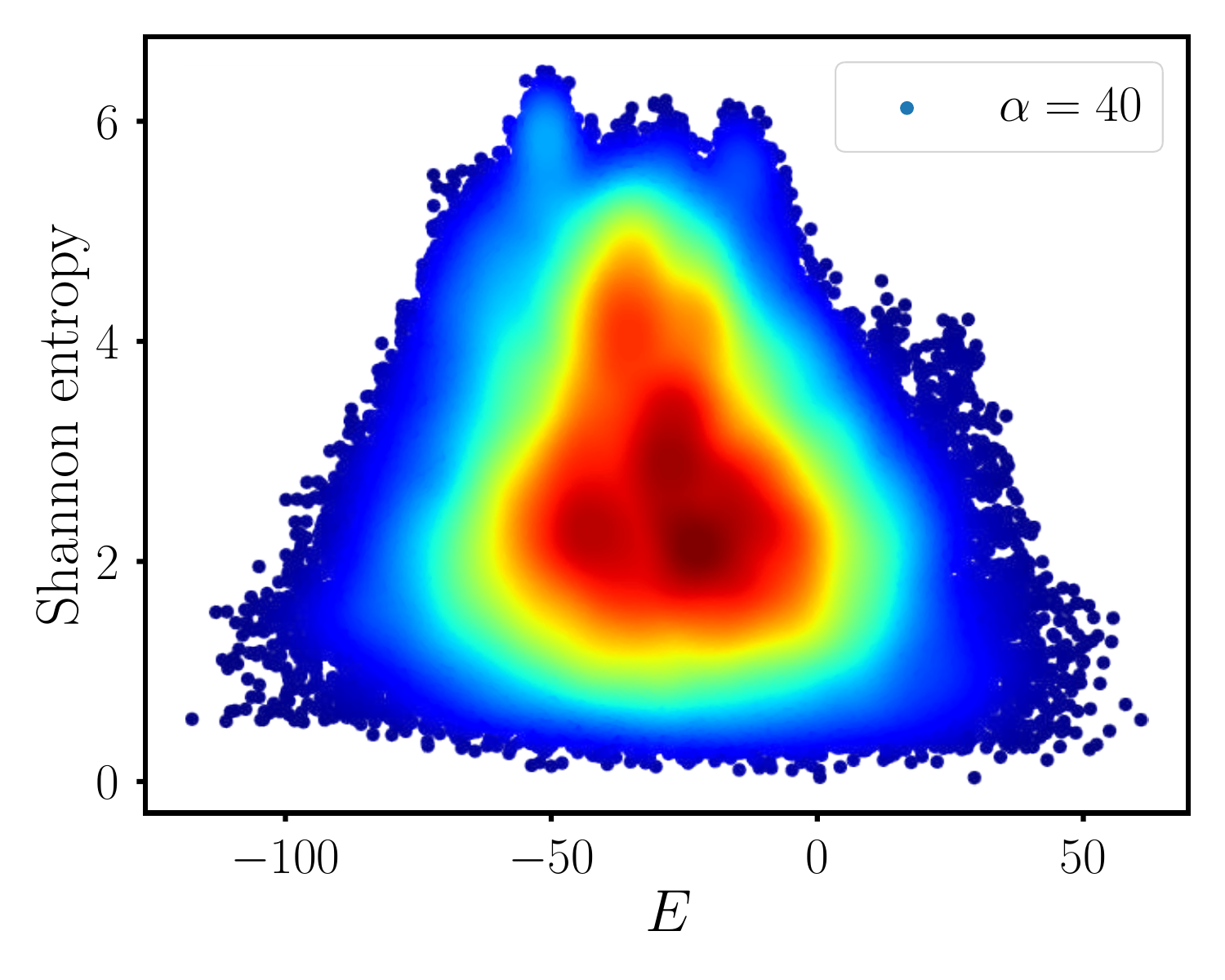}
        \includegraphics[scale=0.2]{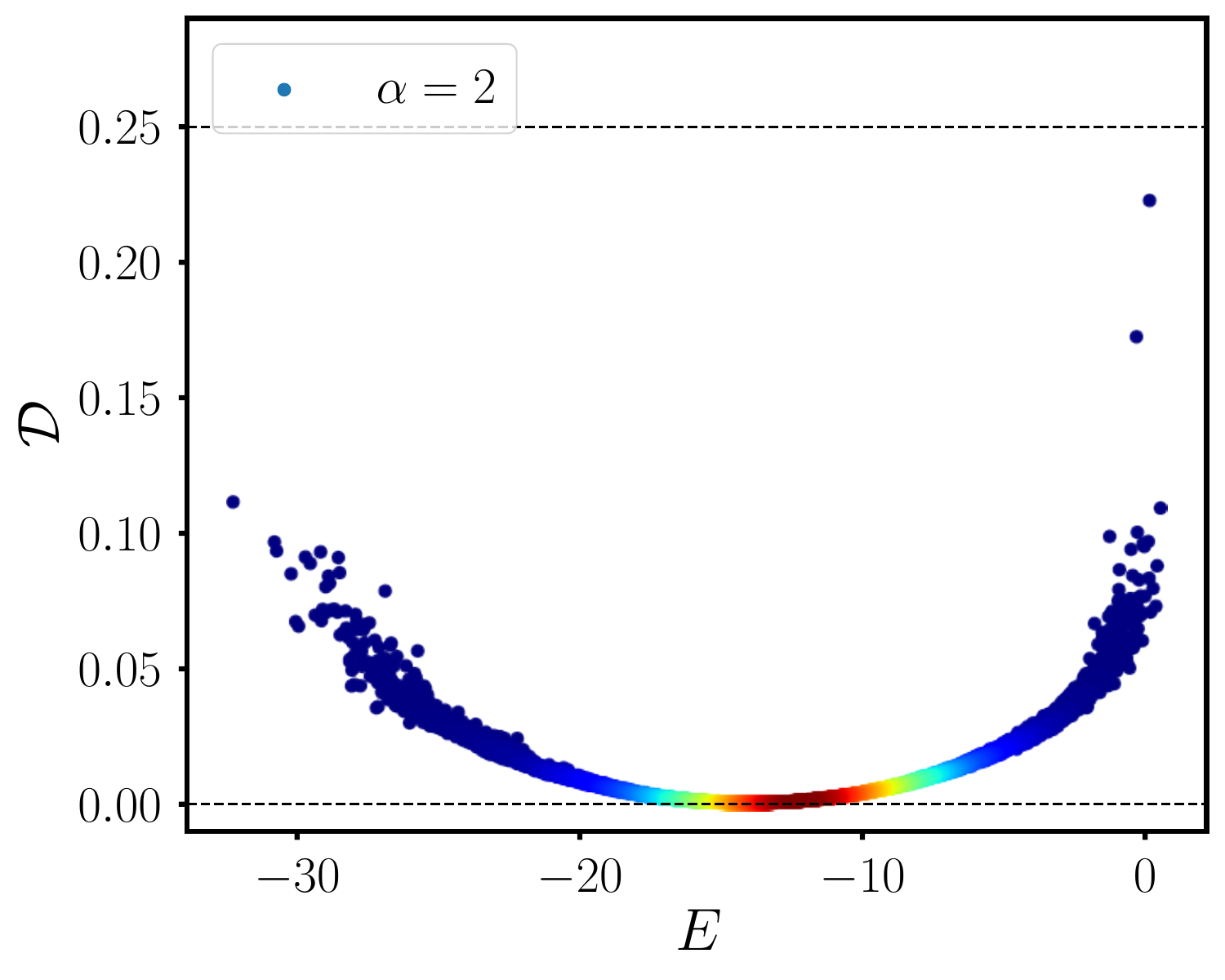}
        \includegraphics[scale=0.2]{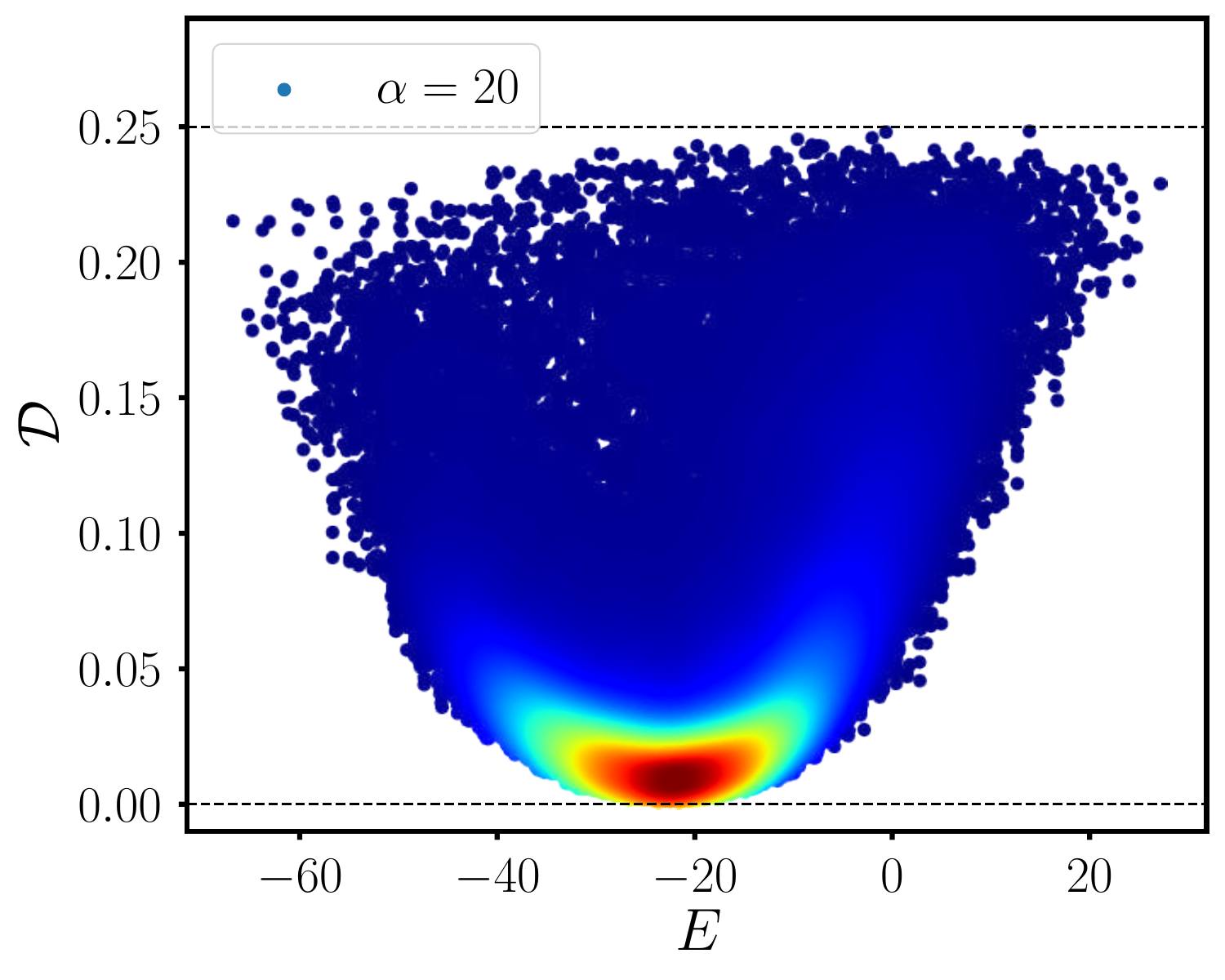}
        \includegraphics[scale=0.2]{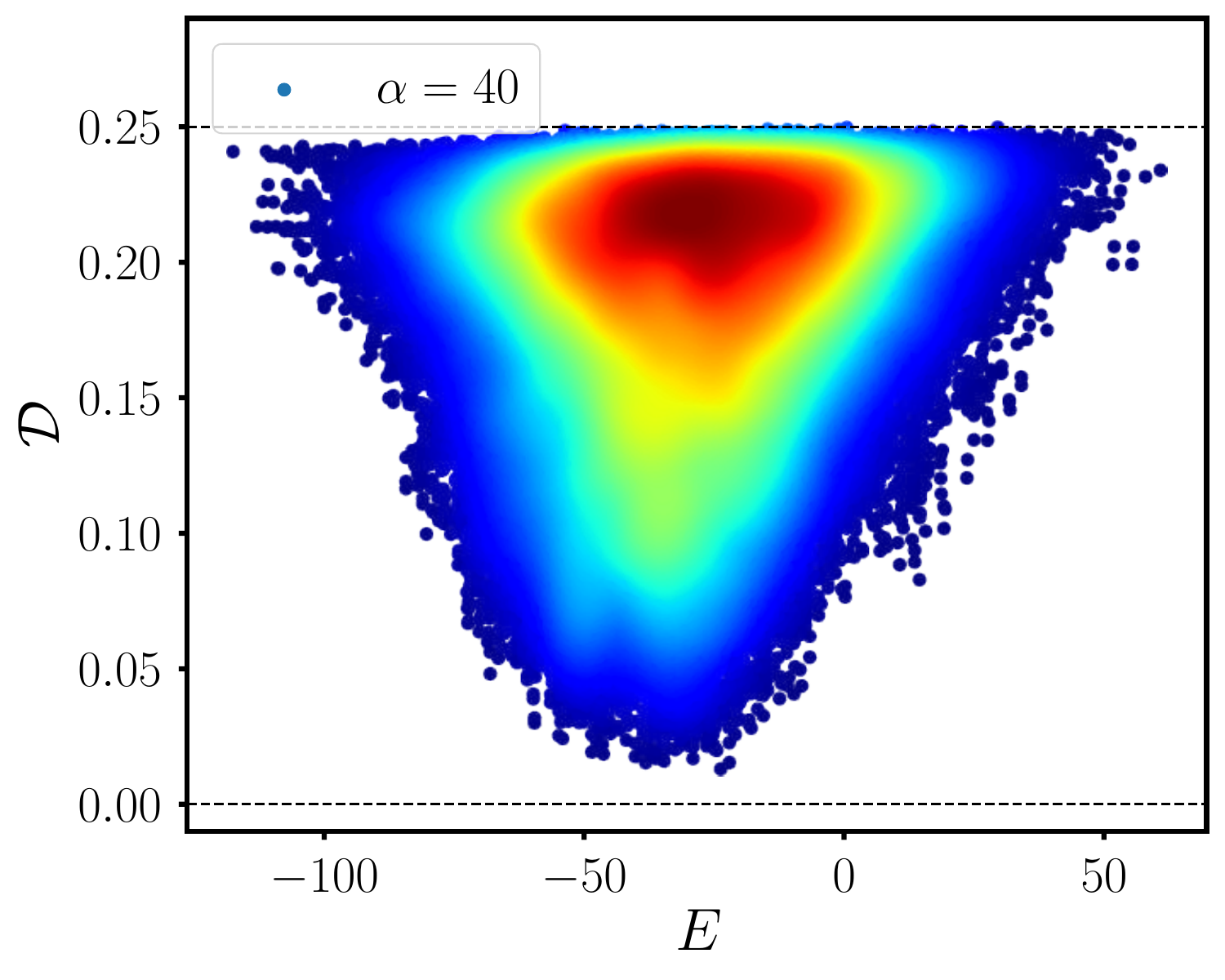}
        
        \caption{(Top row) Shannon entropy $S_1$ (Eq.~\ref{eq:Shannon}) as a function of energy for all the energy eigenstates of a particular disorder realization for a $12 \times 2$ ladder in the $C=-1$ sector at low ($\alpha=2$), intermediate ($\alpha=20$) and large ($\alpha=40$) disorder strengths. The top left panel displays a horizontal dotted line at the value of $\ln (\mathrm{HSD})$ for comparison. (Bottom row) $\mathcal{D}$ (Eq.~\ref{eq:formulaD}) as a function of energy for all the energy eigenstates for the same system. In all panels, the density of states is indicated by a color map where warmer color corresponds to a higher density of states.}
        \label{fig:shanEn12X2}
    \end{figure}
    
\end{widetext}
Since $\mathcal{D}_{\mathrm{th}} \ll 1$ while $\mathcal{D}=1/4$ for strictly localized states in the electric flux Fock states, it seems plausible that the value of $\mathcal{D}$ acts as a direct estimator of the concentration of an eigenstate in Fock space. We show numerical evidence that this is indeed the case in Fig.~\ref{fig:shanEn12X2} where the top three panels display the Shannon entropy $S_1$ for all energy eigenstates in the $C=-1$ sector for one disorder realization of a $12 \times 2$ for three different disorder strengths, while the bottom three panels show $\mathcal{D}$ calculated for each eigenstate from the same data sets. The density of states is indicated by the same color map, where warmer colors signify higher density of states, in all the panels. It is clear from the panels in Fig.~\ref{fig:shanEn12X2} that $\mathcal{D}$ \emph{mirrors} the Shannon entropy in all cases, i.e., weak, intermediate and strong disorder, with lower values of $\mathcal{D}$ for the mid-spectrum states implying higher values of $S_1$ and, hence, increased delocalization in Fock space. \\

\subsection{$\mathcal{D}$ as estimator of elementary plaquettes being active/inert in eigenstate}
\label{subsec:activeornot}

From Table.~\ref{tab:Opot}, we see that $\mathcal{D}_\square$ acts as a quantifier for whether a plaquette is active or inert in a given mid-spectrum eigenstate since $\mathcal{D}_{\mathrm{th}} \ll 1$ while $\mathcal{D}_\square$ should be close to $1/4$ deep in the many-body localized phase. For small $\alpha$, we have verified that $\mathcal{D}_\square$ is close to $\mathcal{D}_{\mathrm{th}}$ (apart from finite-size fluctuations) in the mid-spectrum eigenstates, and thus all plaquettes are active as expected.

The behavior of $\mathcal{D}_\square$ for different elementary plaquettes of a ladder for mid-spectrum eigenstates is far more interesting for intermediate and large $\alpha$ (see more details in Sec.~\ref{subsec:FSS}). In Fig.~\ref{fig:opotspAlph100_12x2} and Fig.~\ref{fig:opotspAlph100_6x4}, we display certain chosen mid-spectrum eigenstates from a given disorder realization of a $12 \times 2$ ladder and a $6 \times 4$ ladder at a large disorder strength of $\alpha=100$. While certain eigenstates indeed have all plaquettes to be inert, there is a hierarchy of \emph{thermal regions} of varying sizes starting from a few plaquettes all the way up to a system-spanning region of active plaquettes that are connected to each other in other neighboring mid-spectrum eigenstates. We see that mid-spectrum eigenstates that have a bigger number of active plaquettes also have a smaller value of $\mathcal{D}$ at large $\alpha$. Deep inside a many-body localized phase, these thermal regions should be finite and should not scale with system size in any typical mid-spectrum eigenstate to ensure stability of MBL. The fact that there is a small, albeit significant, probability to have large thermal regions in mid-spectrum eigenstates even at large $\alpha$ makes the distribution $p(\mathcal{D})$ to be non-trivial even when $\alpha \gg 1$ and this will be discussed in the next section.

While $n$ connected plaquettes, each with a small $\alpha |R_\square|$ at large $\alpha$ compared to the bulk, can arise from purely statistical reasons for uniformly distributed random numbers and act as \emph{thermal regions} because of an effectively smaller disorder locally, the probability of such events scale as $O(1/\alpha^n)$ and thus decrease very rapidly with increasing $n$ at large $\alpha$. The actual values of $R_\square$ for the particular disorder realizations shown in the top panels of both Fig.~\ref{fig:opotspAlph100_12x2} and Fig.~\ref{fig:opotspAlph100_6x4} only show certain $n=1$ regions with a low $\alpha |R_\square|$ compared to the bulk and rule out this simple interpretation.  

\begin{widetext}

    \begin{figure}
        \centering
        \includegraphics[scale=0.195]{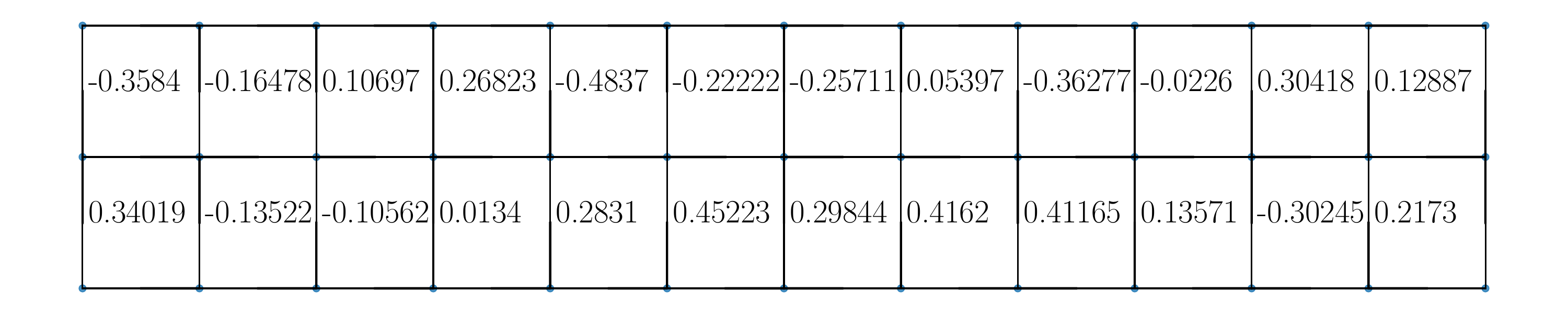}
        \includegraphics[scale=0.175]{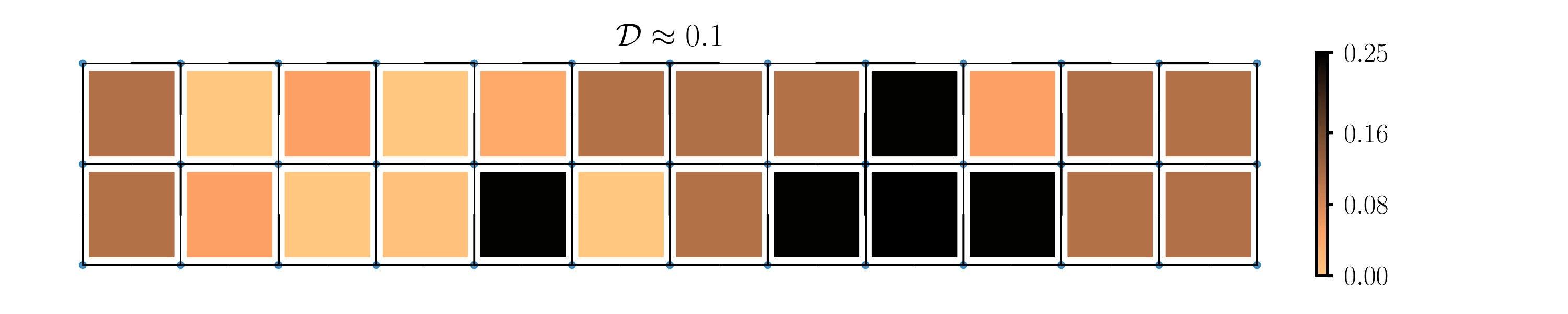} 
        \includegraphics[scale=0.175]{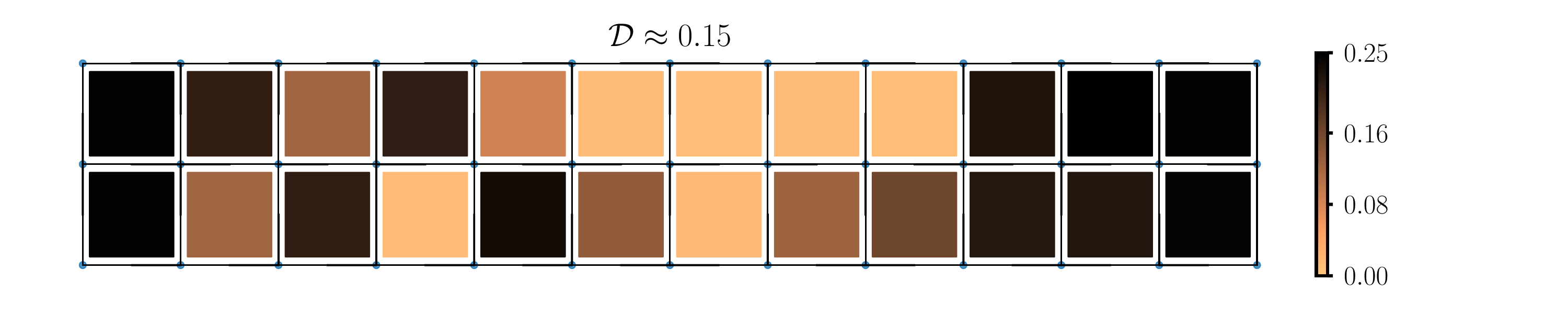} 
        \includegraphics[scale=0.175]{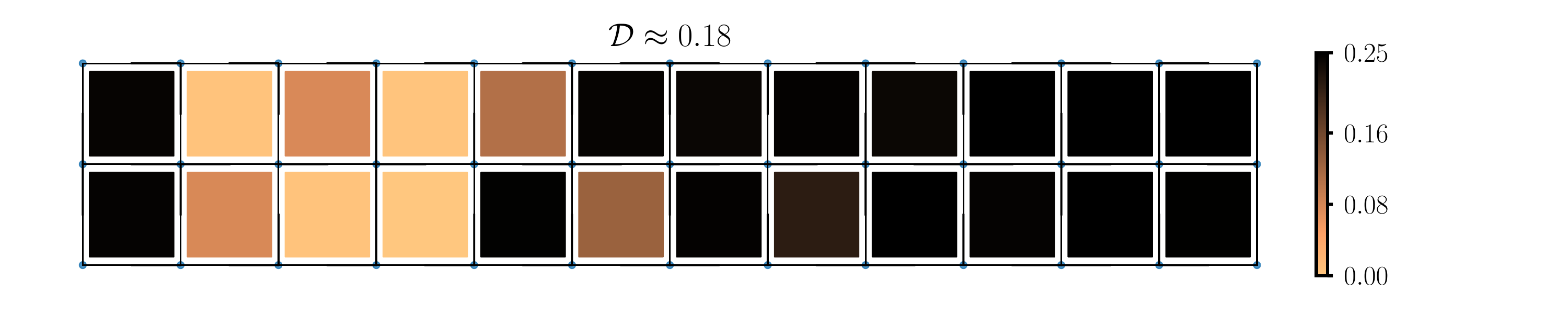} 
        \includegraphics[scale=0.175]{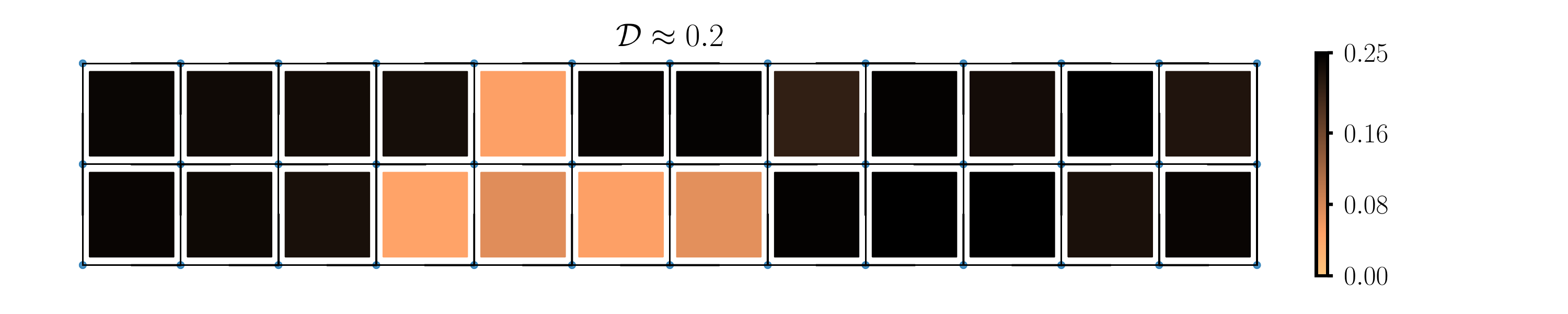} 
        \includegraphics[scale=0.175]{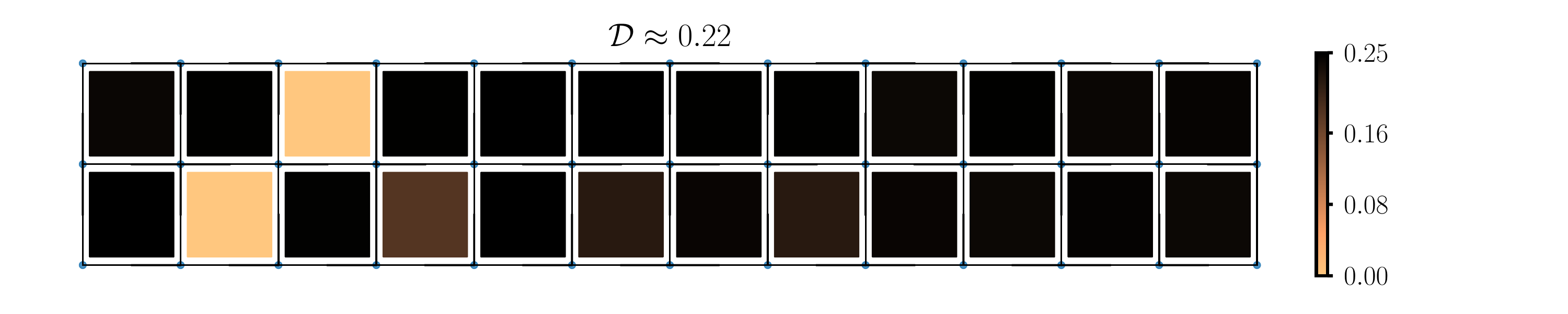} 
        \includegraphics[scale=0.175]{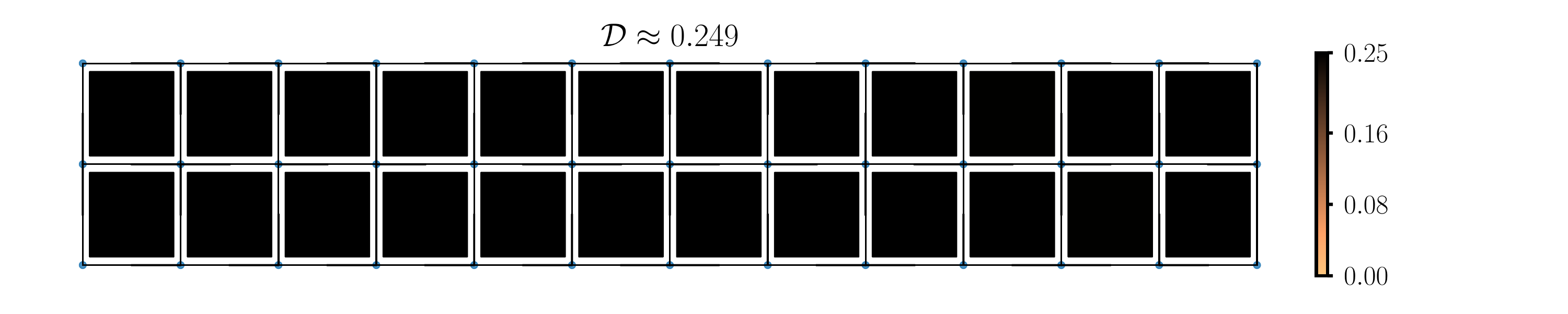} 
        \caption{$\mathcal{D}_\square$ shown for a few selected mid-spectrum eigenstates from a particular disorder realization for a ladder of $12 \times 2$ in the $C=-1$ sector at a large value of disorder, $\alpha=100$. The $R_\square$ for each plaquette is specified in the top panel for the particular disorder realization used while in the remaining $6$ panels, darker colors indicate $\mathcal{D}_\square$ to be closer to $1/4$ while lighter colors indicate progressively more active plaquettes. For these $6$ panels, the value of $\mathcal{D}$ for the eigenstate is also indicated. }
        \label{fig:opotspAlph100_12x2}
    \end{figure}

\begin{figure}
        \centering
        \includegraphics[scale=0.16]{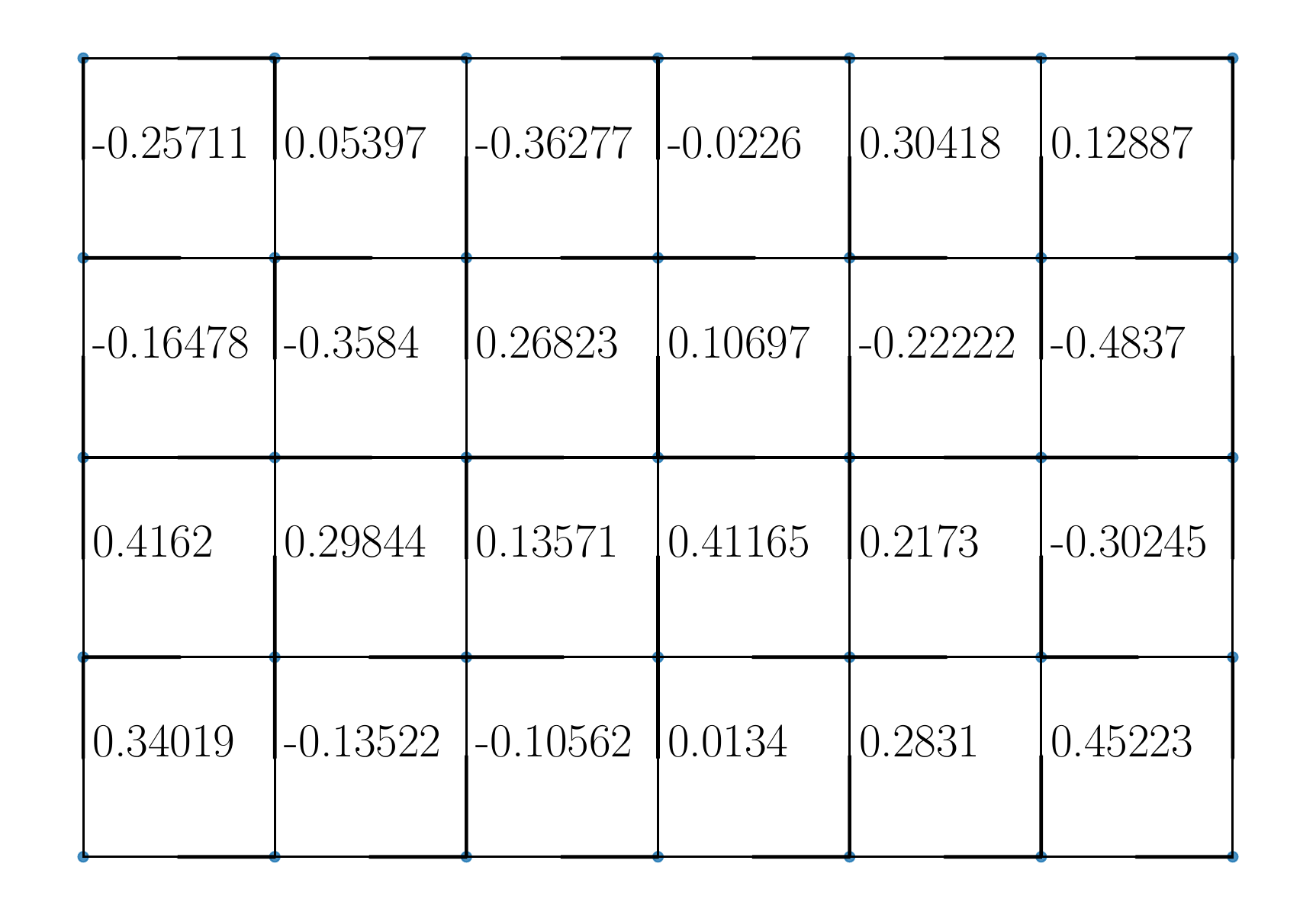}\\
        \includegraphics[scale=0.185]{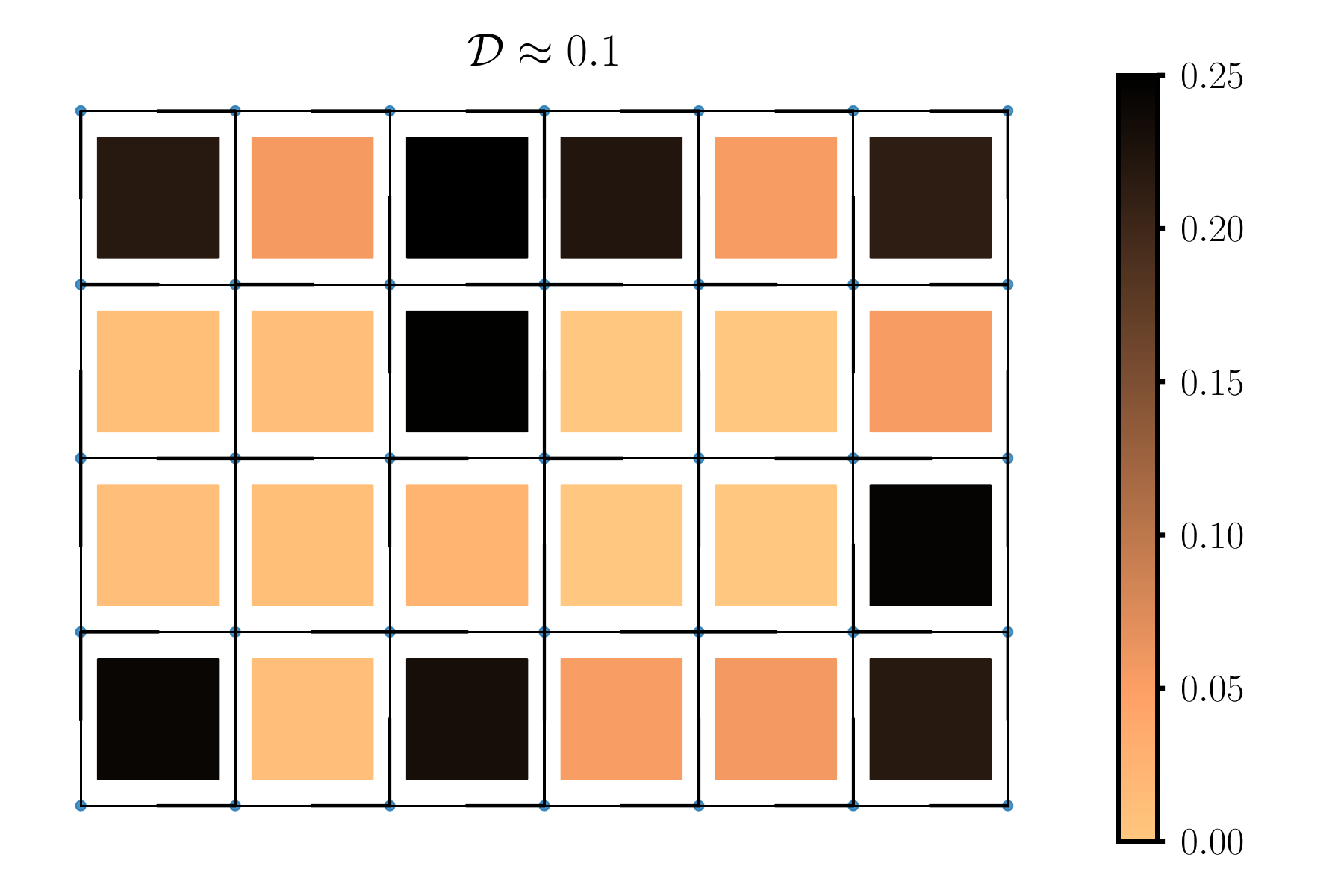} 
        \includegraphics[scale=0.185]{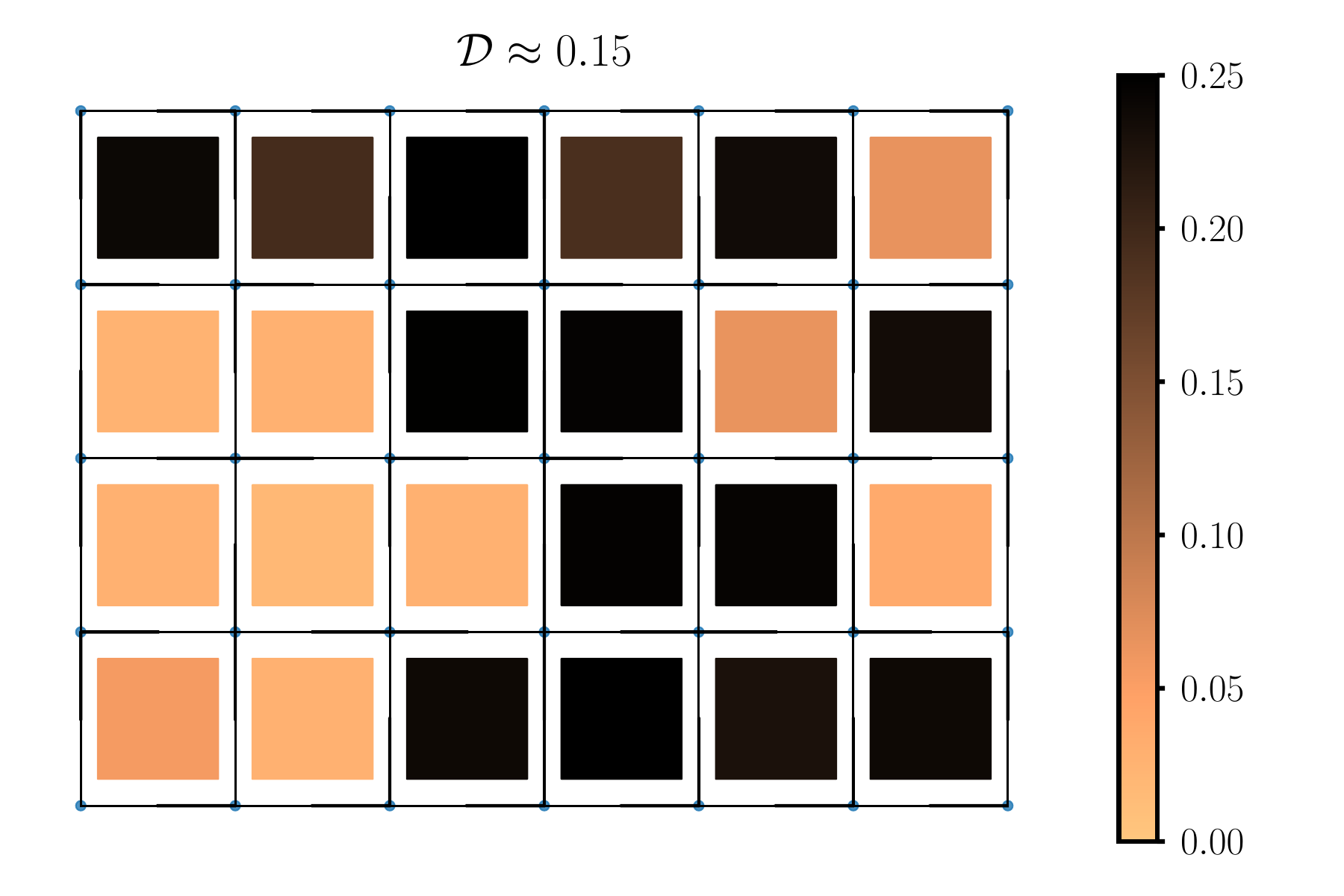} 
        \includegraphics[scale=0.185]{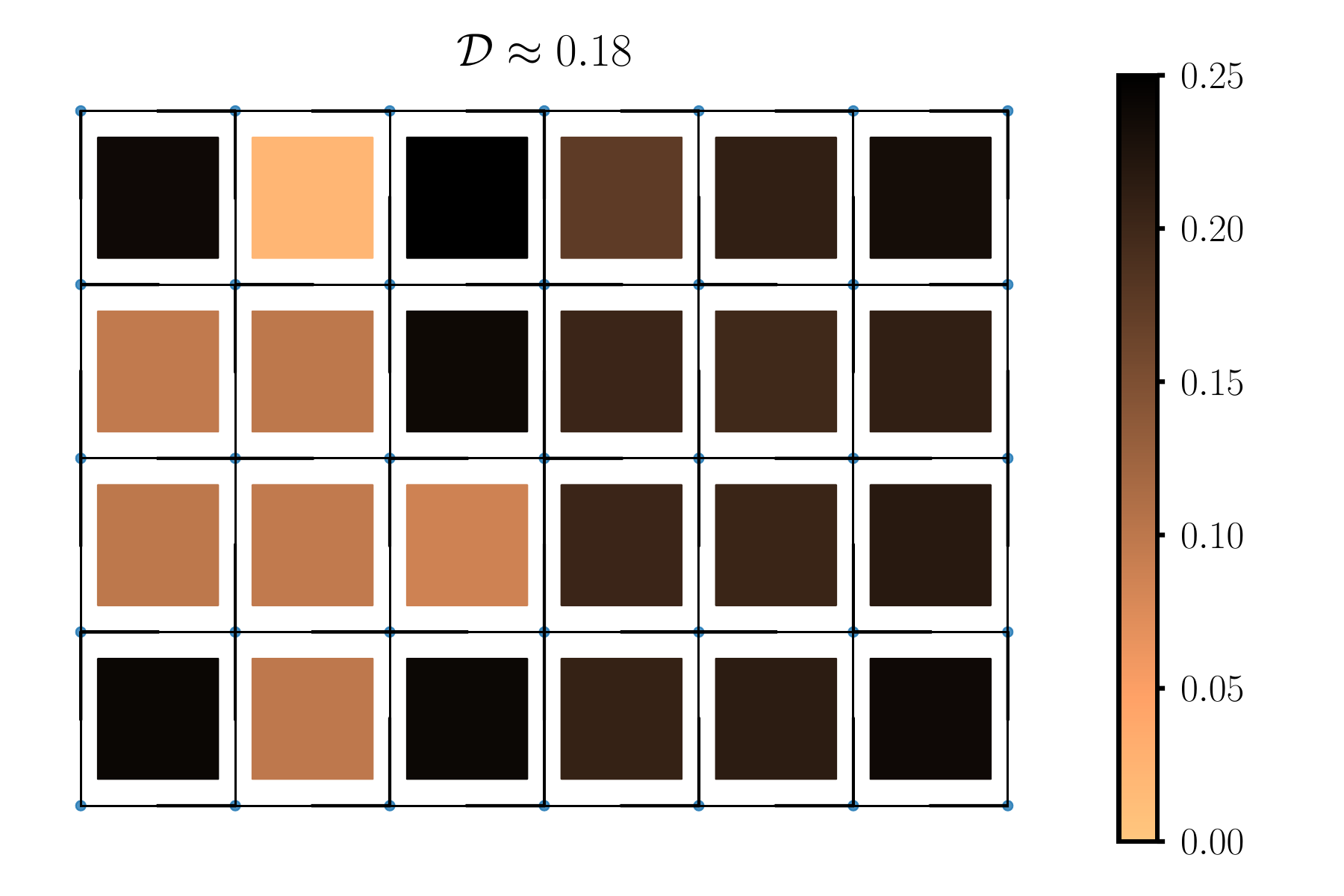} 
       \includegraphics[scale=0.18]{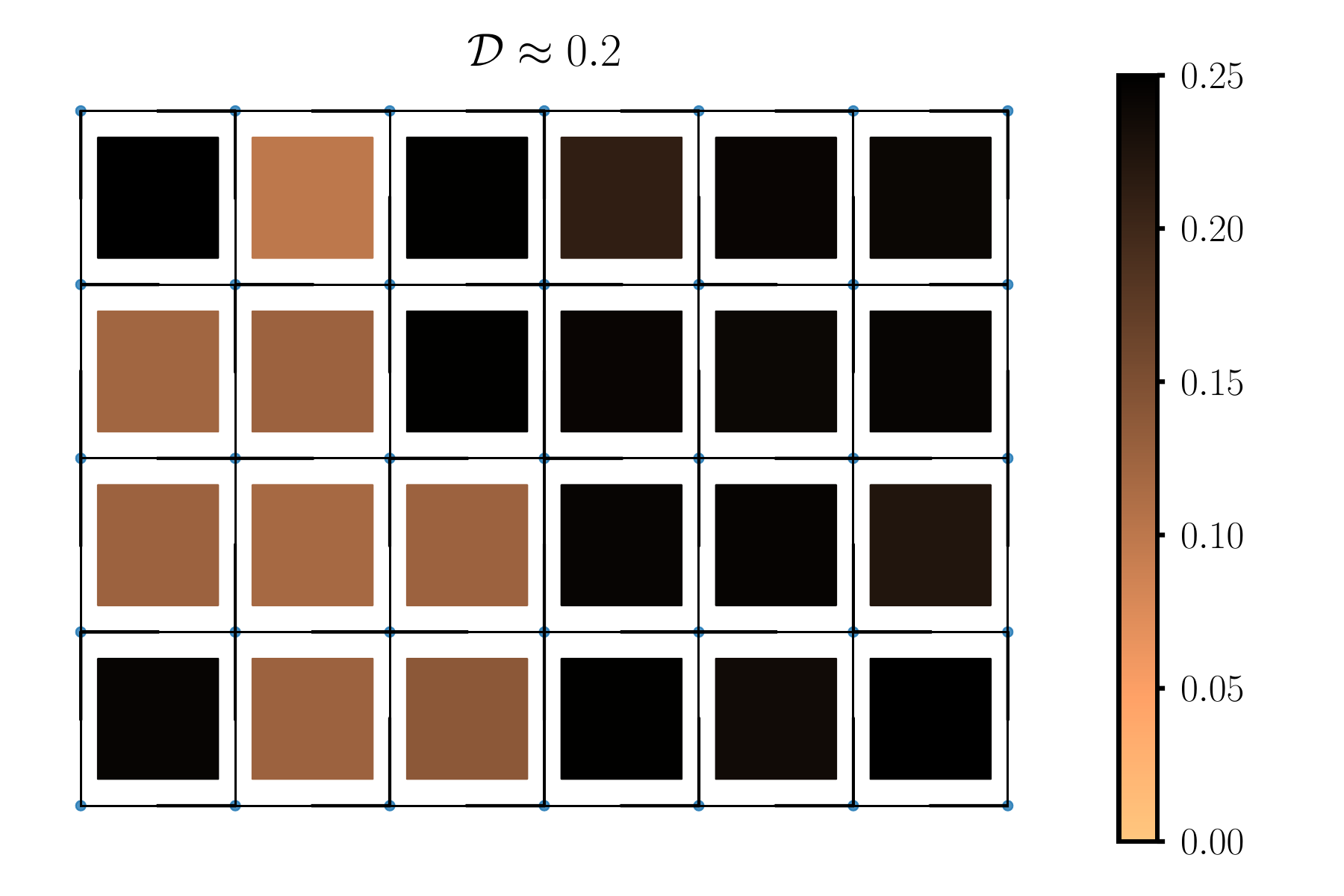} 
        \includegraphics[scale=0.18]{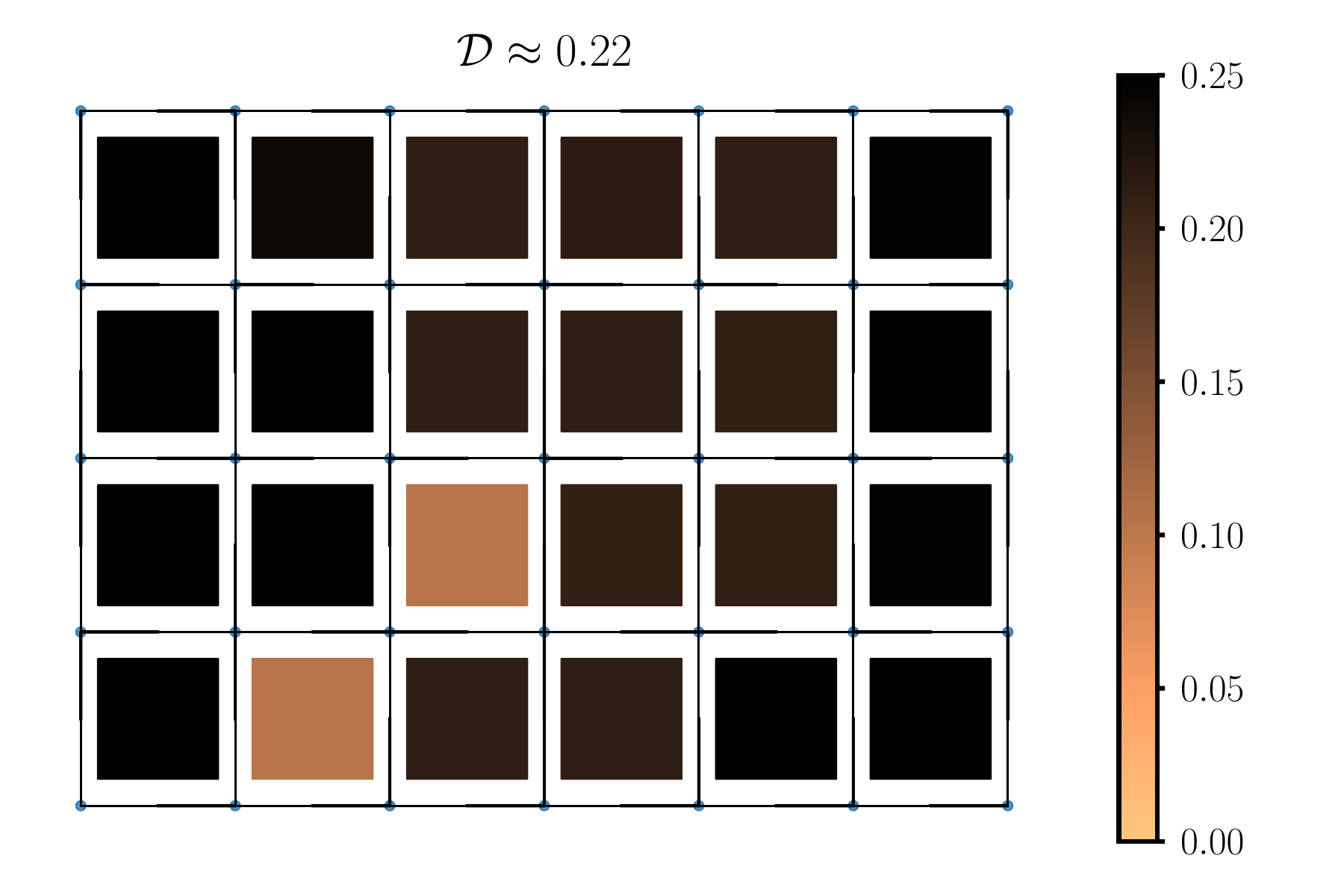} 
        \includegraphics[scale=0.18]{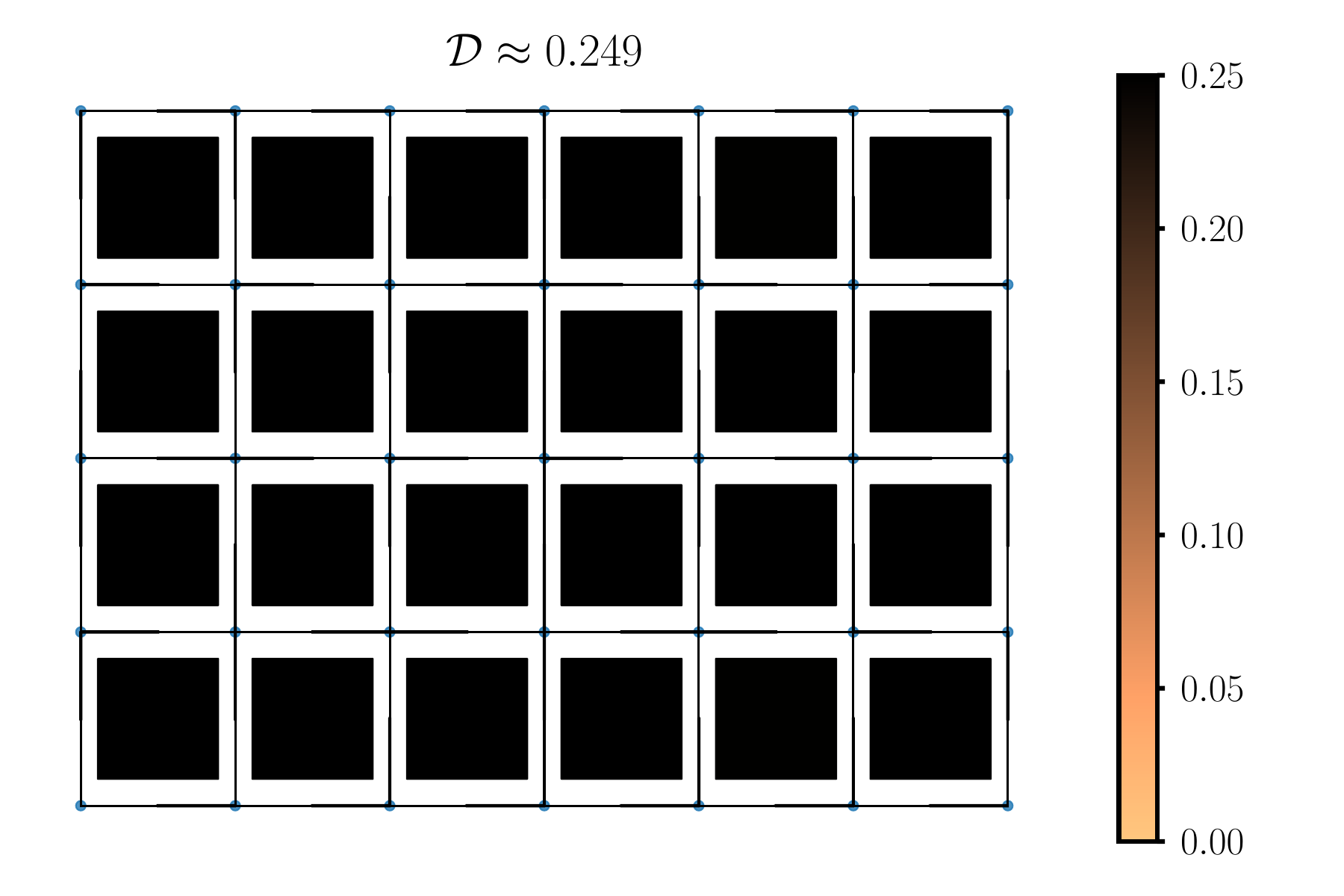} 
        \caption{$\mathcal{D}_\square$ shown for a few selected mid-spectrum eigenstates from a particular disorder realization for a ladder of $6 \times 4$ in the $C=-1$ sector at a large value of disorder, $\alpha=100$. The $R_\square$ for each plaquette is specified in the top panel for the particular disorder realization used while in the remaining $6$ panels, darker colors indicate $\mathcal{D}_\square$ to be closer to $1/4$ while lighter colors indicate progressively more active plaquettes. For these $6$ panels, the value of $\mathcal{D}$ for the eigenstate is also indicated. }
        \label{fig:opotspAlph100_6x4}
    \end{figure}

\end{widetext}

\subsection{Estimating MBL transition using finite-size behavior of  $p(\mathcal{D})$}
\label{subsec:FSS}
In this section, we will consider the disorder-averaged normalized distribution function, $p(\mathcal{D})$, from ED data for a number of disorder realizations for $8 \times 2$, $10 \times 2$, $12 \times 2, 14 \times 2$ and $6 \times 4$ ladders for various values of $\alpha$. For any given ladder dimension and $\alpha$, we consider several independent disorder realizations and calculate the value of $\mathcal{D}$ for each mid-spectrum eigenstate from that realization. As stated earlier, we divide the total energy bandwidth in $25$ equal bins and choose the bin that contains the maximum number of eigenstates from each disorder realization for this purpose. While we use $500$ disorder realizations for $8 \times 2$ and $10 \times 2$ ladders and $50$ disorder realizations for $6 \times 4$ ladders at each $\alpha$, for the $12 \times 2$ ladder with the largest HSD, we use $10$ disorder realizations for $\alpha < 20$, $20$ disorder realizations for $\alpha$ between $20$ and $30$ and $30$ disorder realizations for higher values of $\alpha$. The entire dataset for the values of $\mathcal{D}$ for the mid-spectrum eigenstates of all the disorder realizations for a given $L_x \times L_y$ and $\alpha$ is then divided into $40$ bins to construct the normalized distribution $p(\mathcal{D})$. For ladders of size $14 \times 2$, we calculate $p (\mathcal{D})$ at various $\alpha$ using shift-invert techniques as explained in Appendix.~\ref{app:midspectrum}. We check for the consistency of the whole procedure for smaller ladders of dimension $12 \times 2$ by comparing $p(\mathcal{D})$ obtained from this method to that obtained from full ED (Appendix~\ref{app:midspectrum}). It is useful to note that distributions of local operators such as local magnetization in mid-spectrum eigenstates~\cite{Luitz_dis1, Luitz_dis2} of the random-field XXZ $S=1/2$ model on finite chains has been studied previously to understand MBL in unconstrained systems.

\begin{figure}
    \centering
    \includegraphics[scale=0.2]{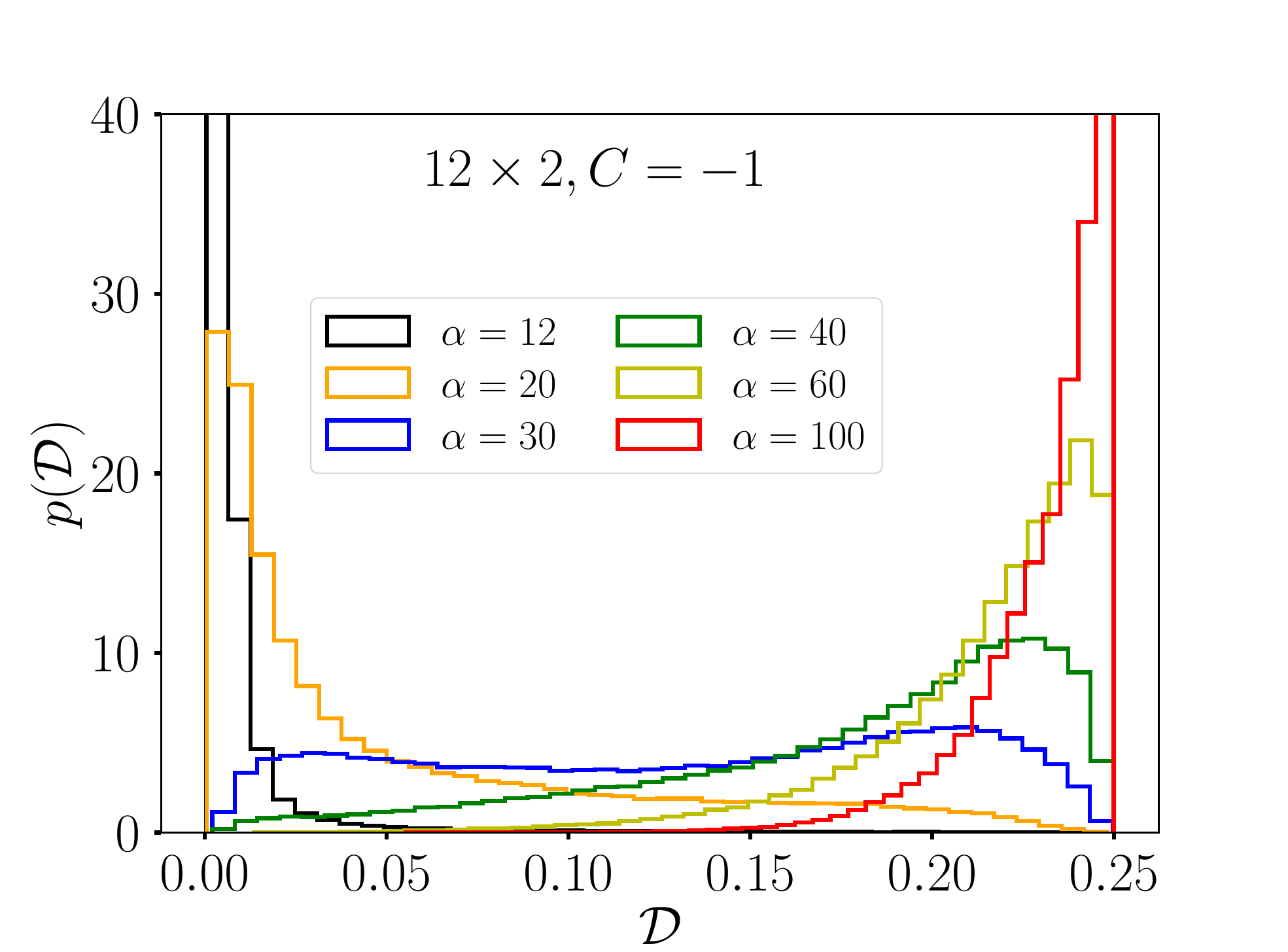}
    \includegraphics[scale=0.23]{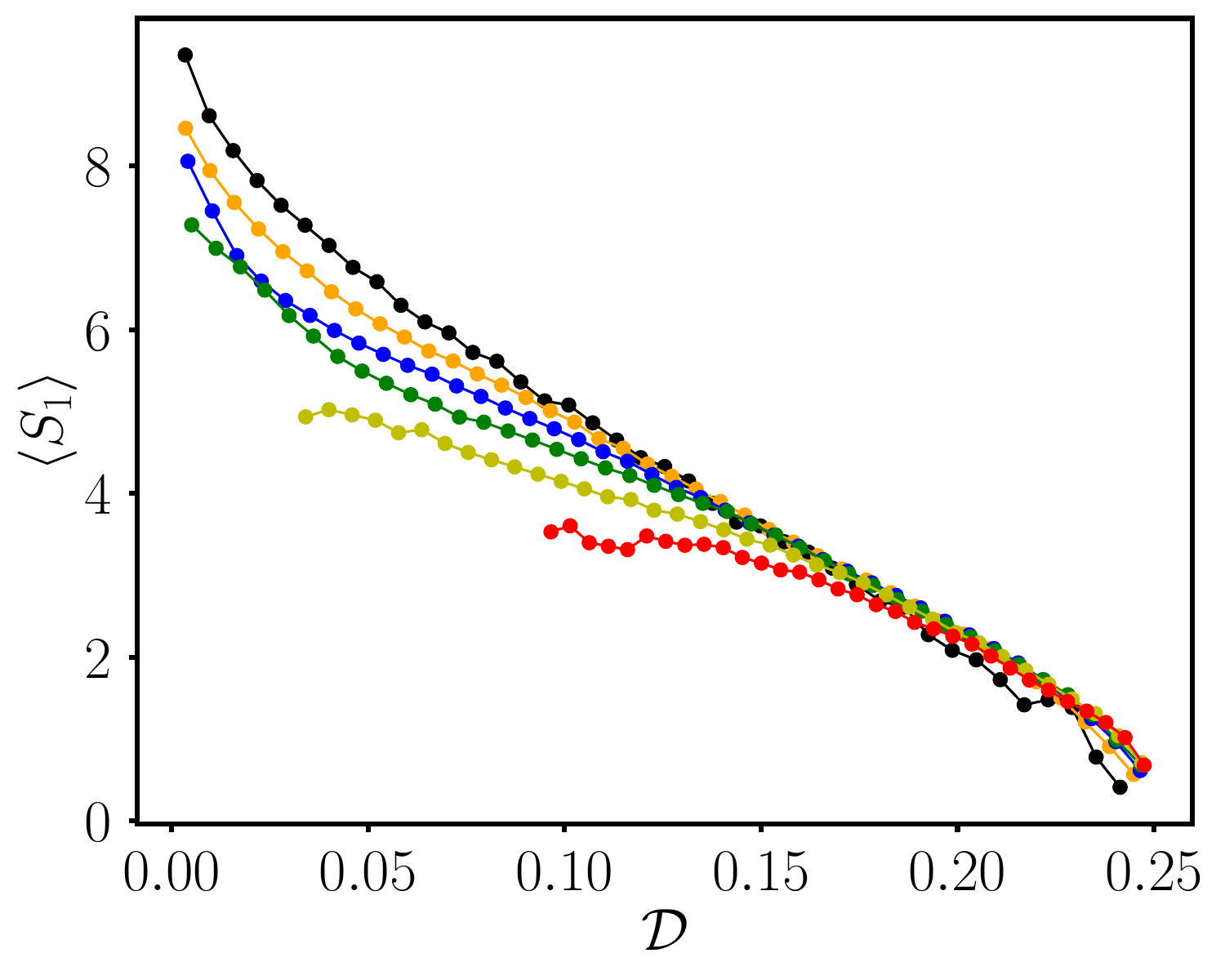}
    \caption{(Top panel) Evolution of $p(\mathcal{D})$ as a function of $\alpha$ for a ladder of dimension $12 \times 2$. The y-axis is cut off at $40$ for clarity. (Bottom panel) Average Shannon entropy $\langle S_1 \rangle$ for the mid-spectrum eigenstates in each bin used to construct $p(\mathcal{D})$ for a given $\alpha$ in the top panel. The same colors are used to distimguish different values of disorder strength, $\alpha$, in both the panels.}
    \label{fig:pD12times2}
\end{figure}

\begin{figure}
    \centering
    \includegraphics[scale=0.2]{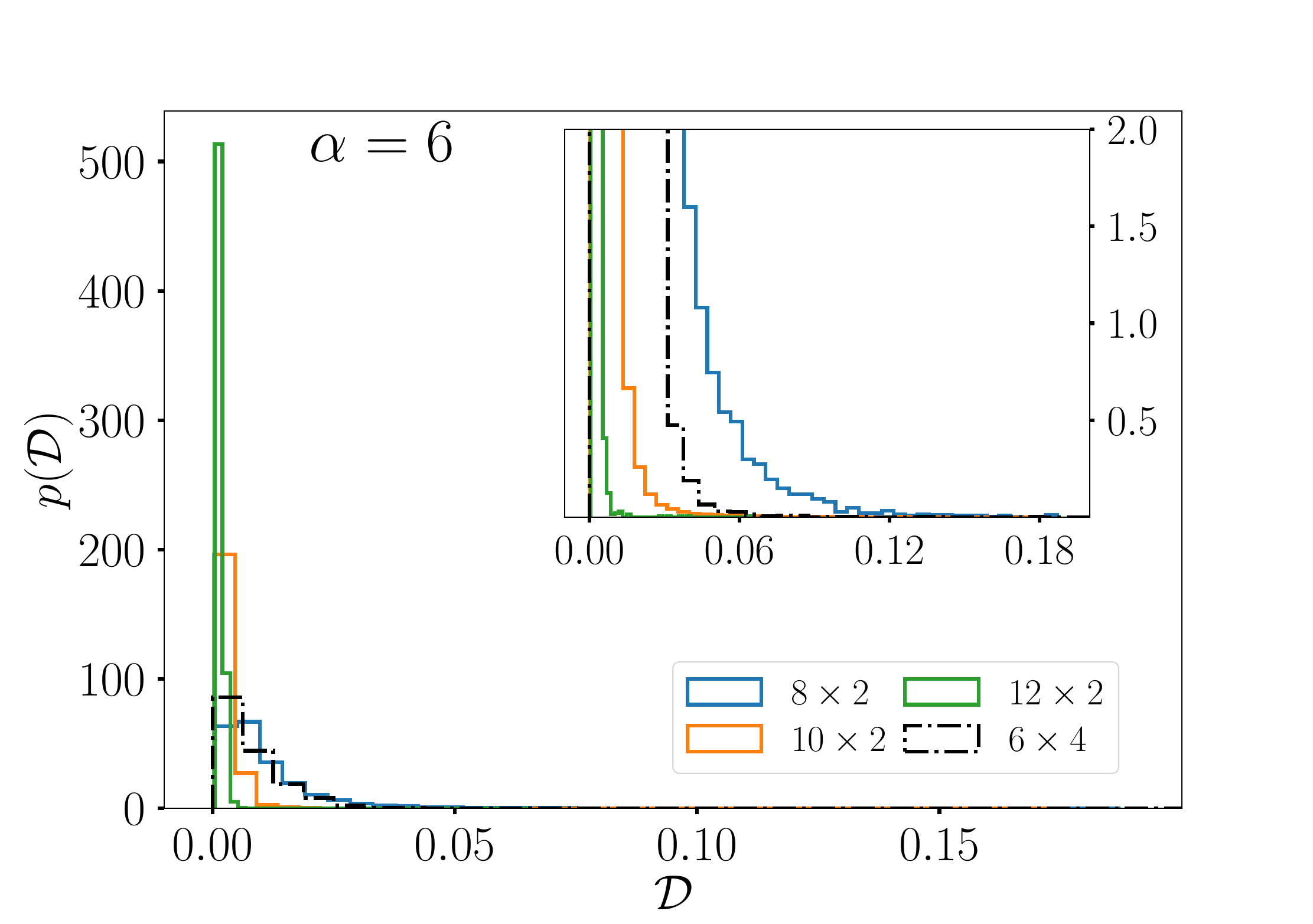}
    \includegraphics[scale=0.2]{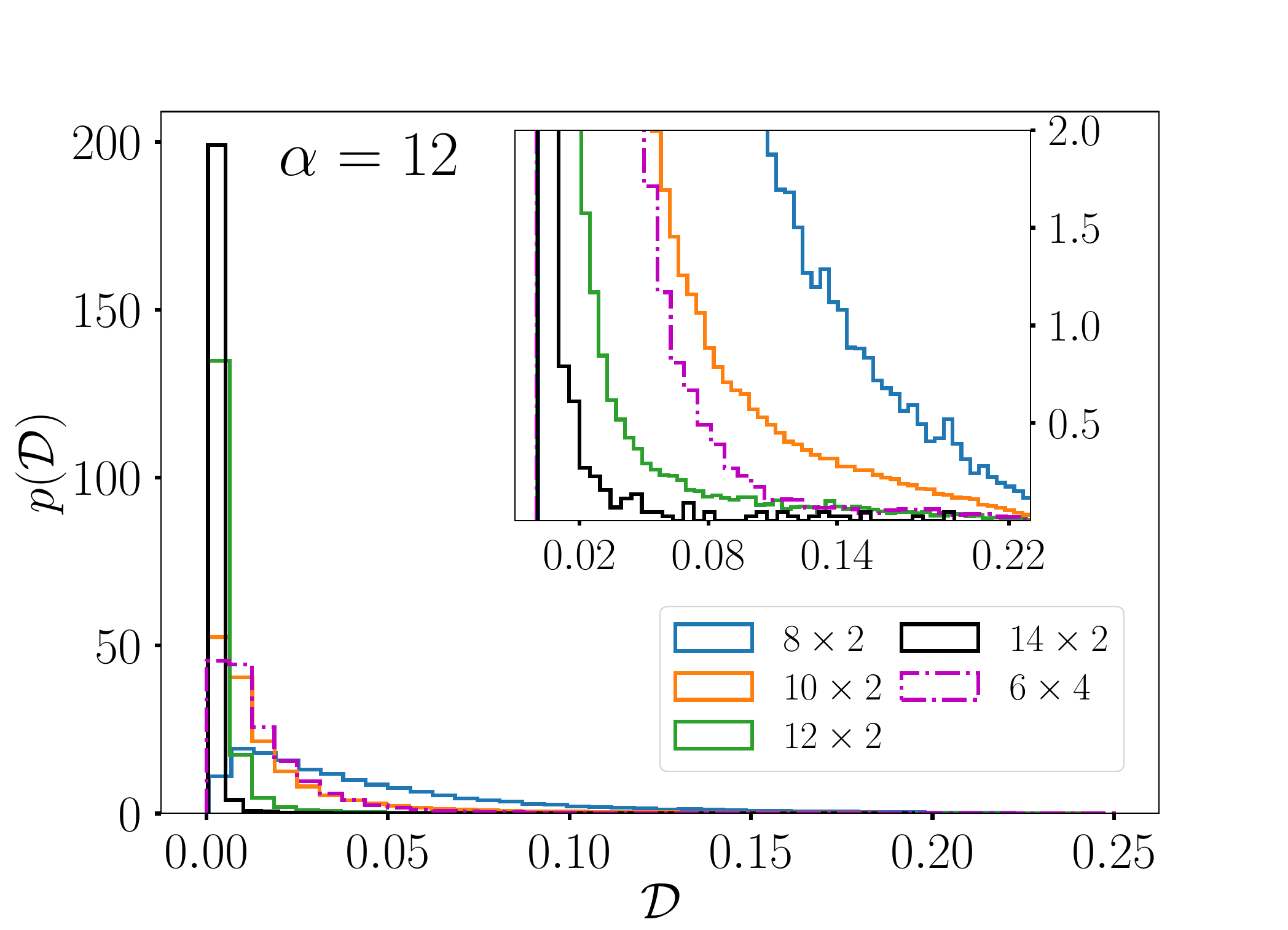} 
    \caption{Behavior of $p(\mathcal{D})$ for $8 \times 2$, $10 \times 2$, $12 \times 2$ and $6 \times 4$ ladders for $\alpha=6$ (top panel) and $\alpha=12$ (bottom panel). The insets in both panels show the behavior of the tails of $p(\mathcal{D})$ prominently.}
    \label{fig:pDsmalla}
\end{figure}
\begin{figure}
  \centering
  \includegraphics[scale=0.2]{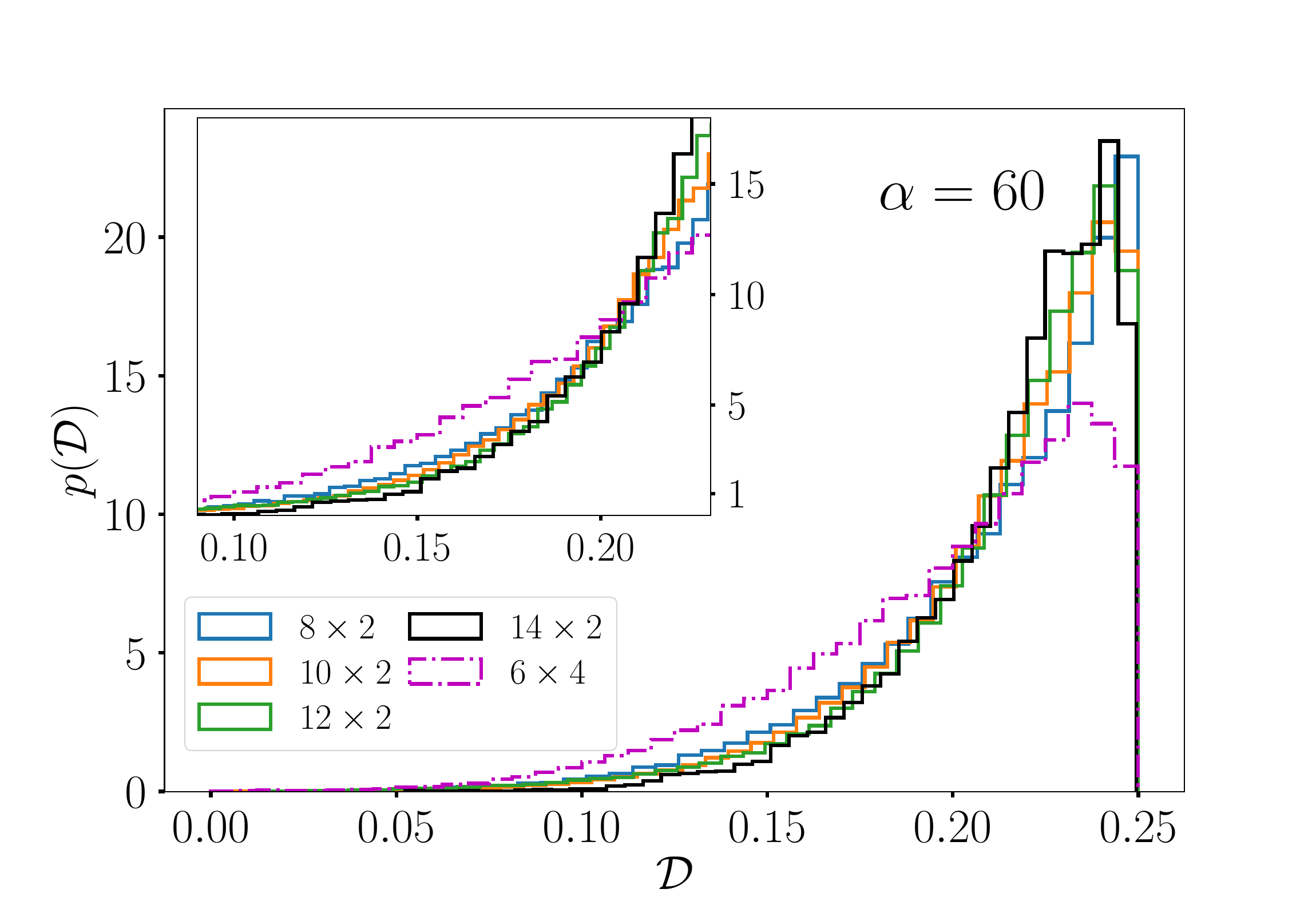}
    \includegraphics[scale=0.2]{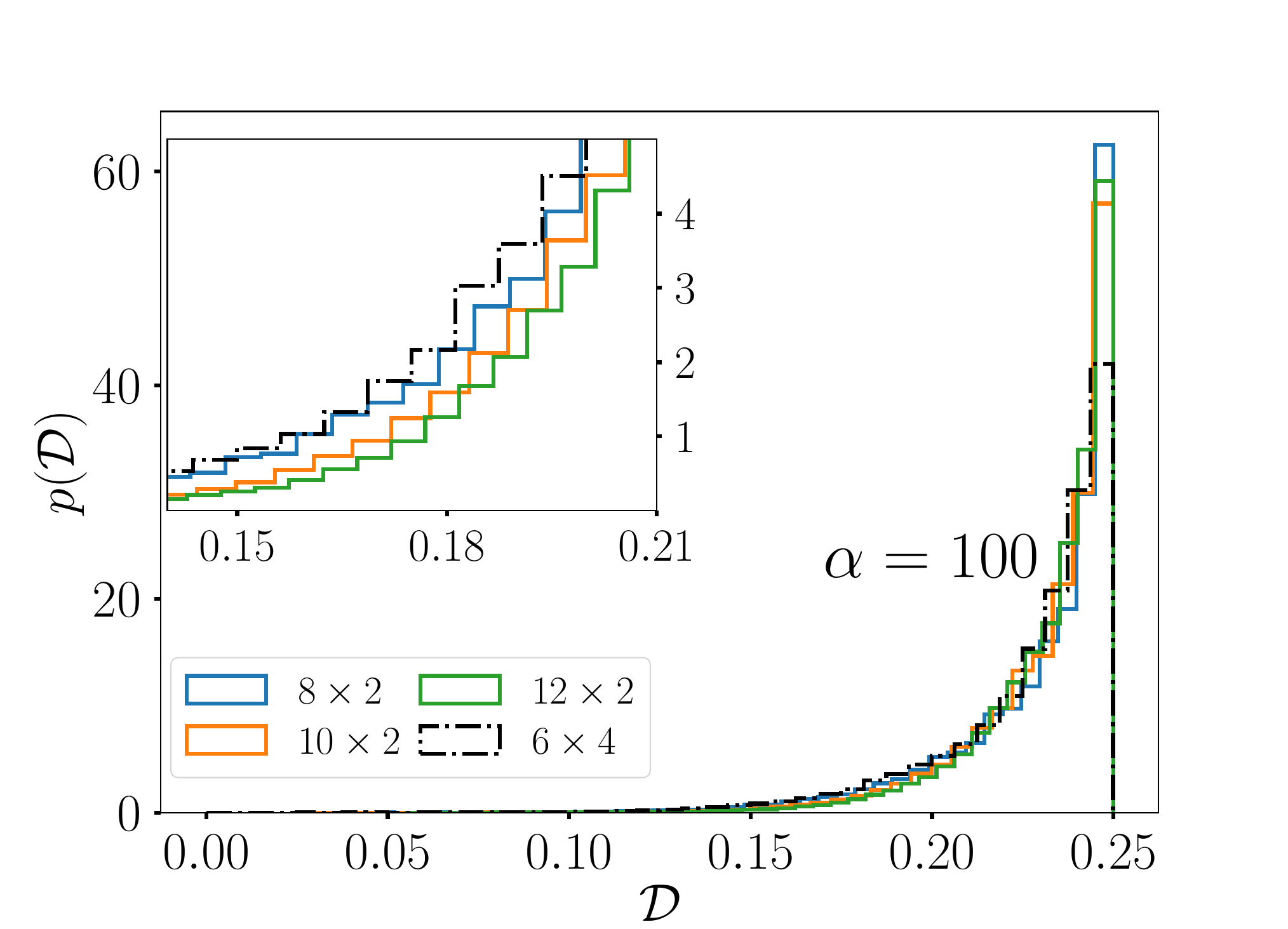}
    \caption{Behavior of $p(\mathcal{D})$ for $8 \times 2$, $10 \times 2$, $12 \times 2$ and $6 \times 4$ ladders for $\alpha=60$ (top panel) and $\alpha=100$ (bottom panel). The insets in both panels show the behavior of the tails of $p(\mathcal{D})$ prominently.}
    \label{fig:pDlargea}
\end{figure}

Let us first consider how $p(\mathcal{D})$ is expected to behave when $\alpha \ll 1$ and when $\alpha \rightarrow \infty$ (i.e., infinite disorder limit) for fixed ladder widths $L_y=2$ or $4$ if $L_x \gg 1$. In both limits, $p(\mathcal{D})$ is
  expected to approach a delta function distribution for large systems which motivated us to look at this particular estimator. For $\alpha \ll 1$, and in fact for all $\alpha$ where ETH holds, we expect $p(\mathcal{D}) \rightarrow \delta(\mathcal{D}-\mathcal{D}_{\mathrm{th}}(L_y))$ where $\mathcal{D}_{\mathrm{th}}(L_y)$ $\approx 0.0031 (0.0070)$ for $L_y=2 (4)$ by extrapolating the values given in Table.~\ref{tab:Opot}. For $\alpha \rightarrow \infty$ (infinite disorder limit), we instead get $p(\mathcal{D}) \rightarrow \delta(1/4-\mathcal{D})$. Assuming adiabatic continuity for $\alpha \gg 1$, which is expected deep in the many-body localized phase (if it exists), $\mathcal{D}$ will decrease from $1/4$ for typical mid-spectrum eigenstates due to perturbatively small quantum fluctuations at large, but finite, $\alpha \gg 1$.
We will see below that while the finite-size behavior of $p(\mathcal{D})$ indicates a rapid convergence to $\delta(\mathcal{D}-\mathcal{D}_{\mathrm{th}}(L_y))$ for a range of $\alpha$ (e.g., see Fig.~\ref{fig:pDsmalla}),
the finite-size behavior of $p(\mathcal{D})$ for $\alpha \sim 60-100$ seems more subtle (e.g., see Fig.~\ref{fig:pDlargea}) from data for the available system sizes and suggests a many-body localized phase with significant quantum fluctuations.

In Fig.~\ref{fig:pD12times2} (top panel), we show the behavior of $p(\mathcal{D})$ for a $12 \times 2$ ladder as a function of disorder strength $\alpha$. While the distribution has a maximum in the neighborhood of $\mathcal{D}=0$ both for $\alpha=12$ and $\alpha=20$, the tail of the distribution is far more extended at $\alpha=20$ as compared to $\alpha=12$ due to mid-spectrum eigenstates with larger thermally inactive regions becoming more probable at larger $\alpha$. The distribution becomes extremely broad for $\alpha=30$ indicating an instability towards MBL at this system size before developing a pronounced maximum in the neighborhood of $\mathcal{D}=1/4$ for $\alpha \geq 40$. The weight in the tail of the distribution decreases slowly as one increases the disorder from $\alpha =40$ to $\alpha=100$. However, even at a large disorder of $\alpha=100$, there is significant weight in the tail of $p(\mathcal{D})$, consistent with the presence of thermally active regions at various length scales as seen for the mid-spectrum eigenstates in Fig.~\ref{fig:opotspAlph100_12x2} for one particular disorder realization (as well as for the case of $6 \times 4$ ladder, see Fig.~\ref{fig:opotspAlph100_6x4}). In Fig.~\ref{fig:pD12times2} (bottom panel), we show the average Shannon entropy $\langle S_1 \rangle$ for the mid-spectrum eigenstates in each of the bins that were used to construct $p(\mathcal{D})$ for a given $\alpha$ in Fig.~\ref{fig:pD12times2} (top panel). The behaviour of this quantity illustrates that the lower (higher) $\mathcal{D}$ eigenstates have a higher (lower) average Shannon entropy and are thus more delocalized (localized) in Fock space irrespective of the value of $\alpha$.

To understand whether a ladder of width $L_y$ satisfies ETH or demonstrates MBL for a fixed $\alpha$ in the thermodynamic limit of $L_x \gg 1$, one needs to compare the behavior of $p(\mathcal{D})$ at that $\alpha$ for different ladder dimensions and use finite-size scaling to decipher whether $p(\mathcal{D})$ develops a maximum at $\mathcal{D} \ll 1$ ($\mathcal{D} \sim 1/4$) as the system size is increased. Since $\mathcal{D}$ is an intensive estimator, $p(\mathcal{D})$ for different system sizes can be directly compared to each other.

We first consider $\alpha=6$ and $\alpha=12$ as shown in Fig.~\ref{fig:pDsmalla} and focus on $p(\mathcal{D})$ for the $L_x \times 2$ ladders. It is clear from both panels in Fig.~\ref{fig:pDsmalla} that the weight in $p(\mathcal{D})$ rapidly shifts to the vicinity of $\mathcal{D} \approx 0$ as a function of increasing $L_x$. This is more clearly visible from the inset of both the panels. The insets also show that the tails of the distributions have non-vanishing weights for much larger values of $\mathcal{D}$ at $\alpha=12$ (Fig.~\ref{fig:pDsmalla}, bottom panel) compared to $\alpha=6$ (Fig.~\ref{fig:pDsmalla}, top panel). This can be interpreted as the emergence of bigger locally inert regions in the mid-spectrum eigenstates as the disorder strength, $\alpha$, is increased from $6$ to $12$. The appearance of a long tail in $p(\mathcal{D})$ for
  $\alpha=12$ for these finite-sized systems has a dynamical consequence when the time evolution of
  $\Opots$ is probed from typical Fock states as will be discussed in the next section. However, the probability of finding such regions with larger inert regions (leading to larger values of $\mathcal{D}$) in mid-spectrum states rapidly decreases with the linear dimension of the ladder, $L_x$, as can be seen from the inset of Fig.~\ref{fig:pDsmalla} (bottom panel) by varying $L_x$ from $8$ to $14$. It is also interesting to note that for these disorder strengths, a wider ladder of dimension $6 \times 4$ has more weight in the tails away from $\mathcal{D} \approx 0$ compared to a $12 \times 2$ thin ladder composed of the same number of elementary plaquettes (insets of both panels in Fig.~\ref{fig:pDsmalla}) indicating the probability of finding bigger inert regions in typical mid-spectrum eigenstates of the wider ladder.    
\begin{figure}
    \centering
    \includegraphics[scale=0.2]{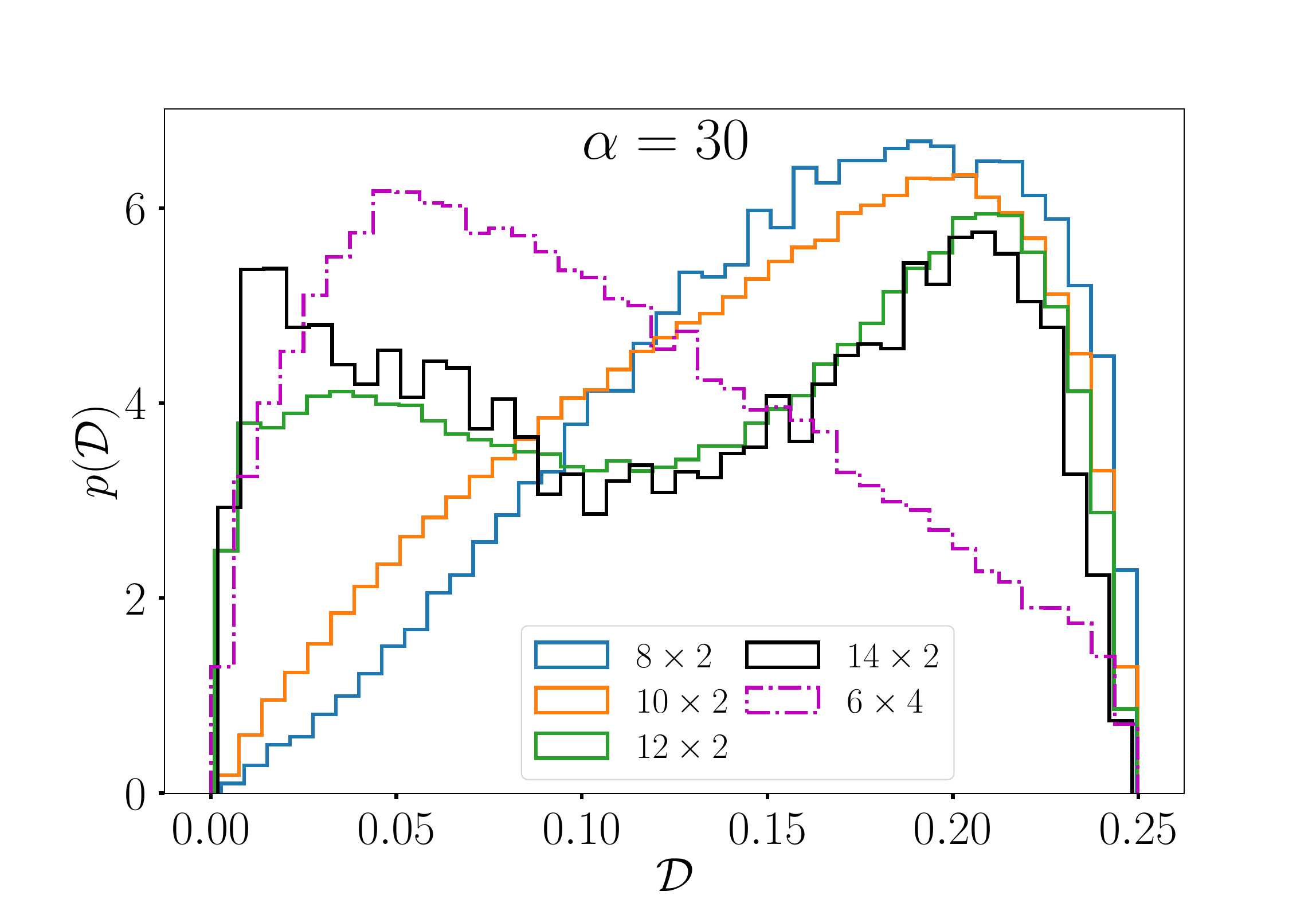}
    \includegraphics[scale=0.2]{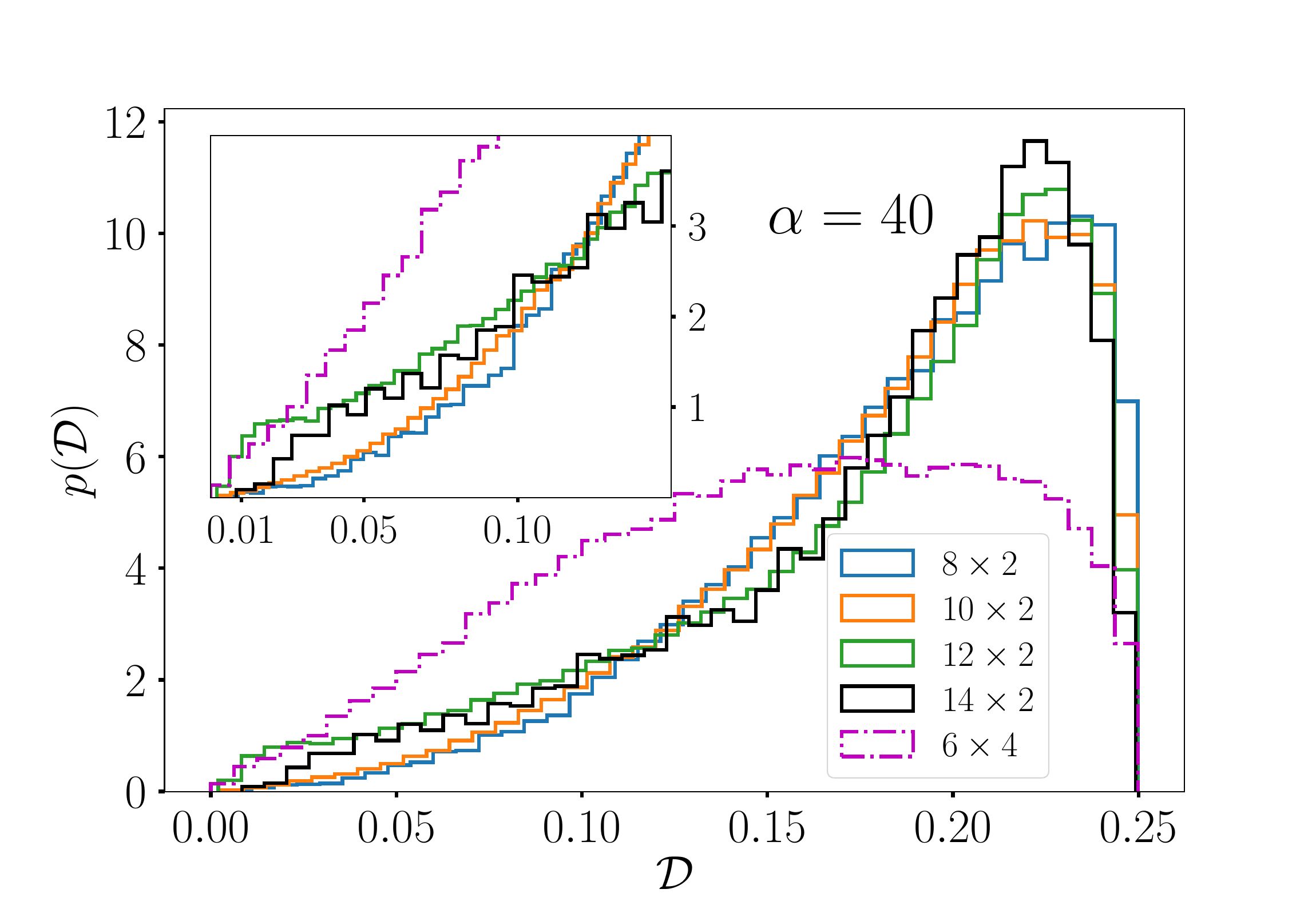}
    \caption{Behavior of $p(\mathcal{D})$ for $8 \times 2$, $10 \times 2$, $12 \times 2$, $14\times 2$ and $6 \times 4$ ladders for $\alpha=30$ (top panel) and $\alpha=40$ (bottom panel). The inset in the bottom panel shows the behavior of the tails of $p(\mathcal{D})$ prominently.}
    \label{fig:pDintermediatea}
\end{figure}  

We then look at still higher disorder values of $\alpha=30$ (Fig.~\ref{fig:pDintermediatea}, top panel) and $\alpha=40$ (Fig.~\ref{fig:pDintermediatea}, bottom panel). For $\alpha=30$, $p(\mathcal{D})$ displays a global maximum in the neighborhood of $\mathcal{D}=1/4$ for $8 \times 2$, $10 \times 2$ and $12 \times 2$ ladders with all three distributions being very broad (Fig.~\ref{fig:pDintermediatea}, top panel) implying that typical mid-spectrum eigenstates have a significant probability to have large inert regions in real space. Interestingly, the distribution becomes symmetric for the largest ladder of dimension $14 \times 2$ which suggests that $\alpha = 30$ may be close to the ETH-MBL transition. The broad nature of $p(\mathcal{D})$ again leaves an imprint on the dynamics of $\Opots$ for finite-sized systems starting from typical Fock states as will be discussed in the next section. However, as the size of a thin ladder with $L_y=2$ is increased from $L_x=8$ to $L_x=14$, we see that the probability to have large active regions in mid-spectrum eigenstates increases, while the weight of the distribution for larger $\mathcal{D}$ (and thus, larger inert regions) decreases. The increase in $p(\mathcal{D})$ for small $\mathcal{D}$ is particularly significant when increasing the ladder dimension from $10 \times 2$ to $12 \times 2$. 
Let us now focus on a higher disorder of $\alpha=40$ for the thin ladders with $L_y=2$ (Fig.~\ref{fig:pDintermediatea}, bottom panel). Here, $p(\mathcal{D})$ has an even more pronounced maximum in the neighborhood of $\mathcal{D} =1/4$ reflecting that the probability of encountering a large inert region in a typical mid-spectrum eigenstate has increased with disorder. Focusing on the weight of the distribution for small $\mathcal{D}$ (see inset of Fig.~\ref{fig:pDintermediatea}, bottom panel for a zoomed version) shows an instability towards thermalization when $L_x$ is increased from $10$ to $12$ due to an increased probability to have larger active regions in real space. However, for $14 \times 2$, we see a small decrease in $p(\mathcal{D})$ at small $\mathcal{D}$ suggesting that $\alpha=40$ is in the MBL phase but close to the transition. 
Comparing $p(\mathcal{D})$ for a $6 \times 4$ ladder to that of a $12 \times 2$ ladder for these two cases (Fig.~\ref{fig:pDintermediatea}) clearly shows that the wider ladder is more efficient at resisting MBL due to a much larger value of $p(\mathcal{D})$ at lower $\mathcal{D}$, and hence an enhanced probability of getting large active regions in typical mid-spectrum eigenstates.

We finally show $p(\mathcal{D})$ for even higher disorder, i.e., $\alpha=60$ (Fig.~\ref{fig:pDlargea}, top panel) and $\alpha=100$ (Fig.~\ref{fig:pDlargea}, bottom panel). While both cases show that $p(\mathcal{D})$ has a pronounced maximum at $\mathcal{D}$ close to $1/4$, with the weight being higher for $\alpha=100$, the weight in the tails away from the maximum (see inset of both panels in Fig.~\ref{fig:pDlargea}) decay very slowly with system size, unlike the case of Fig.~\ref{fig:pDsmalla}. 
Furthermore, comparing the $p(\mathcal{D})$ data for $6 \times 4$ ladder with $12 \times 2$ ladder shows that the wider ladder has a higher probability of large active regions in typical mid-spectrum eigenstates (see inset of both panels in Fig.~\ref{fig:pDlargea}). Even at these high disorder values, since $\mathcal{D}$ fluctuates from $\mathcal{D}=1/4$ to around $0.15$ based on the form of the distribution $p(\mathcal{D})$, it seems to represent a {\emph{strongly fluctuating}} MBL regime at these lengthscales.


\begin{figure}
    \centering
    \includegraphics[scale=0.22]{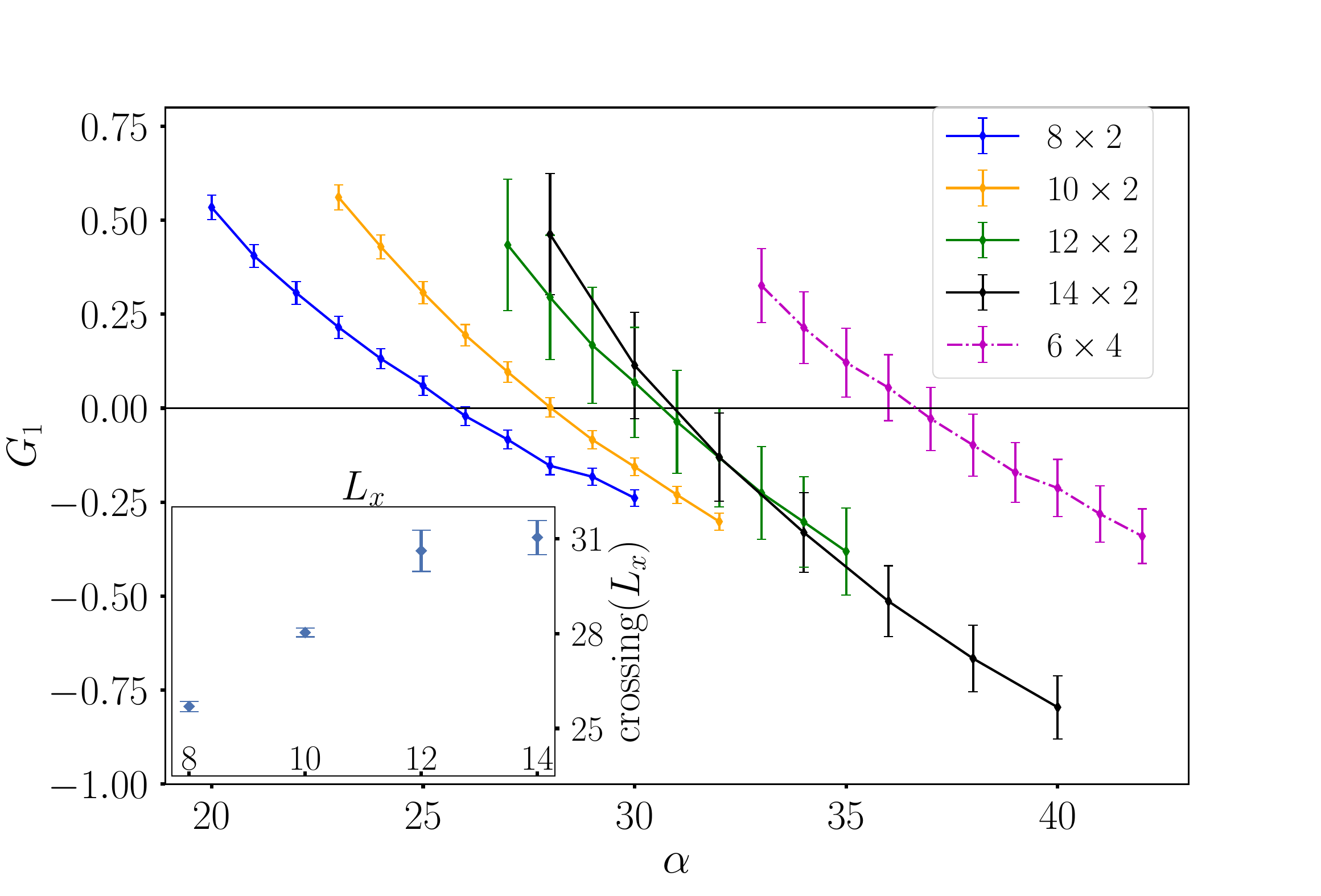}
    \caption{The adjusted Fisher-Pearson coefficient $G_1$ (Eq.~\ref{eq:skewness}) for the different ladders as a function of $\alpha$. For a given ladder dimension, the skewness coefficient crossing from positive to negative values gives a finite-size estimate $\alpha_c(L_x,L_y)$. The inset displays the behavior of this estimator for the thin ladders with $L_y=2$ as a function of ladder length $L_x$ from the available system sizes.}
    \label{fig:skewness}
\end{figure}  
A finite-size estimator for $\alpha_c$ can also be calculated from $p(\mathcal{D})$ for any given $L_x \times L_y$ and $\alpha$. The skewness (which is directly related to the third central moment) of the distribution, $p(\mathcal{D})$, presents a natural finite-size estimator for the location of the transition from ETH to MBL. This is because a positively (negatively) skewed distribution $p(\mathcal{D})$ is indicative of ETH (MBL) while zero skewness indicates a symmetric distribution. The adjusted Fisher-Pearson skewness coefficient is defined~\cite{skewnessarticle} for a data set $\{x_i\}$ of size $n$ with mean $\bar{x}$ and standard deviation $s$ as follows: 
\begin{eqnarray}
    G_1 = \frac{\sqrt{n(n-1)}}{n-2} \frac{\sum_{i=1}^n (x_i -\bar{x})^3/n}{s^3}.
    \label{eq:skewness}
\end{eqnarray}
and measures the asymmetry of the distribution around its mean. E.g., if we consider the evolution of $p(\mathcal{D})$ for a $12 \times 2$ ladder as a function of disorder $\alpha$ (Fig.~\ref{fig:pD12times2}), we see that at low (high) $\alpha$, the distribution has a tail towards higher (lower) values of $\mathcal{D}$ resulting in a positive (negative) $G_1$. The skewness coefficient crosses $0$ around $\alpha \sim 30$ where the distribution becomes broad and symmetric. Thus, the value of $\alpha$ for which $G_1$ crosses from being positive to negative can be taken as a finite-size estimator of $\alpha_c$ for a given ladder $L_x \times L_y$. We compute $G_1$ from the mid-spectrum eigenstates of each disorder realization using Eq.~\ref{eq:skewness} and then use the independent disorder realizations for a given $L_x \times L_y$ ladder and $\alpha$ to compute its average and error bar. We use $500$ disorder realizations each for $8 \times 2$ and $10 \times 2$ ladders, $50$ realizations each for $6 \times 4$ ladders and $30$ realizations each for $12 \times 2$ ladders. {$20$ disorder realizations were used for $14 \times 2$ ladder.} The result of such an analysis is displayed in Fig.~\ref{fig:skewness} from which a finite-size estimator $\alpha_c(L_x,L_y)$ can be directly computed. The inset of Fig.~\ref{fig:skewness} shows that $\alpha_c(L_x,L_y=2)$ first increases linearly with $L_x$ based on the data for $L_x=8, 10, 12$ but then eventually saturates for $L_x=14$ allowing one to estimate $\alpha_c(L_y=2) = 31.04 \pm 0.54$. A comparison of this finite-size estimator for $\alpha_c (L_y=2)$ with level spacing distribution estimators (see Fig.~\ref{fig:levelspacingscriticalcoupling} (bottom panel)) shows that the former has significantly better convergence with increasing $L_x$ compared to the latter. 
The $G_1$ data for the $6 \times 4$ ladder in Fig.~\ref{fig:skewness} clearly shows that the wider ladder with $L_y=4$ localizes at a larger disorder strength compared to a thin $12 \times 2$ ladder with the same number of elementary plaquettes.

\section{Autocorrelation functions for single plaquette diagonal operators}
\label{sec:dyn}

In this section, we focus on the dynamical properties of the disordered $U(1)$ QLM on ladders and particularly consider a range of disorder strengths, $\alpha$, such that it is less than $\alpha_c(L_y)$ (see previous section for estimates of the critical disorder strength to stabilize MBL). While interesting dynamical features including subdiffusion~\cite{MBLdyn2015, MBLdyn2016} have been discussed in thermal systems near a MBL transition, here we probe the autocorrelation functions of the simplest local diagonal operators $\Opots$ starting from typical Fock states for individual disorder realizations for a given $\alpha$ and ladder dimension for this purpose. Since $\alpha_c(L_y)$ is large for both $L_y=2$ and $L_y=4$, the disordered QLM provides us with a setting where the local relaxation of a strongly disordered, yet thermal, system may be studied.

We calculate both the infinite temperature autocorrelation functions as well as autocorrelations starting from typical Fock states whose average energy lies in the bin (of width $4\%$ of the total bandwidth as used in Sec.~\ref{subsec:FSS} to define mid-spectrum eigenstates) that contains the maximum density of states. Somewhat paradoxically, the infinite temperature autocorrelations are featureless and decay monotonically with time both at low and large disorder. However, the autocorrelations from individual Fock states show more structure. While the dynamics at low disorder shows rapid thermalization and negligible dynamic heterogeneity, the situation is different for intermediate and large disorder where interesting spatio-temporal structures emerge in local relaxation starting from randomly sampled typical Fock states. 

We define the autocorrelation functions as follows. Starting from a charge-resolved Fock state $|F\rangle$ (either in the sector $C=+1$ or $C=-1$ using Eq.~\ref{eq:FockCbasis}), the local autocorrelation functions on individual plaquettes in a given disorder realization are defined as 
\begin{eqnarray}
C_\square (|F\rangle, t) = \langle F | \Opots(t) \Opots(0)|F\rangle
\label{eq:localautoplaquette}
\end{eqnarray}
where $\Opots (t)=\exp (+i\mathcal{H}_{\mathrm{dis}} t )\Opots \exp(-i \mathcal{H}_{\mathrm{dis}} t)$, from which a spatially averaged temporal autocorrelation can be defined as 
\begin{eqnarray}
C(|F \rangle, t) =\frac{1}{N_p} \sum_\square C_\square (|F\rangle, t).
\label{eq:autoplaquette}
\end{eqnarray}
An infinite temperature temporal autocorrelation function, that represents the average of $C(|F \rangle, t)$ over all the charge-resolved Fock states, is similarly defined as follows: 
\begin{eqnarray}
C_{\mathrm{inf}} (t) = \frac{1}{\mathrm{HSD}} \frac{1}{N_p} \sum_{\square} \mathrm{Trace} [\Opots (t) \Opots(0)].
\label{eq:autocorrinfT}
\end{eqnarray}
It is useful to note that 
\begin{eqnarray}
  C_\square (|F\rangle, t)=\delta_{\Opots,1} \langle \Opots (t) \rangle
  \label{eq:CtOpot}
\end{eqnarray}
by using the fact that $\Opots|F\rangle = \delta_{\Opots,1} |F\rangle$ in Eq.~\ref{eq:localautoplaquette}, where $\delta_{\Opots,1}$ is a Kronecker delta function. Thus, $C_\square (|F\rangle, t)$ directly probes the temporal evolution of $\langle \Opots \rangle$ and its convergence (or, lack of it) to $\langle \Opots \rangle_{\mathrm{th}}$ (Table~\ref{tab:Opot}) as time increases for elementary plaquettes that have a flippable configuration of electric fluxes at $t=0$. We similarly define
\begin{eqnarray}
  \overline{C}_\square (|F\rangle, t)=\delta_{\Opots,0} \langle \Opots (t) \rangle
  \label{eq:CtbarOpot}
\end{eqnarray}
to probe the temporal evolution of $\langle \Opots \rangle$ for elementary plaquettes that have a non-flippable configuration of electric fluxes at $t=0$.

We also define the following normalized autocorrelators [which approach $1$ ($0$) for $t \rightarrow 0$ ($t \rightarrow \infty$)], $\tilde{C}(|F\rangle,t)$ and $\tilde{C}_{\mathrm{inf}}(t)$:
\begin{eqnarray}
    \tilde{C}(|F\rangle,t) = \frac{C(|F\rangle, t) - \overline{C(|F\rangle)}_{\infty}}{C(|F\rangle,0)-\overline{C(|F\rangle)}_{\infty}}
    \label{eq:normalizedC1}
\end{eqnarray}
and
\begin{eqnarray}
    \tilde{C}_{\mathrm{inf}}(t) = \frac{C_{\mathrm{inf}}(t) - \overline{(C_{\mathrm{inf}})}_{\infty}}{C_{\mathrm{inf}}(0)-\overline{(C_{\mathrm{inf}})}_{\infty}}
    \label{eq:normalizedC2}
\end{eqnarray}
where $\overline{C(|F\rangle)}_{\infty}$ and $\overline{(C_{\mathrm{inf}})}_{\infty}$ represent infinite-time averages of $C(|F \rangle, t)$ and $C_{\mathrm{inf}} (t)$ in the time range $t \in [0, \infty)$ and have the following expressions:
\begin{widetext}
\begin{eqnarray}
  \overline{C(|F\rangle)}_{\infty} &=&\frac{1}{N_p}\sum_\square \sum_m |\alpha^m_F|^2 \langle \Psi_m|\Opots|\Psi_m\rangle \langle F|\Opots|F\rangle \nonumber \\
  \overline{(C_{\mathrm{inf}})}_{\infty} &=& \frac{1}{\mathrm{HSD}} \frac{1}{N_p} \sum_\square \sum_m |\langle \Psi_m| \Opots|\Psi_m \rangle|^2
\label{eq:normalizedC}
\end{eqnarray}
\end{widetext}
with $|F\rangle = \sum_m \alpha_F^m |\Psi_m \rangle$ where $|\Psi_m\rangle$ represents the $m$-th eigenstate of $\mathcal{H}_{\mathrm{dis}}$ for a given disorder realization in the sector $C=\pm 1$.

In this section, we will show results for such autocorrelations for disorder strengths $\alpha=6, 12, 30$ for a $10 \times 2$ ladder as well as for $\alpha=6, 30$ for a wider ladder of dimension $6 \times 4$ for a particular disorder realization (i.e., specification of the random numbers $R_\square$). While $\alpha=6$ can be treated as "weak" disorder in both cases, the dynamics at $\alpha=12$ for the thin ladder shows coherent oscillations for the spatially averaged autocorrelation function $ C(|F\rangle,t)$, which can be traced to the oscillatory dynamics of $\langle \Opots (t) \rangle$ in some elementary plaquettes at intermediate disorder strength (which is still small compared to $\alpha_c(L_y=2)$) in a small fraction of the randomly chosen Fock states. We will finally discuss the case of strong disorder ($\alpha=30$) for both thin and wide ladders where this fraction of Fock states that show coherent oscillations of the diagonal local operator in some plaquettes, as well as in the spatially averaged autocorrelation function, becomes much more significant. Additionally, these oscillations point to the emergence of a plethora of time scales at strong disorder. We note that though $\alpha=30$ is a large disorder strength (since $\alpha$ is dimensionless), it is still smaller than both $\alpha_c(L_y=2)$ ($\approx 0.95 \alpha_c(L_y=2)$) and $\alpha_c(L_y=4)$. These prominent dynamical features at $\alpha=12$ and $\alpha=30$ are present in other disorder realizations as well and likely arise from the presence of weight in $p(\mathcal{D})$ for values of
  $\mathcal{D}$ {\emph{close}} to $1/4$ at these system sizes and couplings. In all the cases, we choose $50$ randomly selected charge-resolved Fock states from the ones whose average energy lies in the bin (with a width of $4\%$ of the total bandwidth) with the highest number of energy eigenstates for the given disorder realization.

\subsection{Autocorrelation functions deep in the thermalizing regime}
\label{subsec:dyntherm}
\begin{figure}[h]
        \centering
        \includegraphics[scale=0.25]{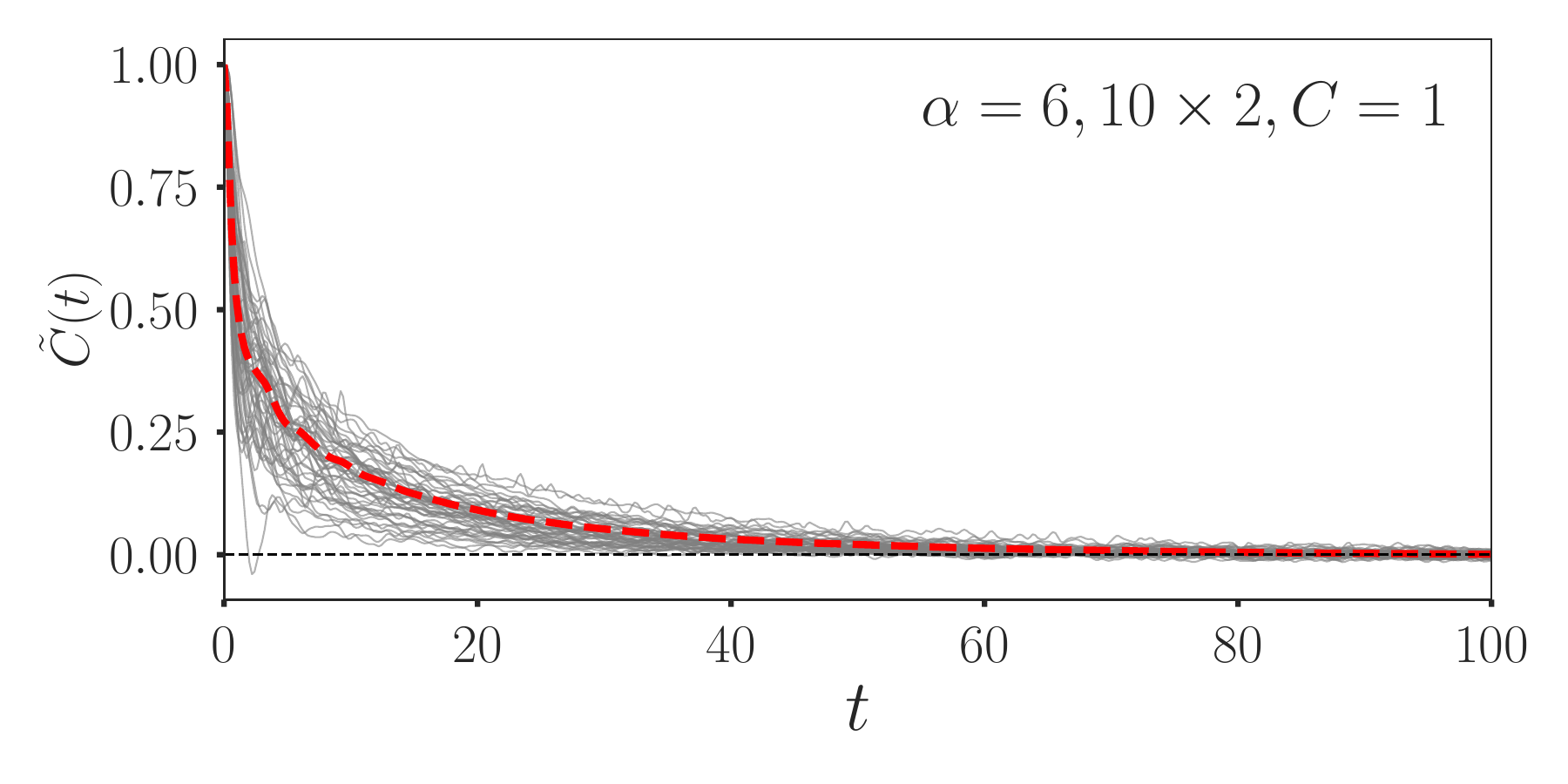}
        \includegraphics[scale=0.25]{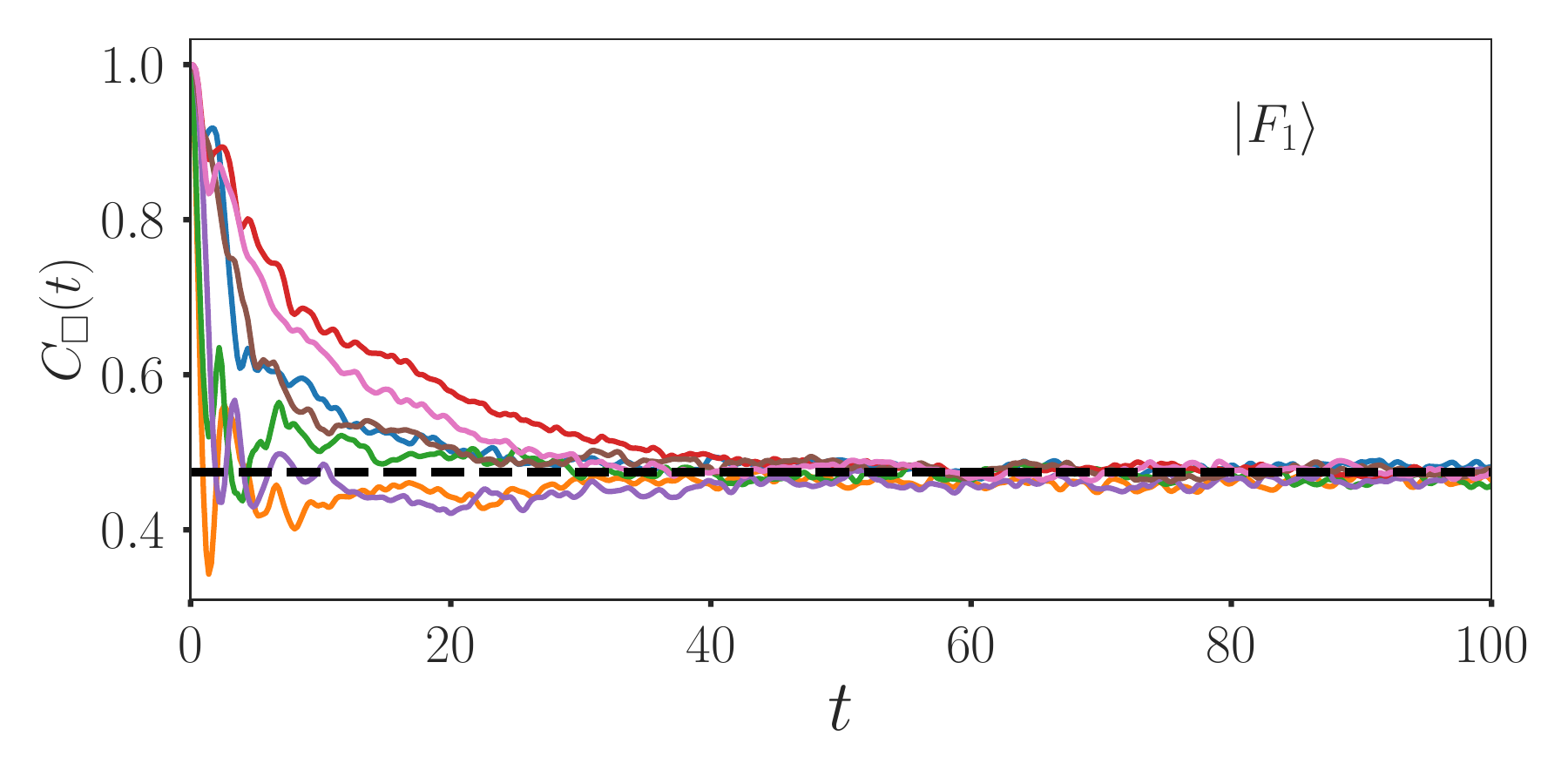}
         \includegraphics[scale=0.25]{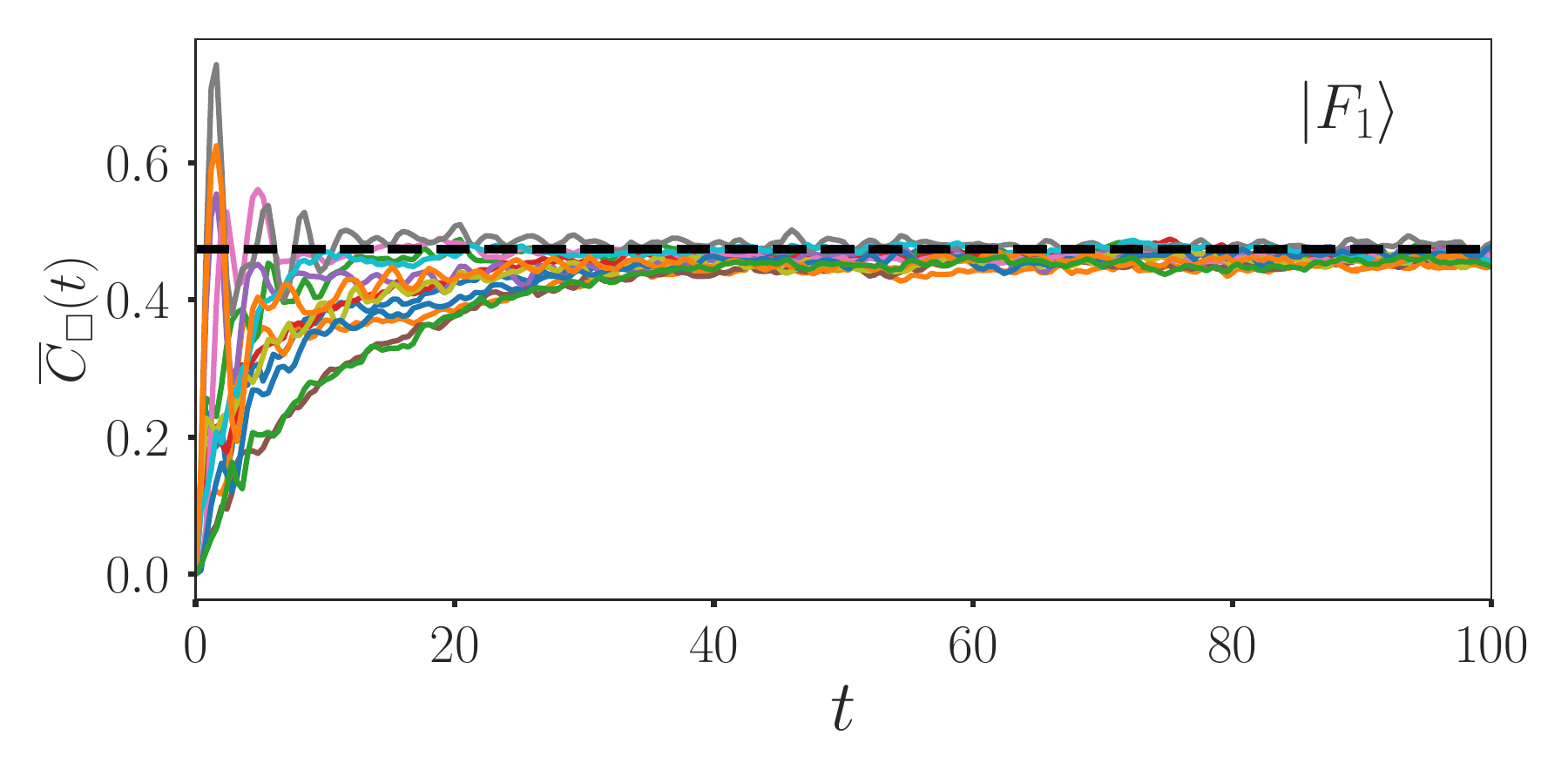}
         \caption{(Top panel) Normalized autocorrelation functions (see Eq.~\ref{eq:normalizedC1} and Eq.~\ref{eq:normalizedC2}) shown for the infinite temperature autocorrelation (dashed red curve) as well as the spatially averaged autocorrelation function starting from $50$ randomly chosen Fock states (shown in grey) with average energies that lie in the bin of width $4\%$ of the total bandwidth and contains the maximum number of eigenstates for a single disorder realization of a $10 \times 2$ ladder for $\alpha=6$. The local autocorrelation $C_\square(|F\rangle,t)$ (Eq.~\ref{eq:localautoplaquette}) (Middle panel) and $\overline{C}_\square (|F\rangle, t)$ (Eq.~\ref{eq:CtbarOpot}) (Bottom panel) shown for one particular Fock state selected from the top panel. The autocorrelation functions for the different elementary plaquettes are shown using different colors. The horizontal dashed line indicates the value of $\langle \Opots \rangle_{\mathrm{th}}$ for a $10 \times 2$ ladder (Table ~\ref{tab:Opot}).}
        \label{fig:autoC10x2allAsmall}
    \end{figure}   
 
\begin{figure}[h]
        \centering
        \includegraphics[scale=0.25]{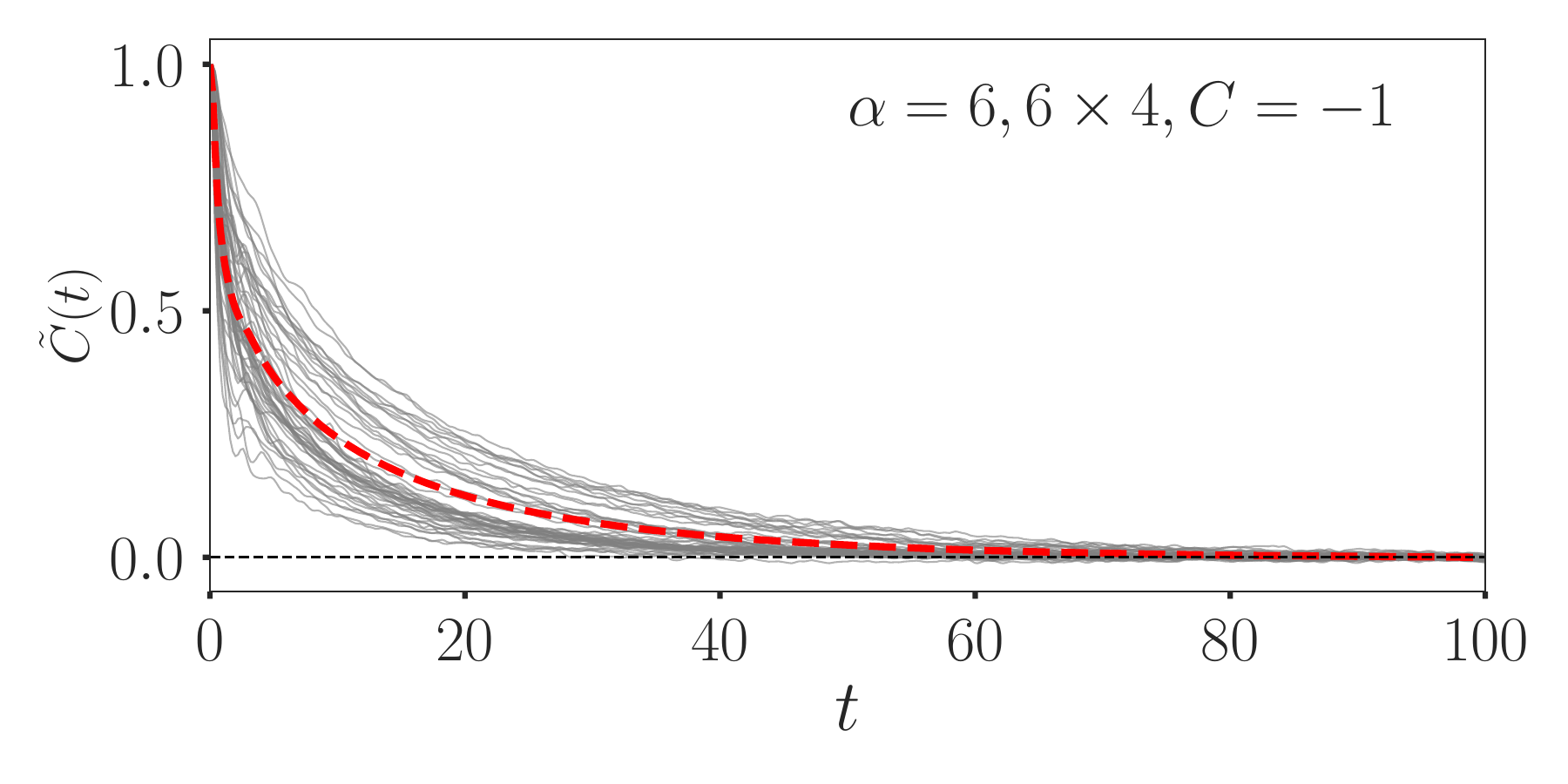}
         \includegraphics[scale=0.25]{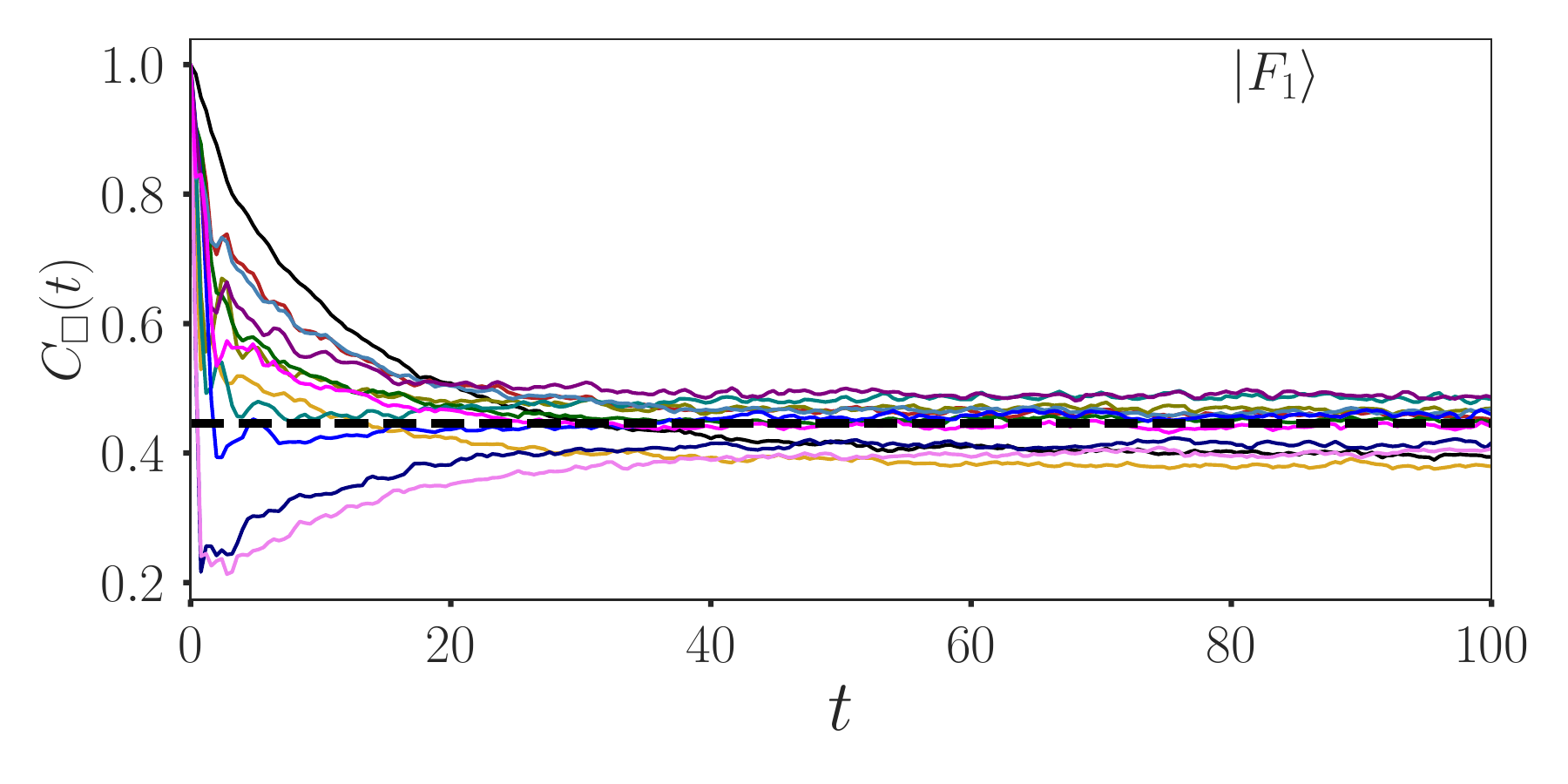}
         \includegraphics[scale=0.25]{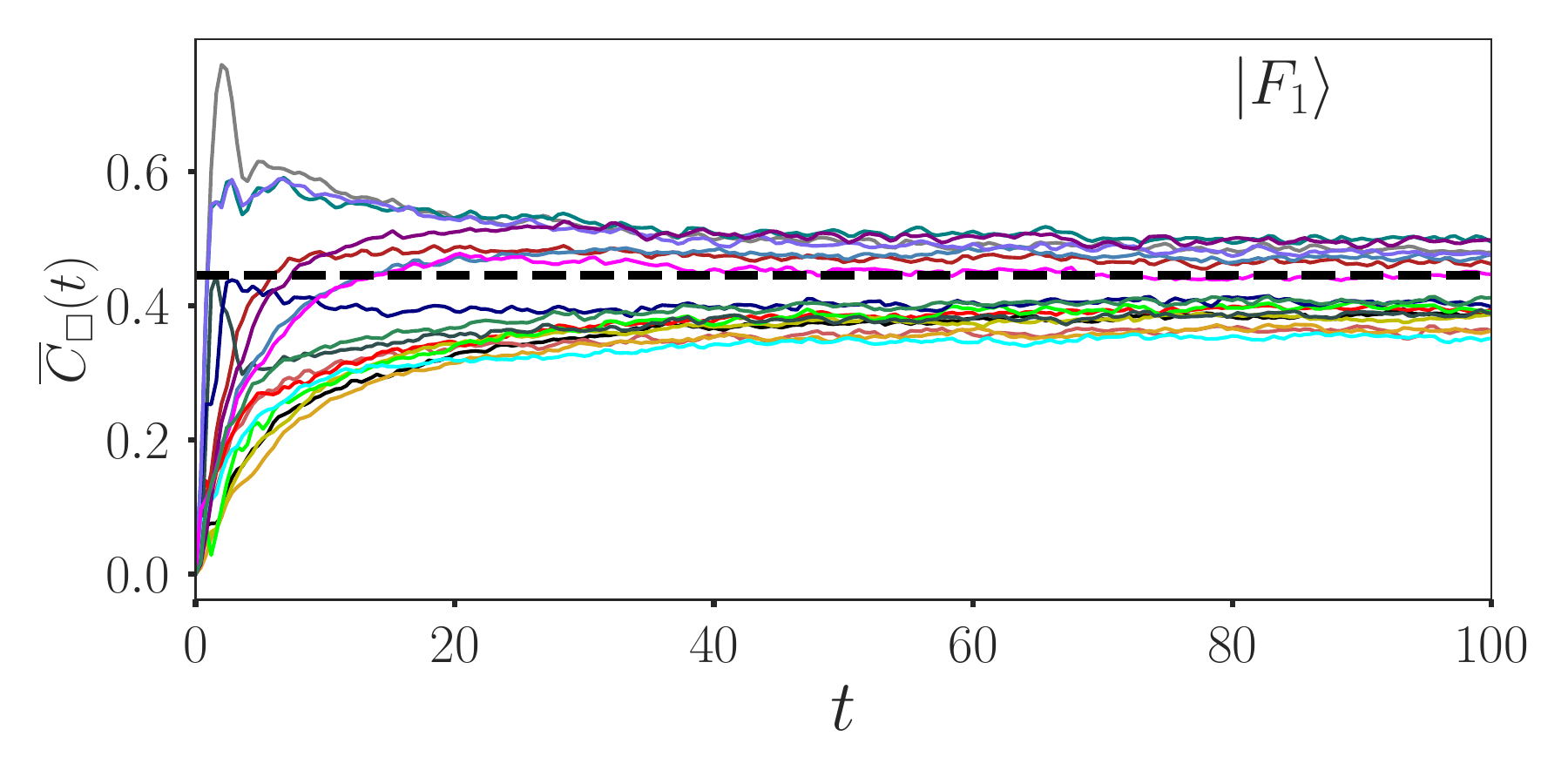}
        \caption{(Top panel) Normalized autocorrelation functions (see Eq.~\ref{eq:normalizedC1} and Eq.~\ref{eq:normalizedC2}) shown for the infinite temperature autocorrelation (dashed red curve) as well as the average autocorrelation function starting from $50$ randomly chosen Fock states (shown in gray) with average energies that lie in the bin of width $4\%$ of the total bandwidth and contains the maximum number of eigenstates for a single disorder realization of a $6 \times 4$ ladder for $\alpha=6$. The local autocorrelation $C_\square(|F\rangle,t)$ (Eq.~\ref{eq:localautoplaquette}) (Middle panel) and $\overline{C}_\square (|F\rangle, t)$ (Eq.~\ref{eq:CtbarOpot}) (Bottom panel) shown for one particular Fock state selected from the top panel. The autocorrelation functions for the different elementary plaquettes are shown using different colors. The horizontal dashed line indicates the value of $\langle \Opots \rangle_{\mathrm{th}}$ for a $6 \times 4$ ladder (Table ~\ref{tab:Opot}).}
        \label{fig:autoC6x4allAsmall}
    \end{figure}

 We first monitor the behavior of the autocorrelations for a disorder strength of $\alpha=6$ for both $10 \times 2$ as well as $6 \times 4$ ladders. We focus on a single disorder realization in both cases and calculate the normalized versions of the infinite temperature autocorrelation (Eq.~\ref{eq:normalizedC2}) as well as the normalized average temporal autocorrelation (Eq.~\ref{eq:normalizedC1}) starting from $50$ randomly selected charge-resolved Fock states. The results are displayed in the top panel of Fig.~\ref{fig:autoC10x2allAsmall} for a $10 \times 2$ ladder and the top panel of Fig.~\ref{fig:autoC6x4allAsmall} for a $6 \times 4$ ladder. Both the (normalized) infinite temperature autocorrelations as well as the ones starting from randomly sampled Fock states relax to $0$ as a function of time $t$ reflecting the approach of the diagonal plaquette operators to their final late-time values.
 The infinite temperature autocorrelation in both cases decay monotonically with $t$; however, the autocorrelation starting from the individual Fock states are not necessarily monotonic at all times, especially at early times. There is a spread of the normalized autocorrelation functions for individual Fock states around the infinite temperature result due to the finite disorder present ($\alpha=6$), with some autocorrelations decaying slower (faster) than the infinite temperature result. However, all the normalized autocorrelations decay to nearly zero beyond a time scale of $t \sim 60$ ($t \sim 80$) for the $10 \times 2$ ($6 \times 4$) ladder.

It is also instructive to look at $C_\square(|F\rangle,t)$ (Eq.~\ref{eq:localautoplaquette}) as well as $\overline{C}_\square(|F\rangle,t)$ (Eq.~\ref{eq:CtbarOpot}) for the individual Fock states to probe the local thermalization of $\langle \Opots (t) \rangle$ directly. Since the problem is disordered, different regions in space can have different relaxational timescales. We look at one particular Fock state from the $50$ randomly selected Fock states in the middle and bottom panels of Fig.~\ref{fig:autoC10x2allAsmall} (Fig.~\ref{fig:autoC6x4allAsmall}) for $10 \times 2$ ($6 \times 4$) ladder. It is clear from both figures that while the transients differ for each plaquette due to their different local environments, all the $\Opots(t)$ curves saturate close to a steady state value after a time scale of $t \sim 40$ in both cases for the chosen Fock states. Furthermore, the steady state values are close to $\langle \Opots \rangle_{\mathrm{th}}$ (indicated by a horizontal dotted line in the middle and bottom panels of Fig.~\ref{fig:autoC10x2allAsmall} and Fig.~\ref{fig:autoC6x4allAsmall}) for different elementary plaquettes for the corresponding ladder dimension in both the cases. We have checked that this picture remains qualitatively true for the other Fock states shown in the top panel of Fig.~\ref{fig:autoC10x2allAsmall} and Fig.~\ref{fig:autoC6x4allAsmall}. 

\subsection{Dynamic heterogeneity at intermediate and strong disorder}
\label{subsec:dynFock}

\begin{figure}
        \centering
        \includegraphics[scale=0.25]{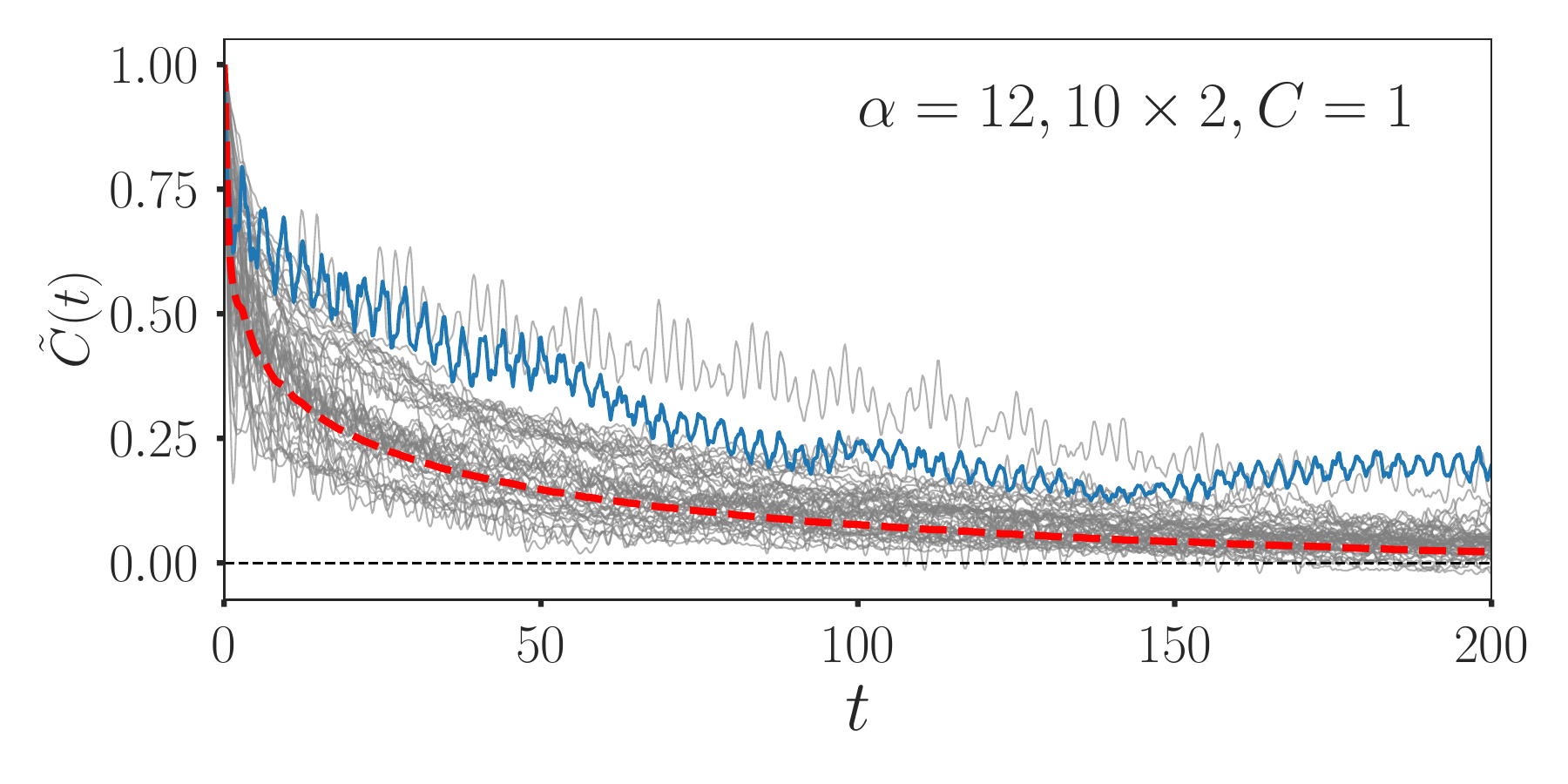}
        \includegraphics[scale=0.25]{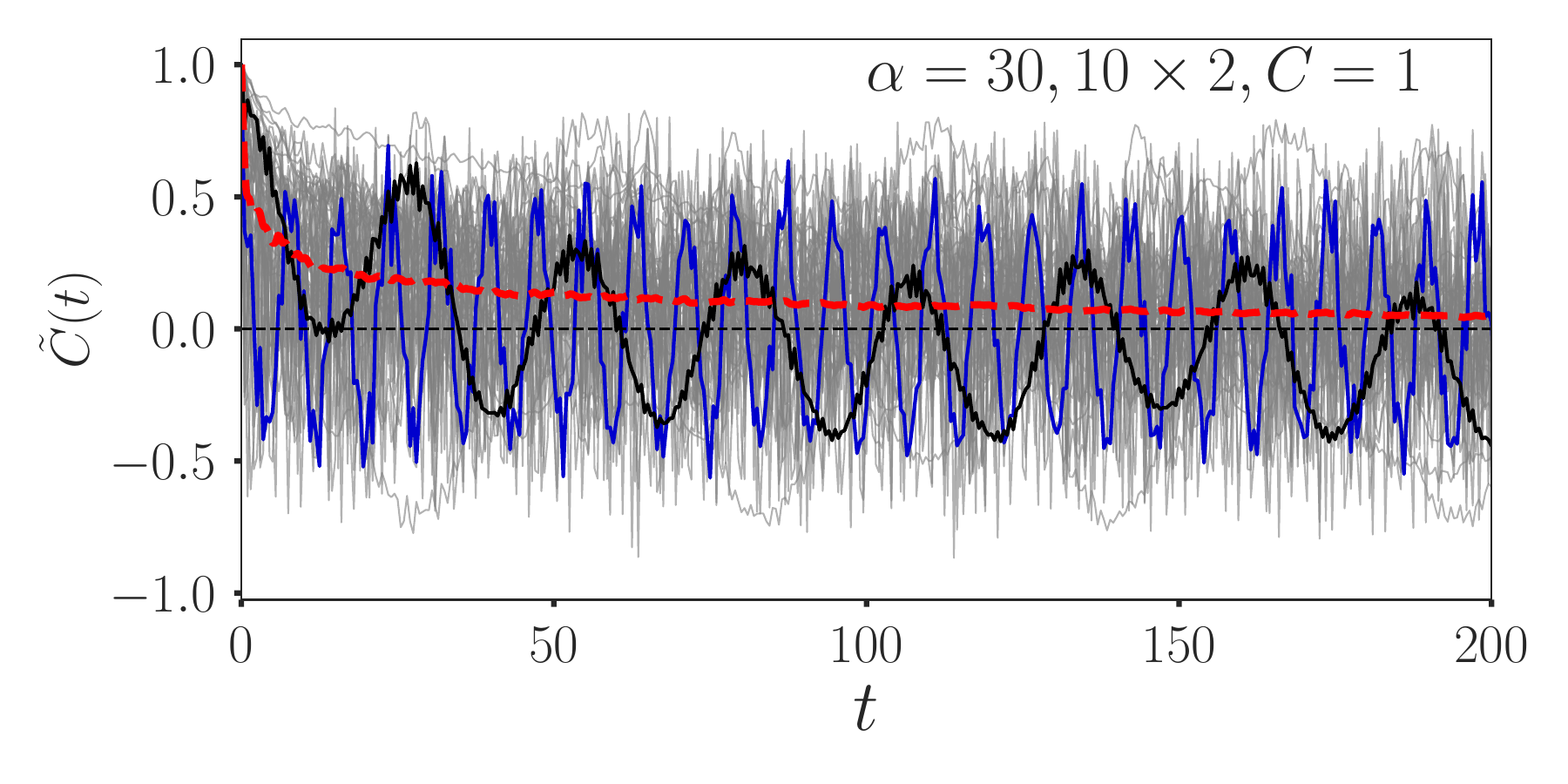}
        \includegraphics[scale=0.25]{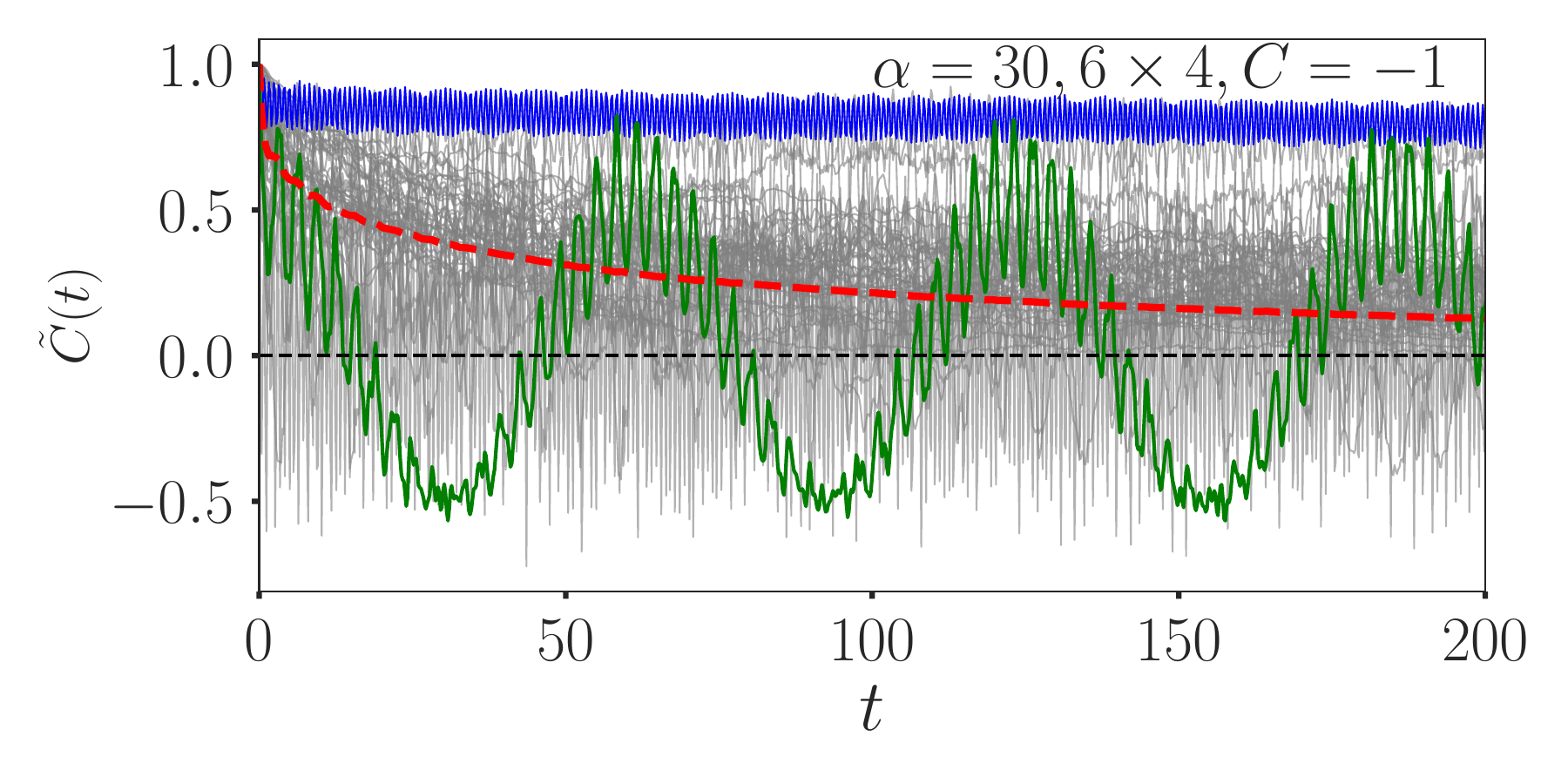}
        \caption{Normalized autocorrelation functions (see Eq.~\ref{eq:normalizedC1} and Eq.~\ref{eq:normalizedC2}) shown for the infinite temperature autocorrelation (dashed red curve in all panels) as well as the spatially averaged autocorrelation function starting from $50$ randomly chosen Fock states (shown in grey in all panels) with average energies that lie in the bin of width $4\%$ of the total bandwidth and contains the maximum number of eigenstates for a single disorder realization of a $10 \times 2$ ladder with $\alpha=12$ (top panel), $\alpha=30$ (middle panel) and a $6 \times 4$ ladder with $\alpha=30$ (bottom panel). Some Fock states that show an oscillatory behavior of the spatially averaged autocorrelation are marked in different colors in all three panels for clarity.}
        \label{fig:autoCintermediate}
    \end{figure}

Let us now consider the nature of the autocorrelation functions for an intermediate disorder strength of $\alpha=12$ for a $10 \times 2$ ladder in a single disorder realization. This value of $\alpha$ is still much lower than $\alpha_c(L_y=2)$, with $\alpha/\alpha_c(L_y=2) \approx 0.4$, based on our estimates in Sec.~\ref{subsec:FSS}. From Fig.~\ref{fig:autoCintermediate} (top panel), we see that while the infinite temperature autocorrelation is still monotonically decaying in time, the spatially averaged autocorrelation functions from $50$ randomly chosen Fock states shows a bigger spread around the infinite temperature autocorrelation with a few Fock states ($3$ out of $50$ for this particular disorder realization) displaying clear oscillatory behavior with one such autocorrelation curve indicated in blue for clarity in Fig.~\ref{fig:autoCintermediate} (top panel).

\begin{figure}
    \centering
    \includegraphics[scale=0.25]{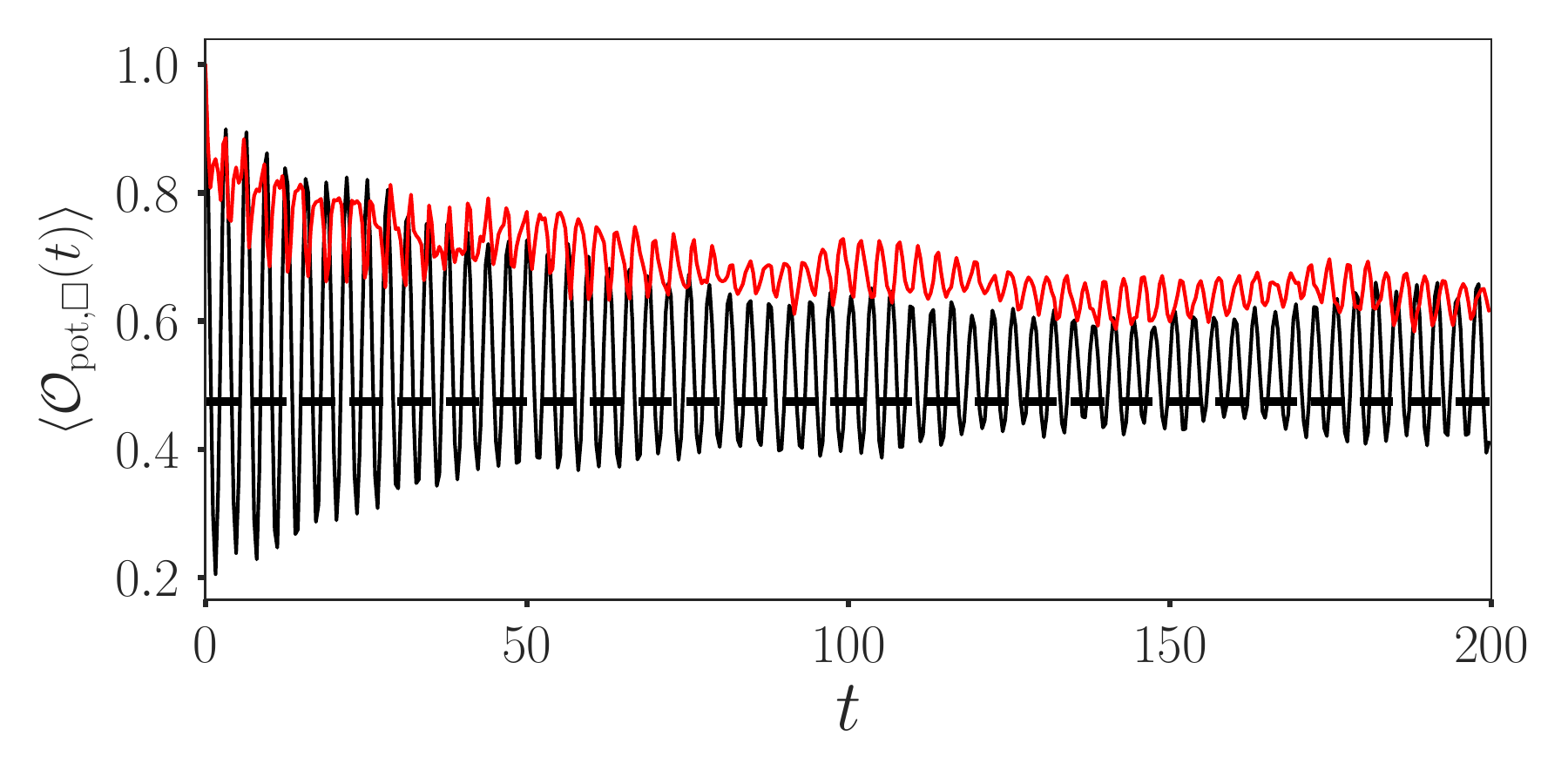}
    \includegraphics[scale=0.25]{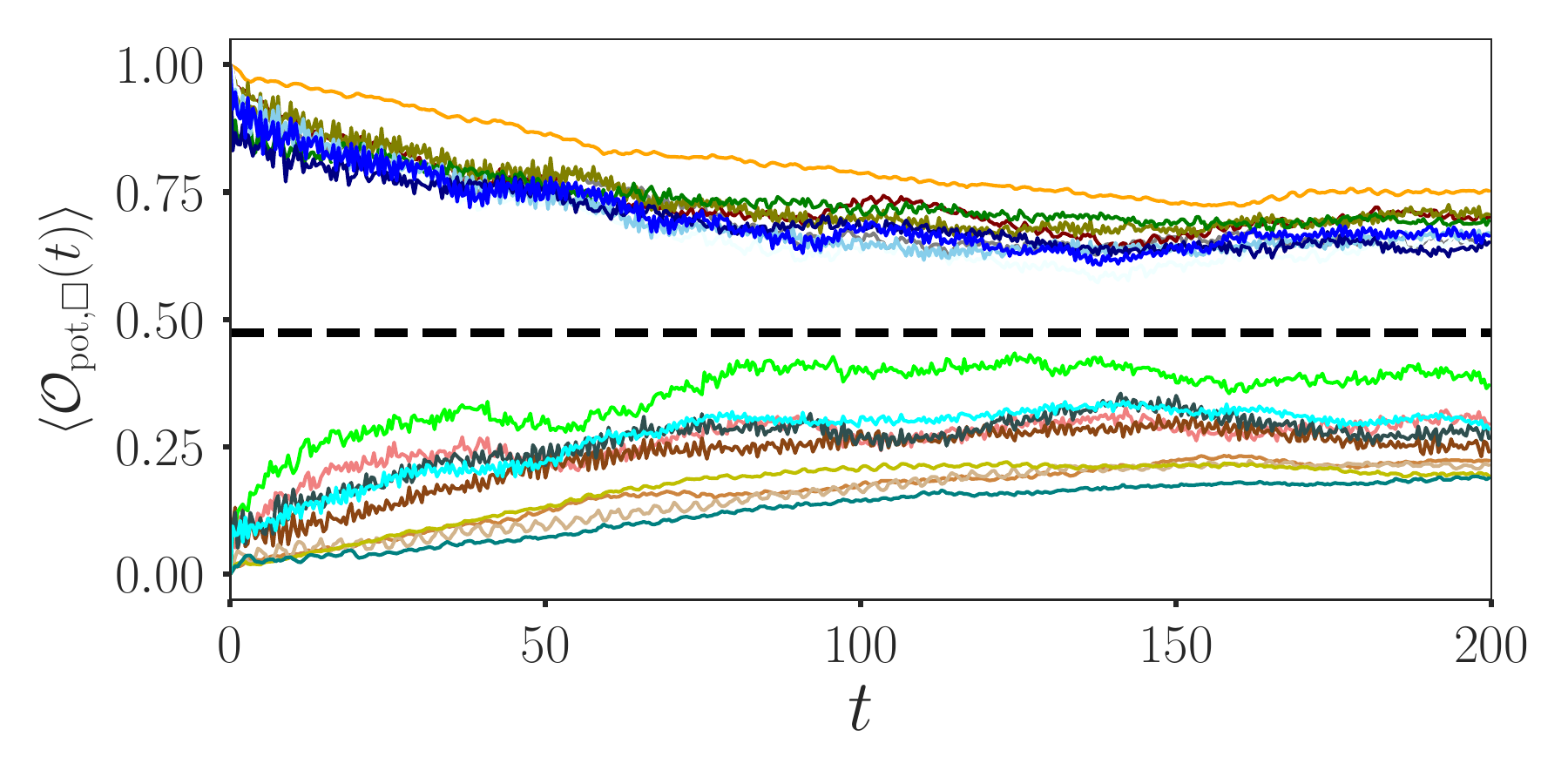}
    \includegraphics[scale=0.16]{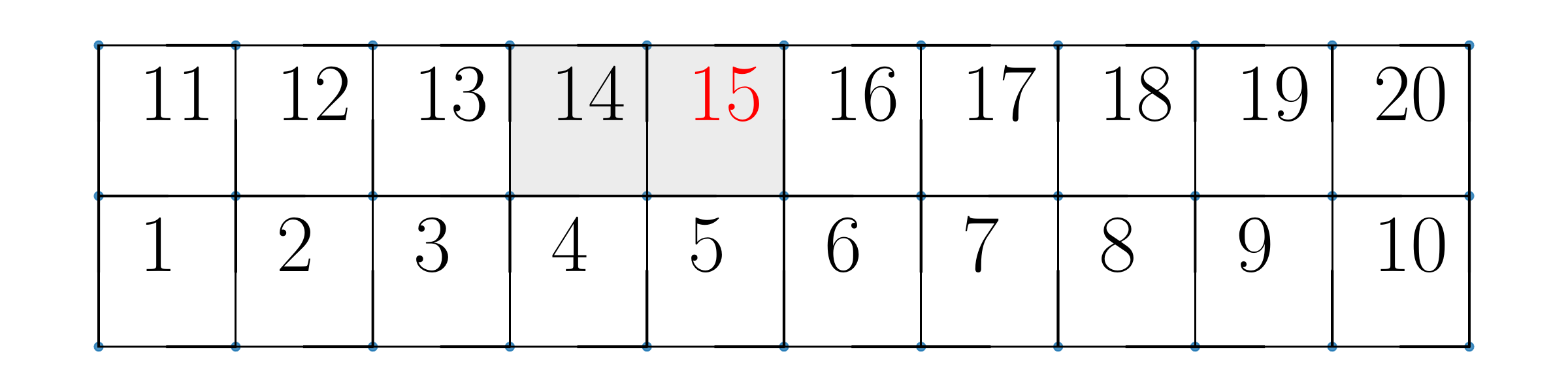}
    \caption{Behavior of $\Opots (t)$ on individual plaquettes for a particular Fock state that shows oscillatory dynamics in the spatially averaged autocorrelation function for a $10 \times 2$ lattice with $\alpha=12$. $\Opots (t)$ on elementary plaquettes that do (do not) display oscillatory dynamics is shown in the top (middle) panel. The real space location of the plaquettes that show oscillatory dynamics is shown in gray in the bottom panel ($14$th and $15$th plaquette) with the color of plaquette number chosen to be the same as used in the top panel to display the temporal behavior of $\langle \Opots \rangle$ on that plaquette.}
    \label{fig:12S1F2}
\end{figure}

For the Fock states that do not exhibit oscillatory average autocorrelations ($47$ out of $50$ for this particular disorder realization), while the individual plaquette operators $\langle \Opots (t) \rangle$ do seem to attain steady state values after a timescale that is longer compared to the case of $\alpha=6$, the steady state values show a much bigger spread compared to $\alpha=6$ around the expected result of $\langle \Opots \rangle_{\mathrm{th}}$ from ETH. However, the behaviour for the $3$ Fock states with oscillatory average autocorrelations is markedly different. We display the behaviour of $\langle \Opots (t) \rangle$ on each elementary plaquette for one such Fock state in Fig.~\ref{fig:12S1F2} (top and middle panels). While all but two plaquettes approach their steady state values at a timescale comparable to the time scales for the other typical Fock states (Fig.~\ref{fig:12S1F2}, middle panels), these values are very different from the one expected from ETH. The remaining two plaquettes, which share an edge with each other, show clear coherent oscillations with a slowly decaying envelope (at least till $t=200$) for the diagonal operator $\langle \Opots \rangle$, in stark contrast to expectation from ETH. Both plaquettes oscillate with almost the same frequency with the amplitude of oscillation of one being larger than the other. The oscillations are initially in phase, then turn out of phase, and finally get in phase again for $t \in [0,200]$. 

    The fraction of such randomly chosen Fock states with oscillatory, instead of decaying, spatially averaged temporal autocorrelations as well as the dynamic heterogeneity in real space increases significantly as one cranks up the disorder strength even further. We now show results for a disorder strength of $\alpha=30$, both for $10 \times 2$ and $6 \times 4$ ladders. This increased disorder is still not enough to localize the system even though $\alpha/\alpha_c(L_y=2) \approx 0.95$ which places this coupling to be close to the critical coupling, but on the thermal side, in the case of thin ladders with $L_y=2$. The increased dynamic heterogeneity is already evident when one calculates the infinite temperature autocorrelation and compares it to the spatially averaged autocorrelation from $50$ randomly chosen Fock states whose average energies lie within the bin with the highest number of energy eigenstates (Fig.~\ref{fig:autoCintermediate} (middle and bottom panels)). While the infinite temperature autocorrelation is still quite featureless and decays monotonically for both the ladders, we see that the randomly chosen Fock states show a very wide range of dynamical behavior. In particular, there are now $23$ ($14$) Fock states with clear oscillatory dynamics in the average autocorrelation for the particular disorder realization used for $10 \times 2$ ($6 \times 4$) ladder to generate Fig.~\ref{fig:autoCintermediate} (middle and bottom panels). Some of these autocorrelations are marked using a different color for clarity in the same figure.      

\begin{widetext}
\begin{figure*}
    \centering
     \includegraphics[scale=0.22]{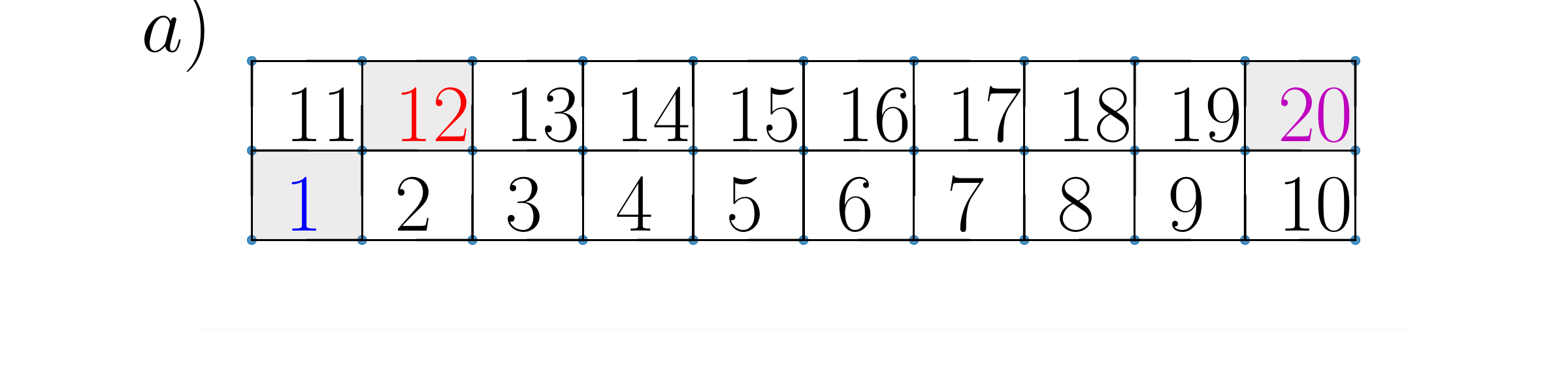}
     \includegraphics[scale=0.25]{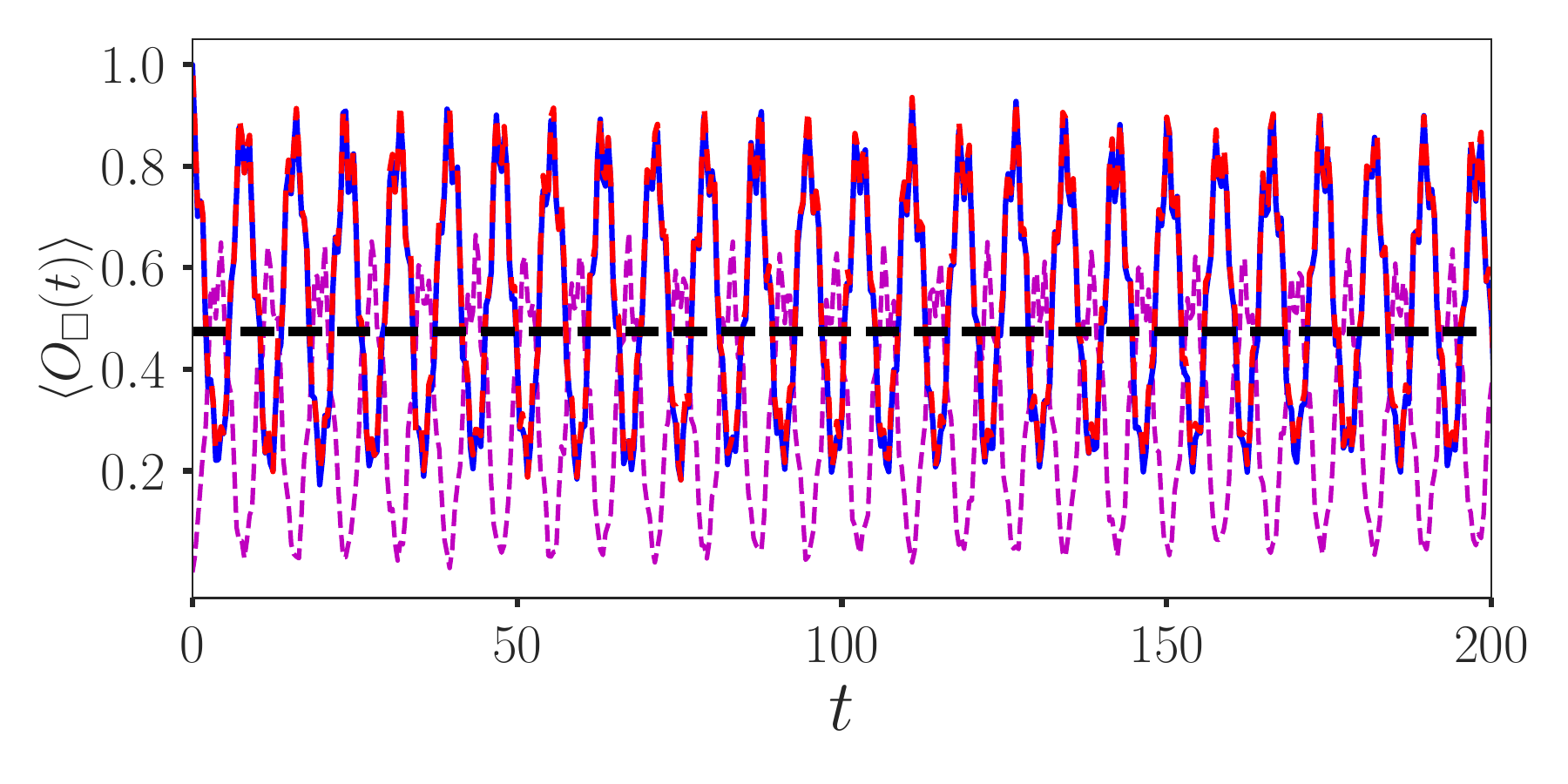}\\
     \includegraphics[scale=0.22]{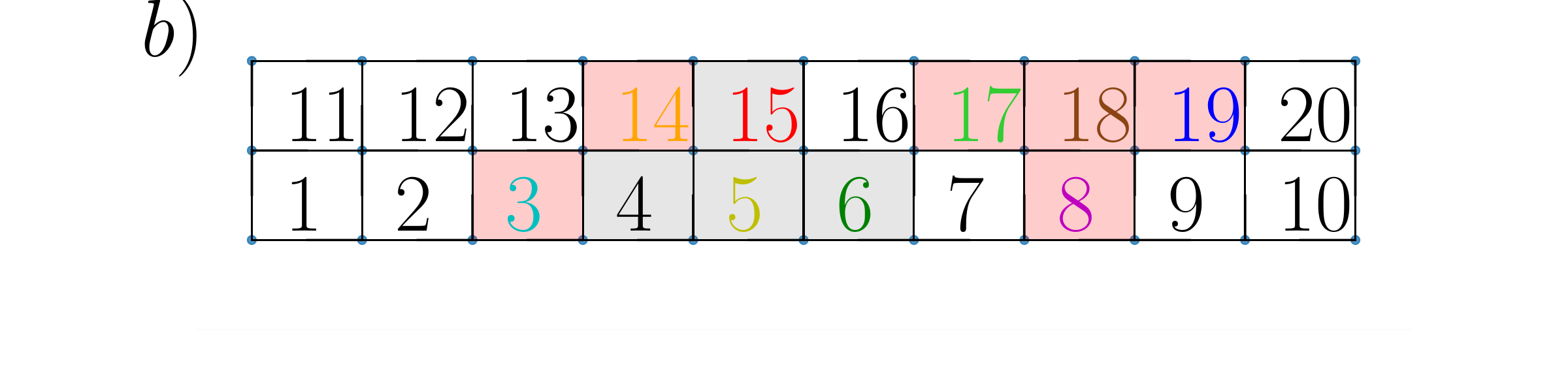}
     \includegraphics[scale=0.25]{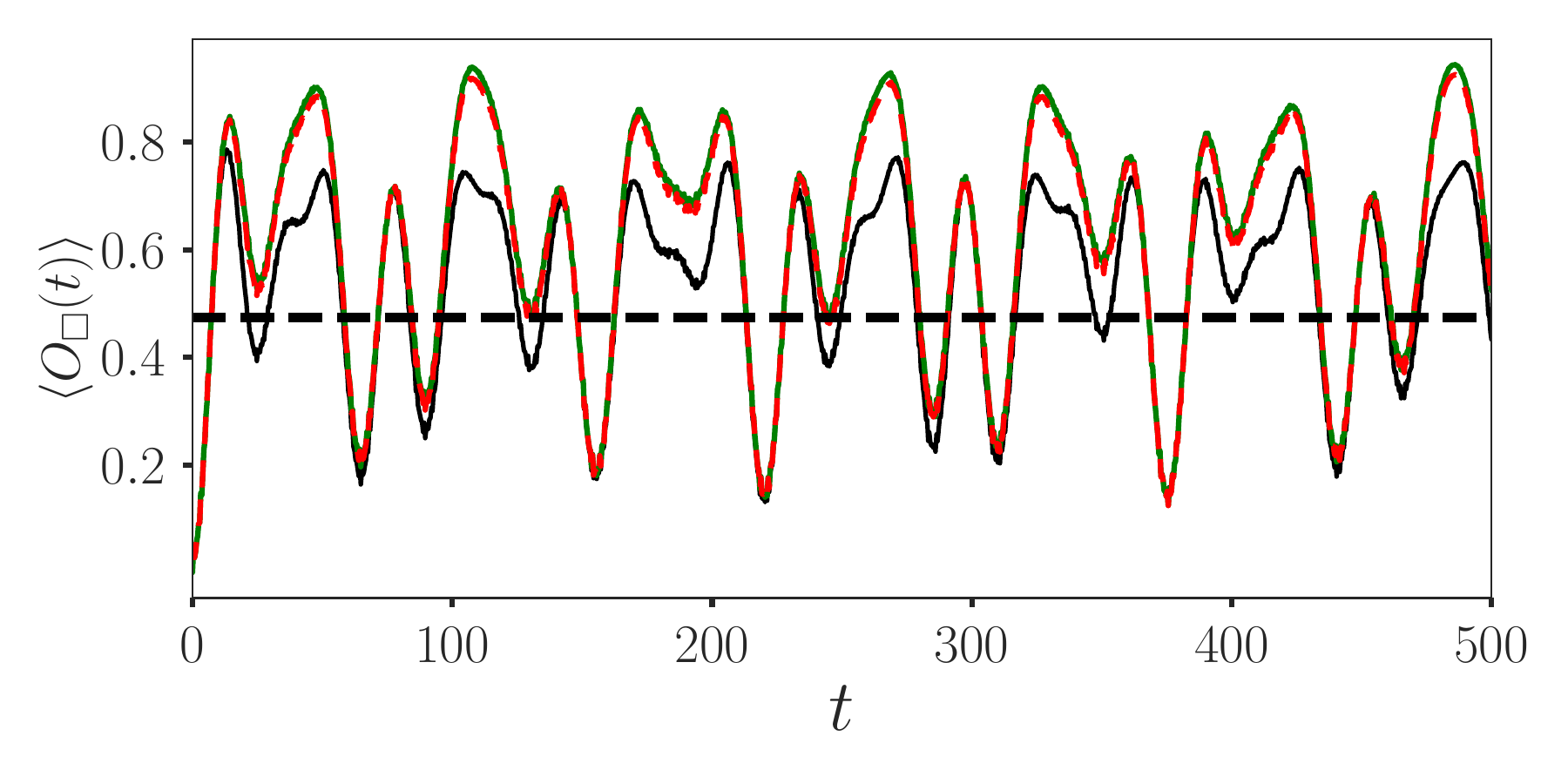}\\
     \includegraphics[scale=0.25]{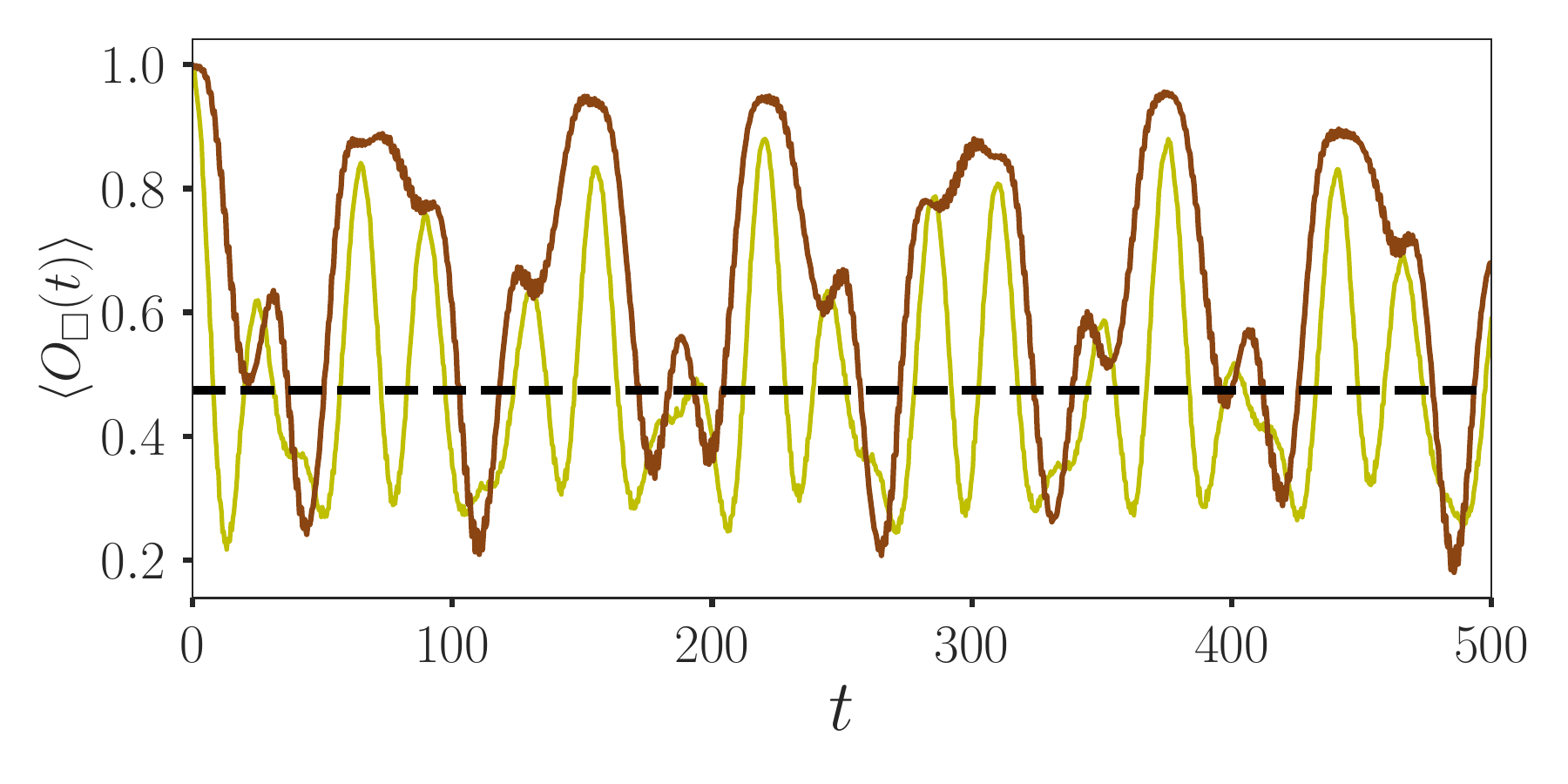}
     \includegraphics[scale=0.25]{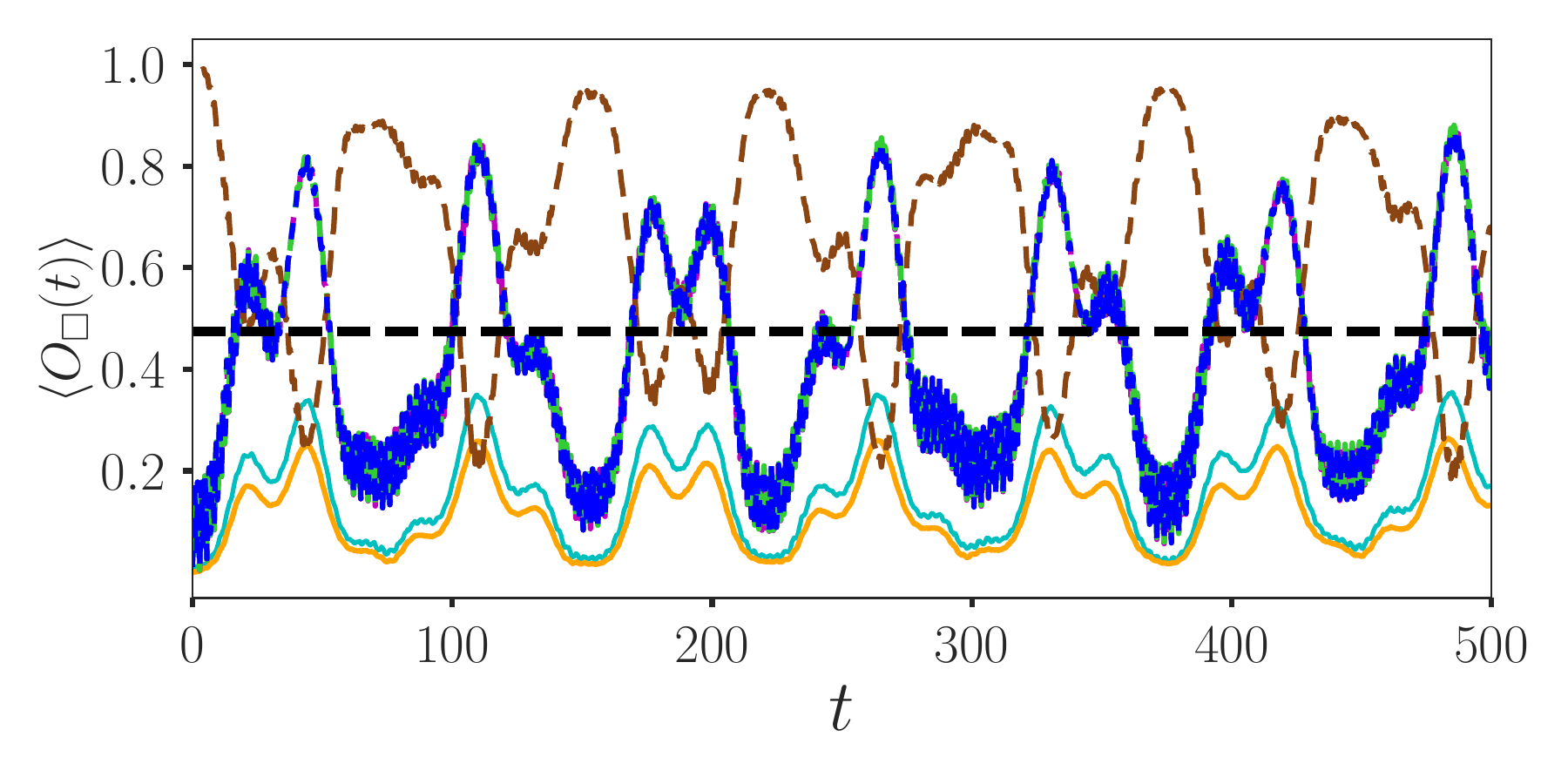}
      \caption{Behavior of $\Opots (t)$ on individual plaquettes for two different Fock state, both with the same disorder realization, that show oscillatory dynamics in the spatially averaged autocorrelation function for a $10 \times 2$ lattice with $\alpha=30$ shown in (a) and (b). Only elementary plaquettes that display oscillations in $\Opots (t)$ shown for clarity. The panels in (a) [(b)] display the behavior of $\langle \Opots (t) \rangle$ on different plaquettes for one [the other] initial Fock state with some plaquettes being repeated in (b) to illustrate the relation between the oscillation of different plaquettes. The extreme left panels in (a) and (b) display the real space location of the plaquettes that show oscillatory dynamics as shaded in either gray or pink. While plaquettes shaded in the same color show similar temporal temporal behavior (while being in or out of phase, see right panel in (a)), plaquettes marked in gray and pink in the extreme left panel in (b) show different temporal behaviors, which are nonetheless strongly correlated to each other. In both the extreme left panels for (a) and (b), the color of the number on the shaded plaquette is chosen to be the same as used in the panels to display the temporal behavior of $\langle \Opots \rangle$ on the corresponding plaquettes.}
      \label{fig:30S10times2}
\end{figure*}
\end{widetext}

A more detailed picture for the dynamics from initial Fock states that show
oscillatory behaviour in the spatially averaged autocorrelation functions emerges when
one monitors $\langle \Opots (t) \rangle$ on all elementary plaquettes. This again reveals the presence of regions formed out of connected elementary plaquettes which show significant oscillations unlike the rest of the system. The decaying envelope of oscillations seen for a lower disorder strength of $\alpha=12$ seems to be absent for $\alpha=30$ for the timescales that we monitor. Furthermore, different Fock states display strikingly different types of oscillatory temporal behaviors for the same disorder realization as can be seen by comparing the dynamics from two such Fock states in Fig.~\ref{fig:30S10times2} for a $10 \times 2$ lattice with $\alpha=30$. The panels in Fig.~\ref{fig:30S10times2} (a) illustrate a reasonably simple oscillatory dynamics in a connected region formed out of $3$ elementary plaquettes ($1$st, $12$th and $20$th plaquette, see extreme left panel of
Fig.~\ref{fig:30S10times2} (a)) where two of the three plaquettes ($1$st and $12$th) oscillate in phase (shown in blue and red, respectively) while the third plaquette ($20$th) oscillates out of phase (shown in magenta) to these two as can be seen clearly from the right panel in Fig.~\ref{fig:30S10times2} (a).

On the other hand, Fig.~\ref{fig:30S10times2} (b) displays a much more complicated temporal behaviour from a connected region formed out of $10$ elementary plaquettes ($3$rd, $4$th, $5$th, $6$th, $8$th, $14$th, $15$th, $17$th, $18$th, and $19$th plaquette, see extreme left panel oh Fig.~\ref{fig:30S10times2}(b)). The $6$th and the $15$th plaquette show similar temporal oscillations (shown in green and red, respectively) which are also followed by the $4$th plaquette (shown in black), though with a somewhat smaller amplitude. The $5$th plaquette oscillates out of phase (shown in yellow) with these three plaquettes with an amplitude that closely follows the $4$th plaquette due to which these four plaquettes have been marked using the same color (gray) in the extreme left panel of Fig.~\ref{fig:30S10times2} (b). The elementary plaquettes marked in pink in the extreme left panel of Fig.~\ref{fig:30S10times2} (b) show a different type of dynamics for $\langle \Opots (t) \rangle$. The $3$rd and $14$th plaquettes, which show similar temporal behavior to each other, oscillate very differently (shown in cyan and orange, respectively) compared to the plaquettes marked in gray. Remarkably, the $8$th, $17$th and $19$th plaquettes show similar oscillations (shown in magenta, lime-green and blue, respectively) to the $3$rd and $14$th plaquettes while being not directly connected to these plaquettes through edges or vertices 
The only difference between the temporal behavior of these three plaquettes compared to the $3$rd and $14$th plaquette is that the amplitude is more than double for the former case compared to the latter and the former's temporal behavior has certain rapid oscillations superposed on the slow ones. The $18$th plaquette oscillates out of phase (shown in brown) to the $8$th, $17$th and $19$th plaquette but lacks the rapid oscillations. The temporal behavior of the plaquettes marked in pink and gray are clearly interrelated (i.e., involve some common frequencies) as can be seen visually from the bottom left panel of Fig.~\ref{fig:30S10times2} (b) where the oscillations of $\Opots$ for the $5$th and $18$th plaquettes are plotted together.

%
\begin{figure}
    \centering
    \includegraphics[scale=0.25]{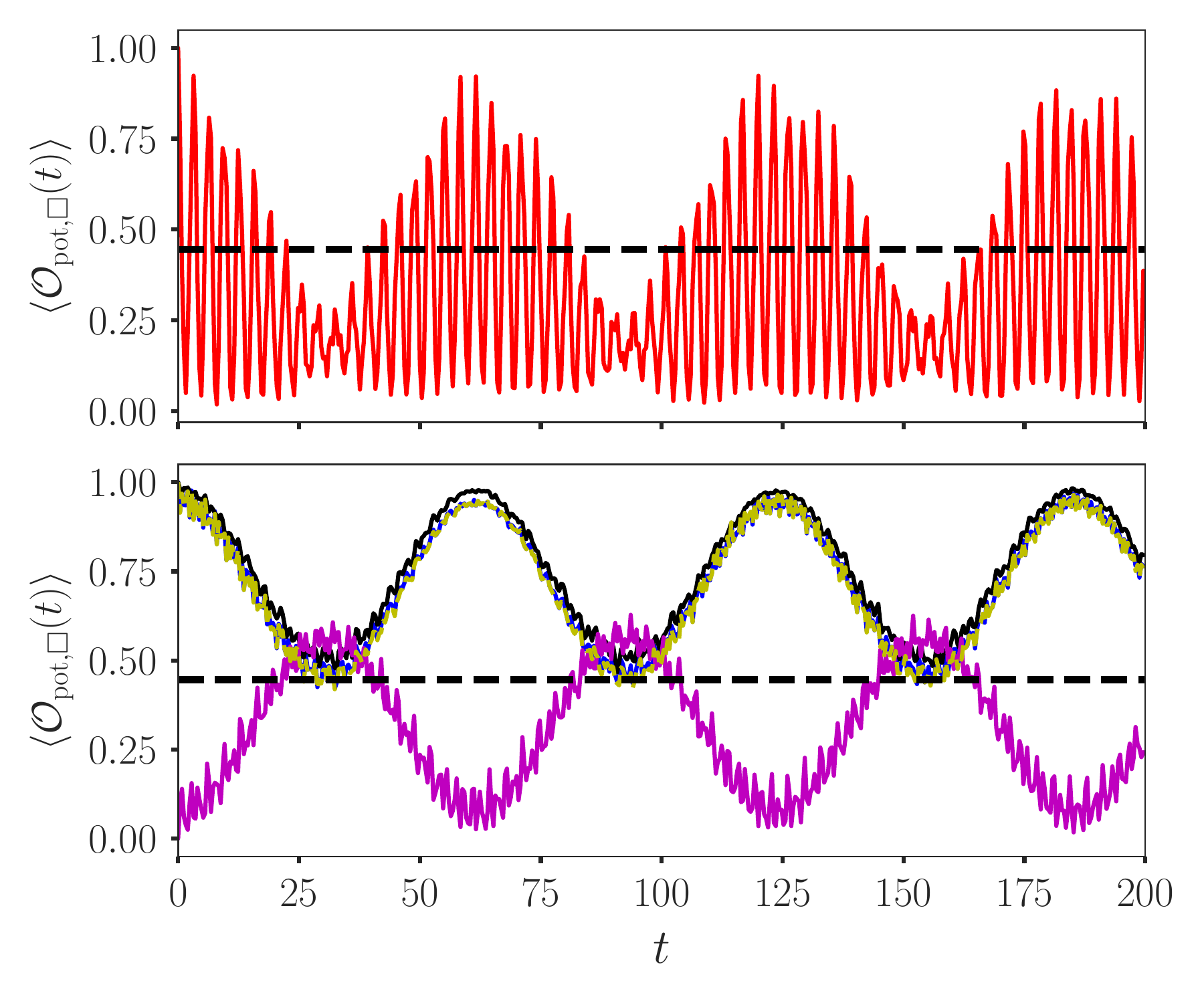}
    \includegraphics[scale=0.18]{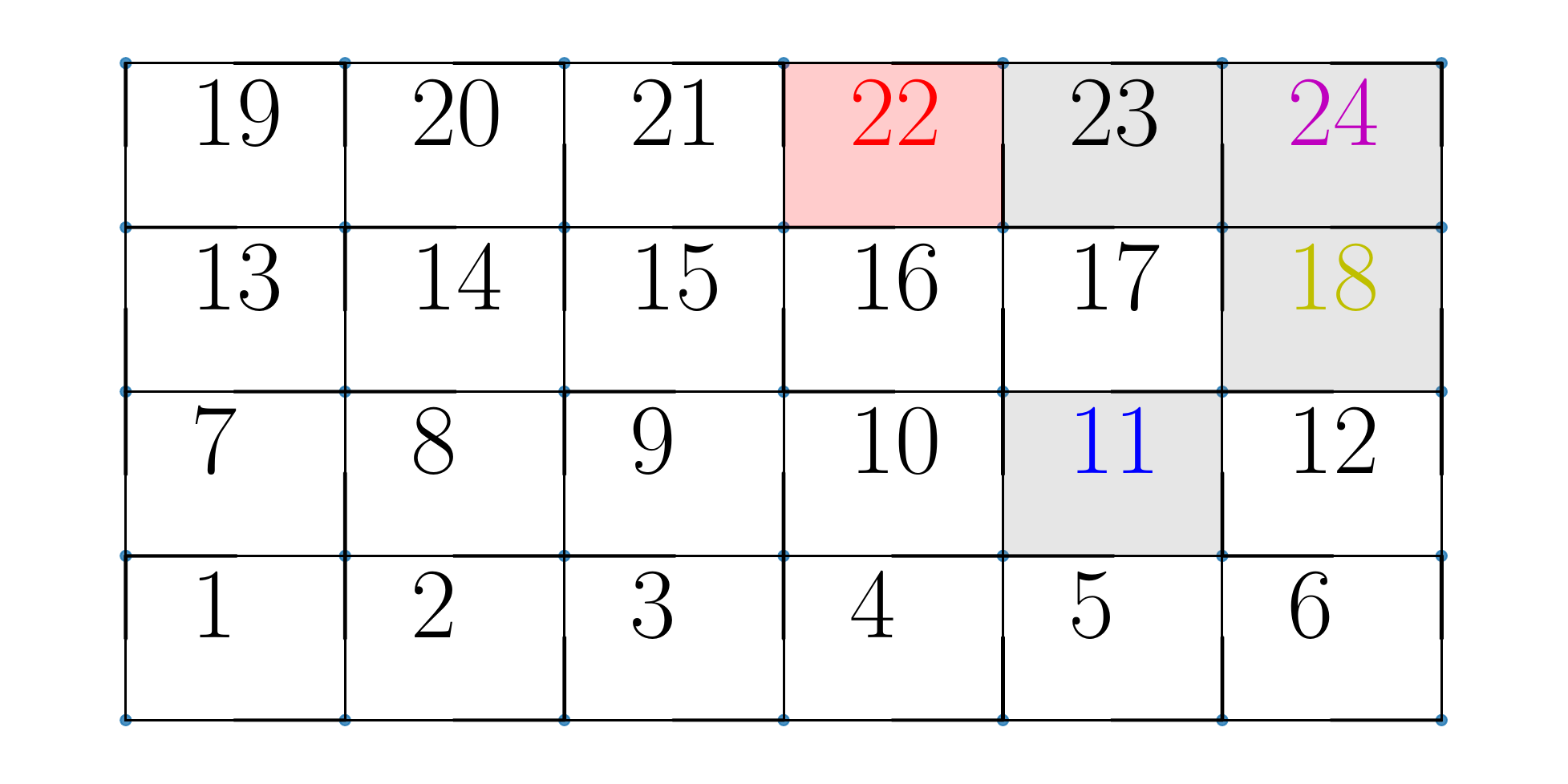}
    \caption{Behavior of $\Opots (t)$ on individual plaquettes for a Fock state that shows oscillatory dynamics in the spatially averaged autocorrelation function for a $6 \times 4$ lattice with $\alpha=30$. Only elementary plaquettes that display oscillations in $\Opots (t)$ shown for clarity. The top and middle panels display the behavior of $\langle \Opots (t) \rangle$ on different plaquettes. The bottom panel displays the real space location of the plaquette.  While plaquettes shaded in the same color essentially have the same temporal behavior (while being in or out of phase), plaquettes marked in gray and pink in the panel showed different temporal behaviors, which are nonetheless strongly correlated to each other. In the bottom panel, the color of the number in the center of the shaded plaquette that indicates its location on the lattice is chosen to be the same as used in the panels to display the temporal behavior of $\langle \Opots \rangle$ on the corresponding plaquette.}
    \label{fig:30S6times4}
\end{figure}

A similar feature where different parts of a connected region of elementary clusters oscillate differently but in a related manner is also evident when the behavior of individual plaquettes is considered for a $6 \times 4$ lattice for $\alpha=30$. For example, this can be clearly seen from the $\langle \Opots (t) \rangle$ behavior of individual plaquettes from a randomly sampled Fock state that displays an oscillatory spatially averaged autocorrelator as shown in Fig.~\ref{fig:30S6times4}. Here, the plaquettes that show significant oscillations form a connected region where four of the plaquettes oscillate in a similar fashion ($11$th, $18$th, $23$rd, and $24$th plaquette marked in gray in bottom panel of Fig.~\ref{fig:30S6times4}) while the fifth one ($22$nd plaquette marked in pink in bottom panel of Fig.~\ref{fig:30S6times4}) follows a different, but related, temporal dynamics. While the $11$th, $18$th and $23$rd plaquette oscillate in a predoninantly sinusoidal fashion and are in phase with each other, the $24$th plaquette oscillates out of phase to these plaquettes (middle panel of Fig.~\ref{fig:30S6times4}). The $22$nd plaquette shows a rather different temporal behavior with fast oscillations, which are absent in the previous four plaquettes, superposed on a slow oscilation that exactly follows the oscillations of the previous four plaquettes (top panel of Fig.~\ref{fig:30S6times4}). 

The oscillatory features in the autocorrelation functions for the diagonal operators at intermediate and strong disorder discussed here for a particular disorder realization in $10 \times 2$ and $6 \times 4$ ladders is also present in other independently chosen disorder realizations. While there are $3$ Fock states from the $50$ randomly sampled 
Fock states with an oscillatory spatially averaged autocorrelation function for the particular disorder realization shown in Sec.~\ref{subsec:dynFock} for a $10 \times 2$ ladder at disorder strength $\alpha=12$, this number changes to $1$ and $8$ respectively for two other independent disorder realizations of $R_\square$. The fraction of Fock states with oscillating average autocorrelations becomes much more significant when the disorder is increased to $\alpha=30$. While we obtained $23$ such Fock states from the randomly chosen $50$ Fock states for the particular disorder realization used in Sec.~\ref{subsec:dynFock} for a $10 \times 2$ ladder at $\alpha=30$, this number changes to $25$ and $33$ respectively for the two other independent disorder realizations of $R_\square$. Similarly, while there are $14$ Fock states from the $50$ randomly sampled Fock states with an oscillatory spatially averaged autocorrelation function for the particular disorder realization shown in Sec.~\ref{subsec:dynFock} for a $6 \times 4$ ladder at disorder strength $\alpha=30$, this number changes to $6$ and $7$ respectively for two other independent disorder realizations. This illustrates the robustness of this phenomenon especially at higher disorder strengths.

The presence of such randomly sampled Fock states which have regions of elementary plaquettes that display rich oscillatory behavior as a function of time seems to be a direct consequence of mid-spectrum eigenstates with small active regions embedded in large inert regions that generate nonzero weight for large $\mathcal{D}$ (close to $1/4$) in $p(\mathcal{D})$ for a given system size and $\alpha$ (e.g. see Fig.~\ref{fig:opotspAlph100_12x2} and Fig.~\ref{fig:opotspAlph100_6x4} for reference). Fock states that can be decomposed into a superposition of predominantly such mid-spectrum eigenstates would naturally show regions in real space where the dynamics of $\Opots (t)$ can be described by using only a few frequencies. While the inset in the top panel of Fig.~\ref{fig:pDsmalla} shows that
  the weight of $p(\mathcal{D})$ is negligible for $\mathcal{D} \approx 1/4$ for
  $\alpha=6$, the inset in the bottom panel of Fig.~\ref{fig:pDsmalla} shows a long tail that gives some contribution to $p(\mathcal{D})$ for $\mathcal{D} \approx 1/4$ for
  $\alpha=12$ for a $10 \times 2$ ladder. This weight increases significantly as one approaches $\alpha=30$ for the same system size (Fig.~\ref{fig:pDintermediatea}, top panel) which explains why the fraction of randomly sampled Fock states with oscillatory spatially averaged autocorrelators increase significantly from $\alpha=12$ to $\alpha=30$. The same is also true for the $p(\mathcal{D})$ of a $6 \times 4$ ladder at $\alpha=30$  (Fig.~\ref{fig:pDintermediatea}, top panel) but the weight in $\mathcal{D} \approx 1/4$ is now reduced compared to the $10 \times 2$ ladder at the same $\alpha$ which is consistent with there being a lesser fraction of randomly sampled Fock states with oscillatory spatially averaged autocorrelations for the three independent disorder realizations we considered for a $6 \times 4$ ladder when compared to a $10 \times 2$ ladder. The intruiging in phase and out of phase temporal oscillations of connected elementary plaquettes in such regions from exact numerics seems to be intimately related to the constrained nature of the Hilbert space as demonstrated by some toy calculations where the presence of small active regions is imply modelled by a two or a three plaquette system governed by $\mathcal{H}_{\rm dis}$ (Eq.~\ref{eq:Hran}) (see Appendix~\ref{app}) and the backreaction from the rest of the inert region is completely ignored which should also renormalize the bare parameters in $\mathcal{H}_{\rm dis}$ in the active regions.

\section{Conclusions and outlook}
\label{sec:con}
In conclusion, we have considered a $U(1)$ quantum link gauge theory Hamiltonian in its $S=1/2$ representation on $L_x \times L_y$ ladders, where both $L_x$ and $L_y$ are taken to be even, with periodic boundary conditions in both directions. This allows us to target the largest superselection sector of such a theory with zero charge at each site and zero winding of electric fluxes in both directions. The Hamiltonian $\mathcal{H}_{\mathrm{dis}}$ (Eq.~\ref{eq:Hran}) is composed of plaquette operators, $\Okins$ and $\Opots$, that are defined on the elementary plaquettes of the lattice. While $\Okins$ is off-diagonal in the electric flux basis and changes a clockwise circulation of electric fluxes on a plaquette to anticlockwise and vice versa, $\Opots$ is diagonal and counts whether a plaquette is flippable. We introduce a disorder field that couples linearly to $\Opots$ and parameterize the strength of the disorder by a dimensionless number $\alpha$ where $\alpha=0$ ($\alpha \rightarrow \infty$) represents no (infinite) disorder. We study the properties of mid-spectrum energy eigenstates of this disordered lattice gauge theory to understand whether such a system exhibits a many-body localized phase or not, both on thin ladders with $L_y=2$ and wider ladders with $L_y=4$, using exact diagonalization techniques. While this specific model is known to be non-integrable for weak disorder, the nature of the mid-spectrum eigenstates for larger disorders has not been explored previously.

In this work, in addition to using standard diagnostics such as level spacing distributions, we introduce an intensive estimator, $\mathcal{D}=(1/N_p) \sum_\square \mathcal{D}_\square \in [0,1/4]$, whose normalized probability distribution $p(\mathcal{D})$ is calculated for mid-spectrum eigenstates using many disorder realizations for a given disorder strength $\alpha$ and ladder dimension. This estimator serves the dual purpose of quantifying how localized a mid-spectrum eigenstate is in Fock space (defined by the electric flux Fock states) and estimating the fraction of elementary plaquettes in an active (thermal) or an inactive (inert) state. This is because while $\langle \mathcal{D}_\square \rangle= (\langle \Opots \rangle-1/2)^2$ is $1/4$ for plaquettes in perfectly localized (in Fock space) electric flux Fock states that become eigenstates at infinite disorder, its infinite-temperature value, $\mathcal{D}_{\mathrm{th}} \ll 1/4$, from explicit calculations which should hold for delocalized (in Fock space) mid-spectrum eigenstates from the eigenstate thermalization hypothesis. The distribution $p(\mathcal{D})$ has a pronounced maximum near $0$ ($1/4$) for small (large) $\alpha$ with a tail whose weight decreases at large (small) $\mathcal{D}$. The analysis of the finite-size behavior of the skewness of $p(\mathcal{D})$ with increasing $L_x$ (for $L_x \leq 14$) for thin ladders with $L_y=2$ allows us to estimate $\alpha_c(L_y=2)=31.04 (0.54)$
where $\alpha_c(L_y)$ is the critical disorder strength for many-body localization for a ladder of fixed width $L_y$ and $L_x \rightarrow \infty$.
The behavior of level statistics indicators, as well as $p(\mathcal{D})$ for a wider $6 \times 4$ ladder, as a function of disorder indicates the weaker tendency of such ladders to localize compared to thin ladders and strongly suggests that $\alpha_c(L_y=4) > \alpha_c(L_y=2)$. This is in line with the prediction that a many-body localized phase is unstable in short-ranged interacting models for dimensions above one~\cite{MBLavalanche1}.

We further probe the local autocorrelation function of the diagonal operators $\Opots$ on elementary plaquettes, starting from randomly sampled typical Fock states whose average energies lie in the bin (of $4\%$ of the total bandwidth) that contains the highest number of energy eigenstates, as well as the infinite temperature autocorrelation that represents the average over the autocorrelations of all the Fock states. While the infinite temperature autocorrelation remains monotonically decaying in time for both weak and strong disorder (but still below $\alpha_c(L_y)$) and is rather featureless, resolving it into autocorrelations of individual Fock states shows increasing dynamic heterogeneity with increasing disorder. Since $\alpha_c(L_y)$ is a large disorder strength for this disordered $U(1)$ quantum link model, it provides an opportunity to study dynamical relaxations for a range of disorder strengths even before a many-body localization may set in. A particularly striking feature is the presence of a fraction of Fock states, obtained from random sampling of such states where the time variation of $\Opots$ on some elementary plaquettes that form connected spatial regions shows a regular oscillatory behavior that is dominated by only a few frequencies. While the probability of encountering such Fock states from a random sampling is found to be small but non-negligible at $\alpha=12$ for thin ladders for different disorder realizations, it becomes much more significant for a larger disorder strength of $\alpha=30$ both for thin and wider ladders. For $\alpha=30$, these oscillations show an emergence of a plethora of time scales even in a single disorder realization which the infinite temperature autocorrelation fails to pick up. Furthermore, different plaquettes in these spatial regions either oscillate in phase or out of phase with respect to each other or have some common frequencies despite displaying different temporal dynamics. The presence of such dynamical behavior seems to be correlated to the weight in $p(\mathcal{D})$ near $\mathcal{D}=1/4$ for a given ladder dimension and disorder strength since that implies the presence of mid-spectrum eigenstates with small active regions embedded in an otherwise inert background. The in phase/out of phase behavior of $\langle \Opots (t) \rangle$ for different plaquettes in these spatial regions seems to be a direct consequence of the constrained nature of the Hilbert space. To the best of our knowledge, such dynamical behavior was not pointed out before in local operators for a strongly disordered system. 

Our study opens up several issues for further exploration. First, whether one-dimensional and two-dimensional models with constrained  Hilbert spaces admit a many-body localized phase is still not understood completely. Our study gives a precise estimate of $\alpha_c(L_y=2)$ based on the skewness of $p(\mathcal{D})$ and also shows that $\alpha_c(L_y=4)>\alpha_c(L_y=2)$.
However, a level spacing estimator shows strong finite-size dependence and an almost linear increase with $L_x$ for $L_x \leq 14$ with the values being below $\alpha_c(L_y=2)$ estimated from the previous approach. It will be important to consider $L_x>14$ to see if both approaches eventually give the same $\alpha_c(L_y=2)$ but this is beyond the scope of the current study. We further show a substantial probability for the presence of thermal regions in the mid-spectrum eigenstates of the system even for ladders with the largest number of elementary plaquettes, i.e., ladders $12 \times 2$ and $6 \times 4$, with disorder strength (without dimensions) as large as $\alpha=100$. Whether this indicates a {\it strongly fluctuating} many-body localized phase even at such large disorders is not clear to us and requires a study on larger systems. 
While we have used computational techniques that access mid-spectrum eigenstates in bigger systems without the need of a diagonalization of the full Hilbert space~\cite{SierantED2020} to calculate $p(\mathcal{D})$ for $14 \times 2$ ladders, it will be instructive to do the same for wider ladders with $L_y=4$ and $L_y=6$. Accessing real space correlations in the local estimators, $\mathcal{D}_\square$, for mid-spectrum eigenstates will be useful to understand the statistics of the distribution of active and inert regions as a function of disorder strength. The possibility of the presence of randomly sampled Fock states, which become statistically more significant for larger disorder, where certain simple local diagonal operators show coherent oscillations in time even in the thermalizing regime of the disordered $U(1)$ quantum link model should be investigated more systematically for other kinematically-constrained systems such as disordered PXP chains and quantum dimer models. A deeper understanding of this intriguing dynamical phenomenon is highly desirable.

In the literature, there has been considerable interest in the experimental realization of this model, particularly regarding the translation-invariant version of the model, where observing string breaking and string dynamics is of particular demand. Experimental realization has been proposed on superconducting qubits \cite{Marcos:2014lda}, while elementary simulations on IBMQ machines have been reported in \cite{Huffman:2021gsi}.
 Proposals for realizing these models on hybrid architectures have been put forward in \cite{Crane:2024tlj},
 and in Rydberg atom systems in \cite{Celi2020}. Solid state experiments to realize artificial spin-ice as a means to
 realize these models have also been suggested in \cite{Skjaervo2019,Zhang2021StringPhase,Sultana2025}.

\begin{acknowledgments}
A.S. acknowledges a useful discussion with Sthitadhi Roy as well as discussions with the participants of the \emph{Topical School of Advanced Condensed Matter Physics} at the Institute of Physics, Bhubaneswar during the writing of this manuscript. We thank the computational resources of the Saha Institute of Nuclear Physics and the Indian Association for the Cultivation of Science, where the bulk of the computation was carried out. We thank the Iridis research computing facility  of the University of Southampton (UK) for the high-memory computer nodes necessary for the $L_x = 14, L_y=2$ systems. 
\end{acknowledgments}

\appendix

\section{Shift-Invert ED and selection of mid-spectrum states for the $14\times2$ lattice}
\label{app:midspectrum}
 As is well-known, ED methods are exponentally expensive and their application becomes challenging as the system size increases. For our case, going to the $14\times2$ lattice is exactly such a challenge: the memory needed to diagonalize a $N = 616227$ exceeds memory available of most medium-sized computing clusters. Consequently, one has to resort to other methods. The Lanczos algorithm is the most straightforward way to attack larger system sizes, but it is typically yields (a specified number of) eigenvalues (and eigenvectors) at the edges of the spectrum. A considerably powerful method is the shift-invert diagonalization, which has been successfully used before in the literature to study localization properties of spin-chains \cite{Pietracaprina2018}. 

\begin{figure}
    \centering
    \includegraphics[width=\hsize]{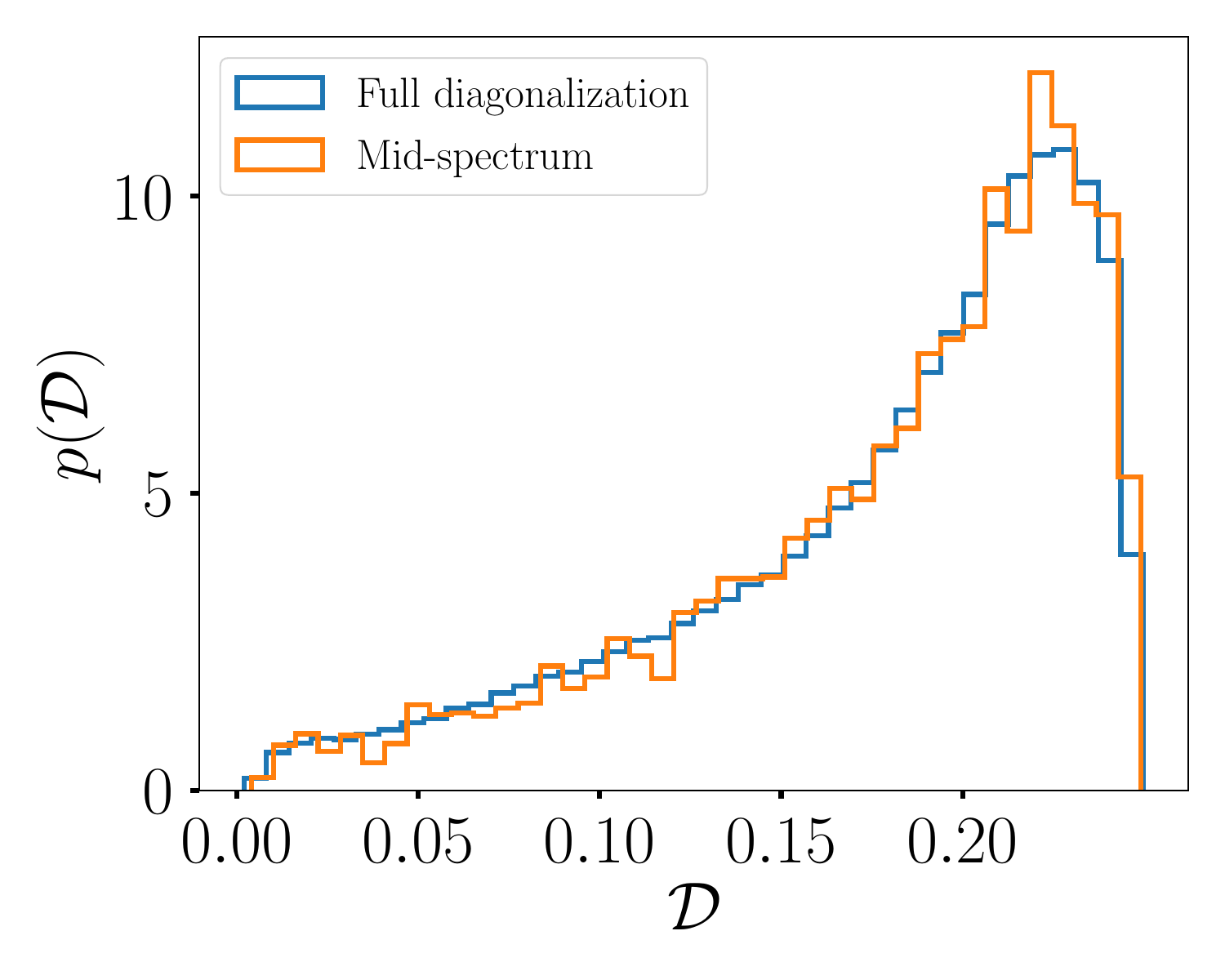}
    \caption{Comparison of $p(\mathcal{D})$ for $12\times2$ lattice at $\alpha=40$ obtained using two different methods, full diagonalization and randomly chosen eigenstates from mid spectrum. $30$ disorder realizations have been used to obtain $p(\mathcal{D})$ in both cases. (For $12\times2$ lattice, we have used $20$ eigenstates around each of the total ten target energies to make the histogram.)}
    \label{fig:EdvsSlepc}
\end{figure}

 We have used a SLEPc suite to implement paralleized shift-invert diagonalization \cite{Hernandez:2005:SSF,Hernandez:2007:PAE}, 
 for the $14 \times 2$ system using which it is relatively easy to find a few ($\sim 100$) eigenstates, around a specified energy from the mid-spectrum region. Our procedure involves the specification of $E_{\rm{min}}$ ($E_{\rm{max}}$) as the minimum (maximum) energy. We divide the spectrum into twenty-five bins of equal width, $w=(E_{\textrm{max}}-E_{\rm{min}})/25$. We randomly choose ten target energies from the middle bin around which we want to find a few eigenstates. Using these eigenstates, we found the distribution $p(\mathcal{D})$ and level spacing statistics. In \cref{fig:EdvsSlepc}, we have shown a comparison of the shift-invert method with the full ED on the smaller $L_x = 12$ lattice to confirm that the results using the shift-invert technique can be taken seriously.

 While the shift-invert method makes the study of the $L_x=14$ at all possible, it nevertheless requires steep resources, especially of computer memory. Our typical runs used approximately $\sim 700$ GB for each instance to compute around $80-100$ eigenstates around a target energy value, and each instance took around $90-120$ mins to find the eigenstates using 12 nodes in parallel. We have varied the number of disorder realizations according to the disorder strength --- for runs deep in the ETH phase ($\alpha \le 20 $) and in the localized phase ($\alpha \ge 60$), we have used 10 disorder realizations, while in the transition region ($\alpha \sim 30 $) we have used 20 or more disorder realizations. 

\section{A toy calculation for a two-plaquette and a three-plaquette active region}
\label{app}
A feature that emerges from exact numerics from initial Fock states that show oscillatory behaviour of the spatially averaged autocorrelation function of $\Opots$ is the in phase/out of phase behavior of $\langle \Opots (t) \rangle$ for the subset of plaquettes that essentially shows similar temporal evolution of oscillations, especially at larger disorder strength of $\alpha=30$ (see Fig.~\ref{fig:30S10times2} and Fig.~\ref{fig:30S6times4}). While we do not have a detailed understanding of this, simple calculations from particular Fock states using $\mathcal{H}_{\rm dis}$ (Eq.~\ref{eq:Hran}) on a two-plaquette and a three-plaquette system seems to bring out similar features for $\langle \Opots  (t) \rangle$ as we show below, which suggests that this may be related to the nature of the constrained Hilbert space for this model.  
\begin{figure}
    \centering
    \includegraphics[width=0.8\hsize]{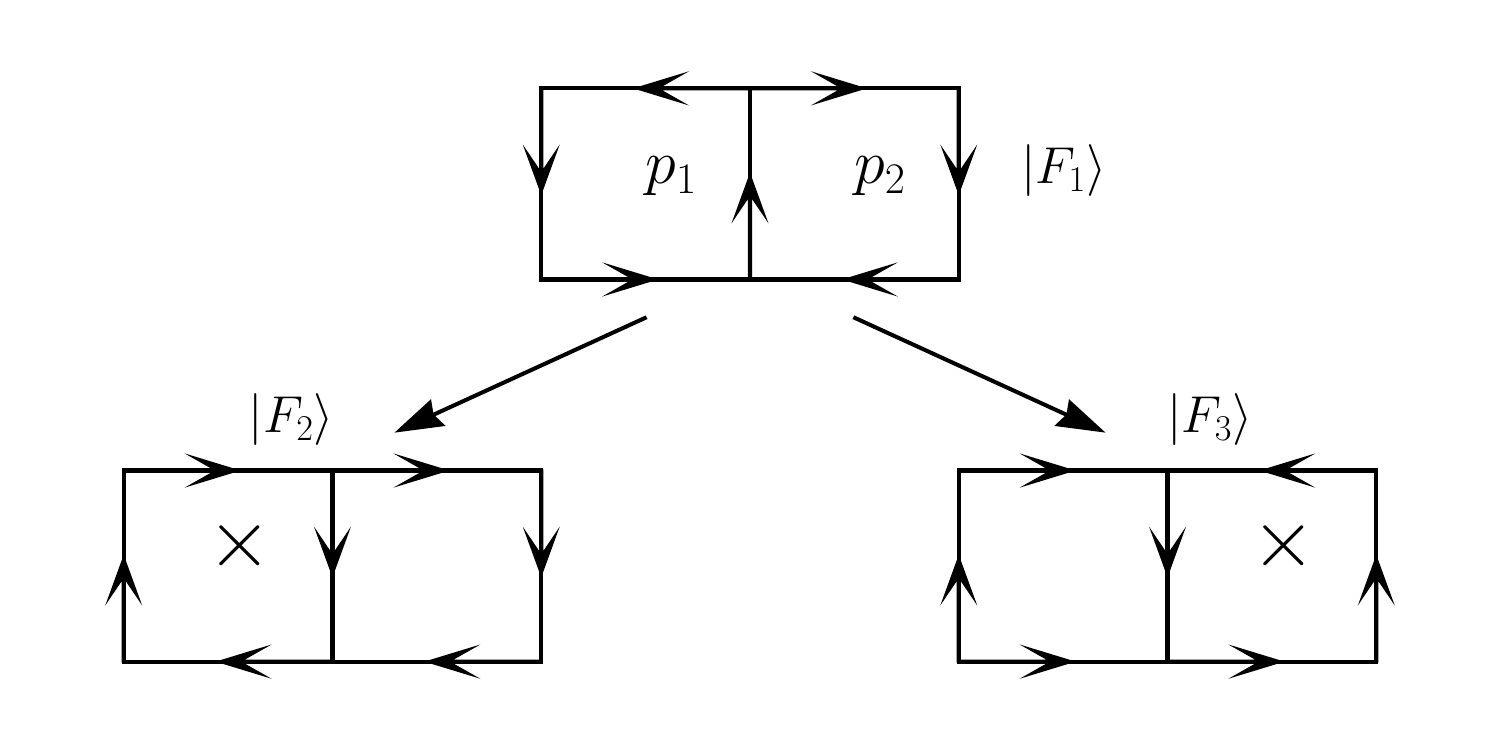}
    \caption{Action of $\mathcal{H}_{\rm dis}$ (Eq.~\ref{eq:Hran}) shown schematically for a
    system with two elementary plaquettes ($p_1$ and $p_2$) that share an edge with the electrix fluxes being shown as arrows. The cross in the center of a plaquette denotes the action of $\Okins$ on that plaquette to generate the corresponding Fock state from $|F_1\rangle$.  }
    \label{app:threestategraph}
\end{figure}  
\begin{figure}
    \centering
    \includegraphics[width=\hsize]{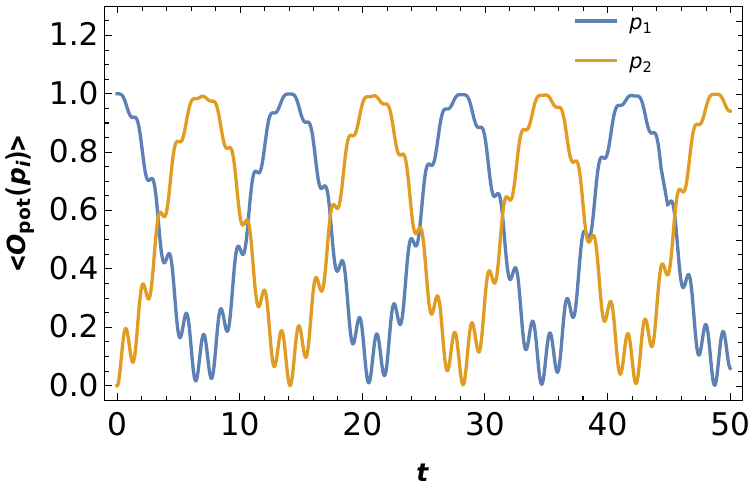}
    \caption{$\langle \Opot (p_i) (t)\rangle$ shown for a system of two elementary plaquettes ($p_1$ and $p_2$) that share an edge starting from the initial state $|F_2\rangle$ (see Fig.~\ref{app:threestategraph}). The parameters used here for $\mathcal{H}_{\rm dis}$ (Eq.~\ref{eq:Hran}) are $\alpha=30$ and $r_1=r_2=0.1$.}
    \label{app:threestatedyn}
\end{figure}  

Let us first consider a two-plaquette system where the plaquettes $p_1$ and $p_2$ share an edge with each other (Fig.~\ref{app:threestategraph}) and start with a Fock state $|F_1\rangle$ such that both the plaquettes have a flippable configuration of the electric fluxes (Fig.~\ref{app:threestategraph}). It can then be easily seen that the repeated action of  $\mathcal{H}_{\rm dis}$ (Eq.~\ref{eq:Hran}) on $|F_1\rangle$ can only lead to two other Fock states $|F_2\rangle$ and
$|F_3\rangle$ (Fig.~\ref{app:threestategraph}). Writing $\mathcal{H}_{\rm dis}$ (Eq.~\ref{eq:Hran}) in the basis of $|F_i\rangle$ where $i=1,2,3$ then leads to the following $3 \times 3$ matrix:
\begin{eqnarray}
  \begin{bmatrix}
-2-\alpha r_1 -\alpha r_2 & -1 & -1 \\
-1 & -1-\alpha r_1 & 0 \\
-1 & 0  & -1-\alpha r_2
\label{app:threestatematrix}
\end{bmatrix}
  \end{eqnarray}
where $r_1$ and $r_2$ refer to the values of the random numbers $R_\square$ on the plaquettes $p_1$ and $p_2$ respectively.

Choosing $\alpha=30$, we find that selecting the random numbers $r_1 \approx r_2$ and starting from the Fock state $|F_2 \rangle$ (or equivalently, $|F_3\rangle$) (Fig.~\ref{app:threestategraph}) leads to oscillations in $\langle \Opot (p_i) (t)\rangle$ as can be seen from Fig.~\ref{app:threestatedyn} where $r_1=r_2=0.1$ such that the two diagonal operator $\Opots$ oscillates out of phase with respect to ecah other on the two plaquettes. Deviating sufficiently from $r_1 = r_2$ leads to significantly weaker oscillations in $\Opots$ as can be seen by putting $r_1=0.1$ and $r_2=0.15$ and again starting with $|F_2\rangle$. 

\begin{figure}
    \centering
    \includegraphics[width=0.8\hsize]{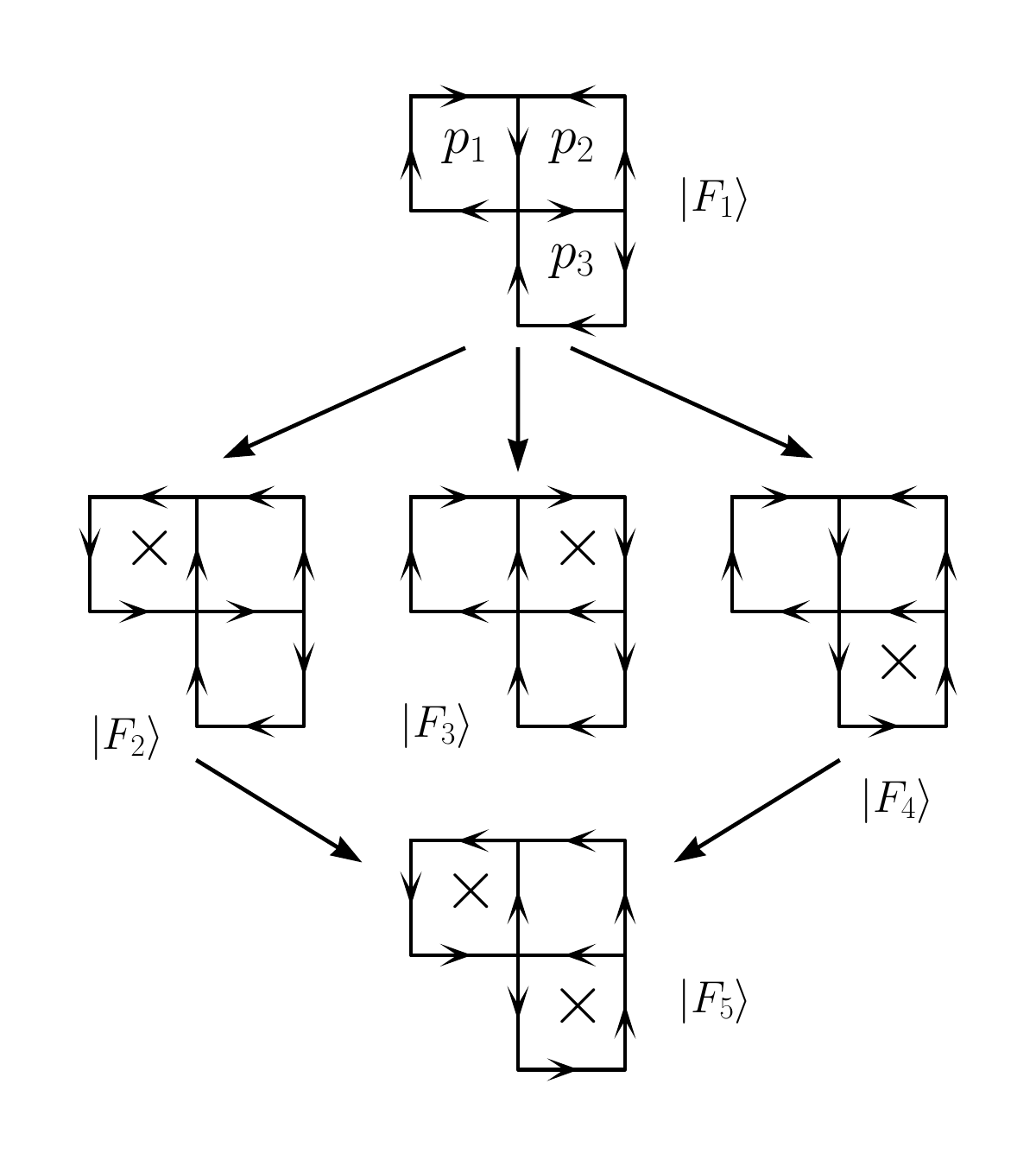}
    \caption{Action of $\mathcal{H}_{\rm dis}$ (Eq.~\ref{eq:Hran}) shown schematically for a
    system with three elementary plaquettes ($p_1$, $p_2$ and $p_3$) where $p_1$ and $p_3$ share a single edge with $p_2$ with the electrix fluxes being shown as arrows. The cross in the center of a plaquette denotes the action of $\Okins$ on that plaquette to generate the corresponding Fock state from $|F_1\rangle$. }
    \label{app:fivestategraph}
\end{figure}  
\begin{figure}
    \centering
    \includegraphics[width=\hsize]{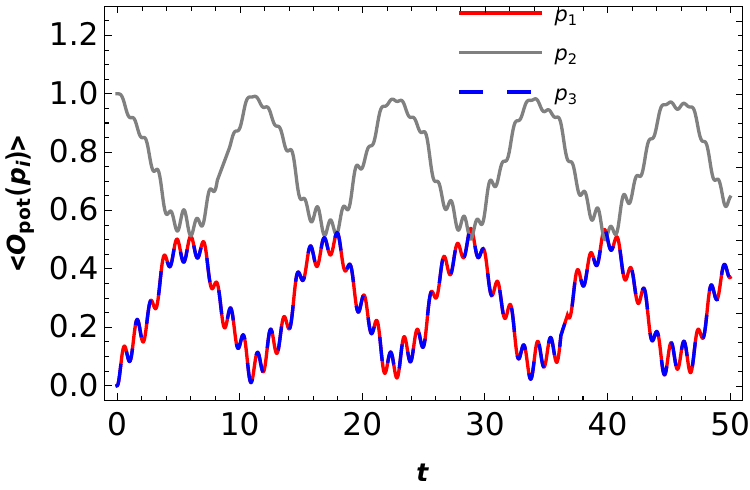}
    \caption{$\langle \Opot (p_i) (t)\rangle$ shown for a system of three elementary plaquettes ($p_1$, $p_2$ and $p_3$) where $p_1$ and $p_3$ share a single edge with $p_2$ starting from the initial state $|F_3\rangle$ (see Fig.~\ref{app:fivestategraph}). The parameters used here for $\mathcal{H}_{\rm dis}$ (Eq.~\ref{eq:Hran}) are $\alpha=30$, $r_1=r_3=0.05$ and $r_2=0.1$.}
    \label{app:fivestatedyn}
\end{figure}  

One can then ask whether it is possible to generate both in phase and out of phase dynamics for the diagonal operator $\Opots$ on different plaquettes as seen in exact numerics. Fortunately, it turns out to be the case if one considers a three-plaquette system instead of a two-plaquette one, where the plaquettes $p_1$ and $p_3$ share a single edge with the plaquette $p_2$ (Fig.~\ref{app:fivestategraph}).  We again start with a Fock state $|F_1\rangle$ such that all the three plaquettes have a flippable configuration of the electric fluxes (Fig.~\ref{app:fivestategraph}). It can then be easily seen that the repeated action of  $\mathcal{H}_{\rm dis}$ (Eq.~\ref{eq:Hran}) on $|F_1\rangle$ can only lead to four other distinct Fock states (Fig.~\ref{app:fivestategraph}). Writing $\mathcal{H}_{\rm dis}$ (Eq.~\ref{eq:Hran}) in the basis of $|F_i\rangle$ where $i=1,2,3,4,5$ then leads to the following $5 \times 5$ matrix:
\begin{widetext}
\begin{eqnarray}
  \begin{bmatrix}
-3-\alpha (r_1+r_2+r_3) & -1 & -1 & -1 & 0 \\
-1 & -2-\alpha (r_1+r_3) & 0 & 0 & -1\\
-1 & 0  & -1-\alpha r_2 &0 &0 \\
-1 & 0 & 0 & -2-\alpha(r_1+r_3) &-1 \\
0 & -1 & 0 & -1 & -2-\alpha(r_1+r_3)
\label{app:fivestatematrix}
\end{bmatrix}
\end{eqnarray}
\end{widetext}
where $r_1$, $r_2$ and $r_3$ refer to the values of the random numbers $R_\square$ on the plaquettes $p_1$, $p_2$ and $p_3$ respectively.

Choosing $\alpha=30$, we find that selecting the random numbers $r_1 + r_3 \approx r_2$ and starting from the Fock state $|F_3 \rangle$ (Fig.~\ref{app:fivestategraph}) leads to oscillations in $\langle \Opot (p_i) (t)\rangle$ as can be seen from Fig.~\ref{app:fivestatedyn} where $r_1=r_3=0.05$ and $r_2=0.1$ such that the two diagonal operator $\Opots$ oscillate in phase with each other on the plaquettes $p_1$ and $p_3$ and out of phase with respect to $p_1$ and $p_3$ on the plaquette $p_2$ that shares an edge both with $p_1$ and $p_3$ (Fig.~\ref{app:fivestatedyn}). Again, deviating rom $r_1 + r_3=r_2$ leads to significantly weaker oscillations in $\Opots$ as can be seen by putting $r_1=r_3=0.05$ and $r_2=0.15$ and again starting with $|F_3\rangle$.

Interestingly, similar features regarding in phase and out of phase oscillations on different active plaquettes are visible at $\alpha=30$ in exact numerics if we identify subunits that have the same arrangement of plaquettes as our toy three-plaquette unit (Fig.~\ref{app:fivestategraph}). E.g., in Fig.~\ref{fig:30S6times4} for a $6 \times 4$ ladder, we see that $\Opots$ in the $24$th plaquette oscillates out of phase with respect to both the $23$rd and the $18$th plaquette which oscillate in phase with each other. Both the $18$th plaquette and the $23$rd plaquette share exactly one edge with the $24$th plaquette just as our toy three-plaquette unit. Similarly, in Fig.~\ref{fig:30S10times2} (b) for a $10 \times 2$ ladder, we see that $\Opots$ in the $4$th, $15$th and $6$th ($17$th, $8$th and $19$th) plaquettes oscillates in phase with respect to each other and out of phase with respect to the $5$th ($18$th) plaquette (see extreme left panel in Fig.~\ref{fig:30S10times2} (b) for the location of these plaquettes).

\bibliography{refs}
\end{document}